\documentclass[3p, review]{elsarticle}

\usepackage{lineno,hyperref}

\usepackage{subcaption}
\usepackage{booktabs}
\usepackage{multirow}

\graphicspath{{./img/}}

\hypersetup{pdfauthor=author}

\journal{Computer Methods and Programs in Biomedicine}

\begin{document}

\begin{frontmatter}

\title{Weakly-supervised detection of AMD-related lesions in color fundus images using explainable deep learning}


\author[citic,varpa-inibic]{José Morano\corref{correspondingauthor}}
\cortext[correspondingauthor]{Corresponding author}
\ead{j.morano@udc.es}

\author[citic,varpa-inibic]{Álvaro S. Hervella}
\ead{a.suarezh@udc.es}

\author[citic,varpa-inibic]{José Rouco}
\ead{jrouco@udc.es}

\author[citic,varpa-inibic]{Jorge Novo}
\ead{jnovo@udc.es}

\author[san-carlos,CIOA]{José I. Fernández-Vigo}
\ead{jfvigo@hotmail.com}

\author[citic,varpa-inibic]{Marcos Ortega}
\ead{mortega@udc.es}

\address[citic]{Centro de Investigación CITIC, Universidade da Coruña, A Coruña, Spain}
\address[varpa-inibic]{VARPA Research Group, Instituto de Investigación Biomédica de A Coruña (INIBIC), Universidade da Coruña, A Coruña, Spain}
\address[san-carlos]{Department of Ophthalmology, Hospital Clínico San Carlos, Instituto de Investigación Sanitaria (IdISSC), Madrid, Spain}
\address[CIOA]{Department of Ophthalmology, Centro Internacional de Oftalmología Avanzada, Madrid, Spain}

\begin{abstract}
\paragraph{Background and Objectives} Age-related macular degeneration (AMD) is a degenerative disorder affecting the macula, a key area of the retina for visual acuity.
Nowadays, AMD is the most frequent cause of blindness in developed countries.
Although some promising treatments have been proposed that effectively slow down its development, their effectiveness significantly diminishes in the advanced stages.
This emphasizes the importance of large-scale screening programs for early detection.
Nevertheless, implementing such programs for a disease like AMD is usually unfeasible, since the population at risk is large and the diagnosis is challenging.
For the characterization of the disease, clinicians have to identify and localize certain retinal lesions.
All this motivates the development of automatic diagnostic methods.
In this sense, several works have achieved highly positive results for AMD detection using convolutional neural networks (CNNs).
However, none of them incorporates explainability mechanisms linking the diagnosis to its related lesions to help clinicians to better understand the decisions of the models.
This is specially relevant, since the absence of such mechanisms limits the application of automatic methods in the clinical practice.
In that regard, we propose an explainable deep learning approach for the diagnosis of AMD via the joint identification of its associated retinal lesions.

\paragraph{Methods}
In our proposal, a CNN with a custom architectural setting is trained end-to-end for the joint identification of AMD and its associated retinal lesions.
With the proposed setting, the lesion identification is directly derived from independent lesion activation maps; then, the diagnosis is obtained from the identified lesions.
The training is performed end-to-end using image-level labels.
Thus, lesion-specific activation maps are learned in a weakly-supervised manner.
The provided lesion information is of high clinical interest, as it allows clinicians to assess the developmental stage of the disease.
Additionally, the proposed approach allows to explain the diagnosis obtained by the models directly from the identified lesions and their corresponding activation maps.
The training data necessary for the approach can be obtained without much extra work on the part of clinicians, since the lesion information is habitually present in medical records.
This is an important advantage over other methods, including fully-supervised lesion segmentation methods, which require pixel-level labels whose acquisition is arduous.

\paragraph{Results}
The experiments conducted in 4 different datasets demonstrate that the proposed approach is able to identify AMD and its associated lesions with satisfactory performance.
Moreover, the evaluation of the lesion activation maps shows that the models trained using the proposed approach are able to identify the pathological areas within the image and, in most cases, to correctly determine to which lesion they correspond.

\paragraph{Conclusions}
The proposed approach provides meaningful information---lesion identification and lesion activation maps---that conveniently explains and complements the diagnosis, and is of particular interest to clinicians for the diagnostic process.
Moreover, the data needed to train the networks using the proposed approach is commonly easy to obtain, what represents an important advantage in fields with particularly scarce data, such as medical imaging.
\end{abstract}


\end{frontmatter}


\section{Introduction}%
\label{sec:introduction}

Age-related macular degeneration (AMD) is a degenerative disorder affecting the macula, a small area near the center of the retina that plays a key role in visual acuity~\cite{Kanski_Elsevier_2011}.
AMD represents the most frequent cause of blindness in developed countries, especially for people over 60 years old~\cite{Wong_LGH_2014,Mitchell_Lancet_2018}.
Worldwide, an estimated 8.7\% of blindness cases are caused by this disorder~\cite{Wong_LGH_2014}.
Furthermore, this proportion is expected to increase in the coming years due to the global population aging.

Conventionally, AMD was divided into two main types: \textit{dry} AMD and \textit{wet} AMD, affecting approximately the 90\% and the 10\% of people diagnosed with the disease, respectively~\cite{Kanski_Elsevier_2011}.
This classification remained in force until 2013, when an expert consensus committee provided a more precise clinical classification of AMD~\cite{Ferris_Ophth_2013}.
This new classification consists of 5 different classes that represent the various stages of development of the disease: (1) no apparent aging changes, (2) normal aging changes, (3) early AMD, (4) intermediate AMD and (5) late AMD.
The characterization of these classes is based on fundus lesions assessed within 2 optic disc diameters of the macula center (of either eye) in people older than 55 years.
Following this classification system, people with no visible drusen or pigmentary abnormalities (PA) should be considered to have no signs of AMD; with only small drusen, normal ageing changes; with medium drusen but not PA, early AM; with large drusen or PA, intermediate AMD; and with neovascular AMD or geographic atrophy (GA, or simply atrophy), late AMD.
More specifically, neovascular AMD is characterized by choroidal neovascularization and pigment epithelial detachment (PED).
PA include any hyper- or hypopigmentary abnormality associated to medium and large drusen but not to other known disease.
Other less common signs of late AMD frequently mentioned in the literature are PED and exudates or hemorrhages in or around the macula~\cite{Tan_AP_2017}.
Furthermore, it has been reported that choroidal neovascularization, with no treatment, occasionally cause fibrosis and/or a disciform scar under the macula.
Thus, the identification and assessment of the lesions in the eye fundus is key towards providing a reliable diagnosis and characterization of AMD.

To assess the presence of these lesions, ophthalmologists commonly use one of the following imaging modalities, if not both: retinography (also called color fundus photography [CFP]) and optical coherence tomography (OCT)~\cite{Ting_PRER_2019}.
These modalities offer unique and complementary information that is useful for detecting AMD~\cite{Mitchell_Lancet_2018,Jain_IOVS_2010,Cheung_NSR_2018,Wu_OR_2021}.
Nevertheless, CFP is still the most used of the two due to its affordability and widespread availability.
For this reason, it is also the predominant modality in large-scale screening and early detection programs.
For AMD, as for so many other ocular diseases, such programs are of great importance, since the detection of the disease at an early stage allows the effective application of certain treatments~\cite{Mitchell_Lancet_2018,Deng_GN_2021}.
For example, recent works have suggested that the progression from early AMD to late AMD can be slowed down with high-dose zinc and antioxidant vitamin supplements~\cite{AREDS2_JAMAO_2014}.
Also, it has been reported that intravitreal anti-vascular endothelial growth factor therapy is effective at slowing down the development of neovascular AMD~\cite{AREDS2_JAMAO_2014}.
Notwithstanding, far more research is needed, since the success of these therapies is limited and there is currently no effective treatment for GA, which represents the most common late AMD variant by a wide margin.
This scenario reinforces the importance of the early detection of the disease.

Despite their convenience, implementing screening programs for AMD on a large scale is usually unfeasible, since the population at risk is large and the analysis of color fundus images is highly challenging.
The inherent difficulty of the diagnosis is compounded by the fact that AMD is characterized by many different lesions that, in many cases, coincide with (or resemble) those of other macular diseases~\cite{Saksens_PRER_2014}.
This forces such analyses to be performed by expert clinicians.
In addition, the visual analysis of images can be subject to interpretation, and there may be relevant differences between the diagnoses of different experts.
All this motivates the research on automatic diagnostic methods~\cite{Ting_PRER_2019,Wu_OR_2021,Pead_SO_2019}.

Of the automatic approaches proposed for AMD diagnosis from CFP, we can distinguish various types depending on the concrete problem they address.
In particular, there are works focused on AMD grading~\cite{Burlina_JAMAO_2018,Tan_FGCS_2018,Grassmann_DL-AMD_O_2018}, AMD diagnosis~\cite{Tan_FGCS_2018,Mookiah_KBS_2015,Li_TMI_2020,Fang_ADAM_TMI_2022}, referable AMD diagnosis (i.e. only late and intermediate AMD, not early AMD)~\cite{Ting_PRER_2019,Pead_SO_2019,Burlina_JAMAO_2017,Gonzalez-Gonzalo_AO_2020} and multi-disease prediction~\cite{Gonzalez-Gonzalo_AO_2020,Ting_JAMA_2017}.
Of these four types, this work is framed in the second: AMD diagnosis.

In the state of the art, the predominant approach for AMD diagnosis is to train a machine learning classifier to discriminate between two classes: AMD and non-AMD.
In early works, the classifier was based on classical methods that rely on ad hoc feature engineering, typically fully connected neural networks~\cite{Pead_SO_2019} or support-vector machines~\cite{Mookiah_KBS_2015}.
In contrast, most recent works are based on convolutional neural networks (CNNs)~\cite{Tan_FGCS_2018,Li_TMI_2020,Burlina_JAMAO_2017,Ting_JAMA_2017,Fang_ADAM_TMI_2022}.
These CNN-based approaches have explored the use of ad hoc CNN architectures~\cite{Tan_FGCS_2018}, ensembles of these networks~\cite{Gonzalez-Gonzalo_AO_2020,Ting_JAMA_2017,Fang_ADAM_TMI_2022}, or standard CNN classification architectures~\cite{Li_TMI_2020,Burlina_JAMAO_2017,Fang_ADAM_TMI_2022}.
Furthermore, while ImageNet pretraining is common when using the standard CNNs~\cite{Ting_PRER_2019,Burlina_JAMAO_2017}, other kinds of self-supervised pretraining were also successfully applied~\cite{Li_TMI_2020, hervella2021}.

All these works provide satisfactory performance in the diagnosis of AMD; in some cases, even similar or superior to those of clinical experts~\cite{Burlina_JAMAO_2018,Grassmann_DL-AMD_O_2018,Burlina_JAMAO_2017,Gonzalez-Gonzalo_AO_2020}.
However, none of them incorporates explainability mechanisms to help the experts to better understand the predictions of the models.
This issue is particularly relevant, since the absence of such mechanisms limits the application of automatic approaches in real-world scenarios~\cite{Yang_IF_2022}.
In diagnostic tasks, as in this case, the need for explainability is even more pronounced~\cite{vanderVelden_XAI_MIA_2022}, since the decision of the model can have a direct impact on the life of the patient.
Furthermore, an increasing number of countries are regulating the right of explanation of algorithm decisions for individuals~\cite{Samek_XAI_Springer_2019}.

Most explainability techniques for CNNs aim to obtain a coarse map indicating the areas of the input image that have been important for the model in making the decision~\cite{Samek_XAI_Springer_2019}.
Ideally, in the particular case of diagnosis, the coarse map should highlight the areas of the image which are indicative of the disease.
Thus, clinicians can examine the model output and the map and check the exactitude of the identified pathological areas (i.e. areas with lesions).
In other words, they can directly check whether the model has used the appropriate information (features) to make the final diagnosis~\cite{Cen_NatureComm_2021}.

In medical imaging, the most commonly used techniques are Class Activation Maps (CAM)~\cite{Zhou_CAM_CVPR_2016}, Gradient-weighted CAM (Grad-CAM)~\cite{Selvaraju_Grad-CAM_ICCV_2017}, and Multiple Instance Learning with Fully Convolutional Networks (MIL-FCN)~\cite{vanderVelden_XAI_MIA_2022,Oquab_CMIL_CVPR_2015,Pathak_FCNMIL_ICLR_2015}.

CAM is a procedure for generating class activation maps (CAMs) using global average pooling (GAP) in CNNs.
A CAM for a particular category indicates the discriminative image regions used by the CNN to identify that category~\cite{Zhou_CAM_CVPR_2016}.
That is, the regions of the image which have ultimately determined the classification.
In order to apply CAM, it is necessary for the CNN to have a GAP operation just after the last convolutional layer, as well as a single linear layer between the GAP output and the final output.
The CAM for a certain category is obtained by means of an element-wise weighted sum of the feature maps of the last convolutional layers.
The weights used in the sum correspond to the weights of the linear layer for the particular category.
An important drawback of CAM is that it limits the architecture of the CNN model.

Grad-CAM is a gradient-based method that uses the gradients of any target concept (e.g. ``AMD'' in a classification network) flowing into the final convolutional layer of a CNN to produce a coarse location map that highlights the important regions of the image for predicting the concept~\cite{Selvaraju_Grad-CAM_ICCV_2017}.
The method is applied \mbox{\textit{a posteriori}} to already trained CNN networks.
Unlike CAM, Grad-CAM does not condition the architecture of the CNN model.
Given an image and a class of interest as input, Grad-CAM forward propagates the image through the CNN part of the model and then through task-specific computations to obtain a raw score for the category.
The gradients are set to zero for all classes except the desired class, which is set to 1. 
This signal is then backpropagated to the rectified convolutional feature maps of interest, which are combined to compute the coarse Grad-CAM localization map which represents where the model \textit{looks} to make the particular decision.

Lastly, MIL-FCN, as its name suggests, is a framework for multiple instance learning using fully convolutional networks~\cite{Oquab_CMIL_CVPR_2015,Pathak_FCNMIL_ICLR_2015}.
With the MIL-FCN framework, each image is cast as a bag of pixel-level or region-level instances.
The FCN predicts the class of all instances, and then integrates all the predictions to determine the class of the bag.
The typical approach consists in adding a $1 \times 1$ convolution with a single output channel at the end of the convolutional trunk of the model, and then to compute the maximum of the resulting feature map in order to obtain the final prediction~\cite{Costa_EyeWeS_MVA_2019,Araujo_MIA_2020}.
Originally, this approach was proposed for weakly-supervised object localization~\cite{Oquab_CMIL_CVPR_2015} and segmentation~\cite{Pathak_FCNMIL_ICLR_2015,Jia_MIL_TMI_2017}.
However, recent studies show that the MIL-FCN approach can increase the explainability of the models in classification tasks~\cite{Araujo_MIA_2020}.

Beyond AMD, there are works addressing other diagnostic tasks based on CFP that have made progresses with respect to the explainability of the learned models by applying these techniques.
Some examples can be found for diabetic retinopathy diagnosis~\cite{Gondal_ICIP_2017,Costa_EyeWeS_MVA_2019,Araujo_MIA_2020,Sun_CVPR_2021}, glaucoma diagnosis~\cite{Martins_CMPB_2020}, multi-disease prediction~\cite{Wang_MICCAI_2019,Meng_JBHI_2020,Chelaramani_WACV_2021,Cen_NatureComm_2021} and multi-disease grading~\cite{Ju_JBHI_2021}.
Most works use CAM over backends of standard classification CNNs~\cite{Cen_NatureComm_2021,Gondal_ICIP_2017,Sun_CVPR_2021,Wang_MICCAI_2019,Meng_JBHI_2020,Chelaramani_WACV_2021,Ju_JBHI_2021} to provide some sort of explainability of the predictions of the models.
However, there are also methods using Grad-CAM~\cite{Martins_CMPB_2020,Meng_JBHI_2020} and MIL-FCN~\cite{Costa_EyeWeS_MVA_2019,Araujo_MIA_2020}.
In the works based on CAM and Grad-CAM, the application of the method is straightforward on standard classification models.
Additionally, works based on MIL-FCN \cite{Costa_EyeWeS_MVA_2019,Araujo_MIA_2020} propose to conform the CNNs to the MIL approach by employing a custom architectural modification.
This modification, based on the use of $1 \times 1$ convolutions, allows to directly obtain a single activation map indicating the patch-level presence of the disease.
Then, the final diagnosis is computed as the maximum of the diagnoses of the different patches of the image.

The mechanisms incorporated by these works certainly improve the explainability of the models, as they allow to identify the areas of the images that are most decisive for them in making the diagnosis.
Moreover, the results of the works demonstrate that, in pathological images, these areas frequently coincide with the regions of the retina affected by a disease.
This indicates that the final diagnosis provided by the CNN models is commonly based on the right features.
Notwithstanding, the explanatory value of these approaches is limited.
Despite the maps can indicate which individual pixels or areas of the input images are important, there is no correlation computed between these regions to more abstract concepts such as the anatomical or pathological structures (e.g. lesions) shown in the image~\cite{Yang_IF_2022}.
More importantly, the explanations---in this case, the provided maps---should be understood by humans to make sense of them and to comprehend the decisions of the model.
Therefore, it is desirable that the models provide higher-level explanations that can integrate the evidence from these low-level activation maps to describe the decisions of the model at a more abstract level~\cite{Yang_IF_2022}.
Such a model would be much more humanly understandable.
In the diagnosis of AMD, a clear example of useful higher-level explanations would be the linking of the different highlighted areas to specific lesions (drusen, atrophy, etc.).
As previously discussed, the identification and localization of lesions are fundamental for performing a proper diagnosis and characterization of AMD.
However, there are currently no works in the state of the art for AMD diagnosis providing that sort of lesion-specific activation maps.
Thus, the explainability of current approaches is limited.

In this work, we propose an explainable deep learning approach for the joint identification of AMD and its associated lesions from color fundus images.
For that purpose, our methodology presents two main novelties with respect to previous alternatives in the state of the art.
First, we propose to simultaneously perform the identification of AMD and its associated retinal lesions using the same CNN.
This is addressed by jointly training the models for both tasks.
To the best of our knowledge, this is the first work jointly addressing these tasks.
Second, we propose a particular architectural setting that directly links the predicted diagnosis to the lesions identified by the network, and these, to independent and specific \emph{lesion activation maps}.
These maps are trained in a weakly-supervised manner using only image-level labels.
Each lesion activation map represents the area or areas of the image where a particular lesion is manifested.
This point clearly differentiates this work from others in the state of the art~\cite{Costa_EyeWeS_MVA_2019,Araujo_MIA_2020,Martins_CMPB_2020}, in which the activation maps can include multiple different lesions; i.e. that the correspondence between the maps and the lesions is not 1 to 1.
Then, from the lesion predictions (i.e. the probabilities of presenting a certain lesion), the final diagnosis is computed, so the diagnosis is ultimately derived from the lesion activation maps, and can be explained by them.
This setting is highly intuitive, as it mimics the manual process followed by clinicians, consisting of localizing and then classifying the retinal lesions.
As in our approach, it is this information from which the diagnosis is ultimately derived.
Furthermore, the proposed approach is architecture-agnostic, so it can be applied, with minor modifications, over any CNN for image classification.
In our case, the proposal is applied on top of a standard VGG~\cite{Simonyan_VGG_ICLR_2015} and it is trained end-to-end using only image-level labels.

This setting has several advantages.
First, it allows to incorporate useful information that conveniently complements the diagnosis.
The lesion presence information provided by the models (lesion identification and lesion-specific activation maps) is of high clinical interest, as it can be indicative of the presence and the severity of AMD.
As we stated at the beginning of the Introduction, the location of lesions and its characterization is decisive for the diagnosis of AMD.
Second, due to the direct and intuitive link between the lesion activation maps, the lesion predictions and the diagnosis, the proposed setting helps to better understand the decisions made by the automatic system, highly improving its explainability.
The proposed approach contrasts markedly with the classical approach for AMD identification, whose only output is the probability of having AMD and does not incorporate any explainability mechanisms.

It is also worth noting that all the extra outputs provided by our approach are achieved by using only image-level labels to train the networks.
In this regard, it should be noted that the extra labels required---image-level lesion labels---are relatively easy to obtain.
Given that the lesion identification is an indispensable part of the diagnostic process, this information can be frequently found in medical records.
This enables the construction of training datasets \textit{a posteriori}, avoiding the ad hoc dedication of clinicians, whose time is commonly limited.
This is a relevant advantage over other methods, especially fully-supervised lesion segmentation methods, whose need for pixel-level labels makes the building of datasets particularly challenging.

To validate the proposed approach, we constructed a private dataset of color fundus images with expert-annotated labels for the diagnosis of AMD and the identification of its associated retinal lesions.
The neural networks are first trained and evaluated on this dataset.
Then, to avoid data bias and to be able to compare our approach with other state-of-the-art methods, the same networks are evaluated on three additional public datasets.
In total, the proposed approach is evaluated for three different tasks: the diagnosis of AMD, the identification of its associated lesions, and, to quantitatively assess the degree of explainability provided by the lesion activation maps, the coarse segmentation of the individual lesions.
In all cases, the models are directly evaluated against the manual annotations of the experts.
Furthermore, in order to validate the adequacy of our approach in the identification of AMD, we compared its performance with that of the traditional approach, which uses a standard CNN and only involves predicting the presence of AMD.
Also, for providing a better understanding of the performance of the proposed approach in AMD diagnosis, we also compared its performance with that of other state-of-the-art methods on a reference public dataset.

The remainder of the manuscript is organized as follows.
In Section~\ref{sec:materials_and_methods}, we present the methodology developed for the simultaneous identification of AMD and its associated retinal lesions; this includes the description of the different approaches to be compared for validating our method, the network architecture, the data, the quantitative evaluation procedure and the experimental details.
Further on, in Section~\ref{sec:results_and_discussion}, we present the results obtained from the comprehensive evaluation of the approaches and their discussion.
Finally, in Section~\ref{sec:conclusions}, we present the main conclusions derived from the results and the potential future work.

\section{Materials and methods}%
\label{sec:materials_and_methods}

\subsection{Overview}

This work provides an explainable deep learning approach for the joint identification of AMD and its associated retinal lesions from color fundus images.
To perform this joint task, we train a CNN end-to-end using image-level labels indicating the presence of AMD and retinal lesions.
Following the proposed approach, the trained network is able to provide individual weakly-supervised activation maps for the different lesions.
We refer to these maps as \textit{lesion activation maps.}

\begin{figure}
    \centering
    \includegraphics[width=\textwidth]{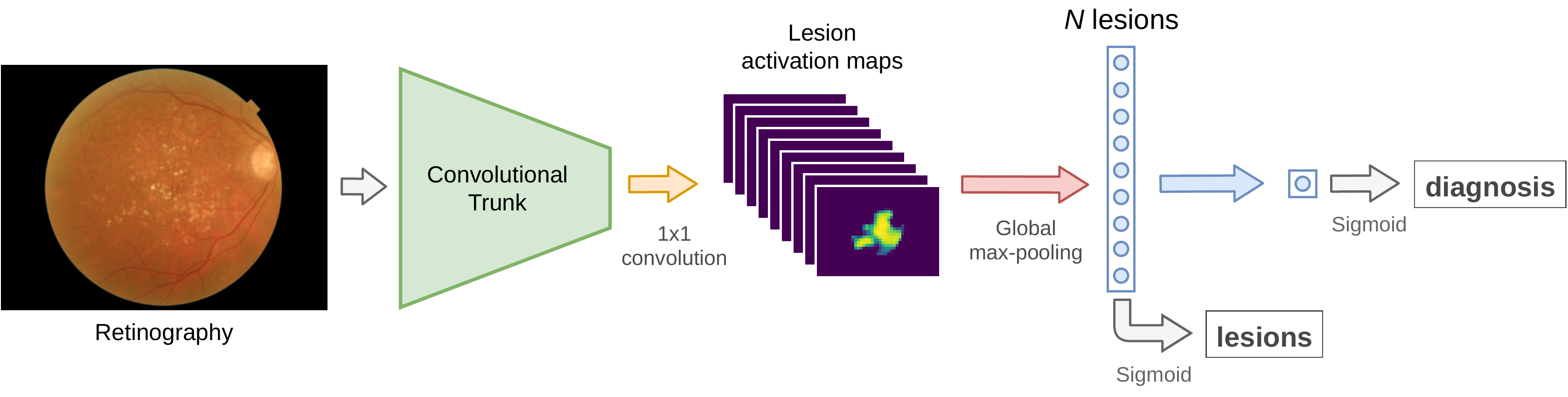}
    \caption{Proposed approach for the joint identification of lesions and AMD diagnosis (AMD+Lesions). The diagnosis is derived solely from lesion predictions, and these, directly from the lesion-specific activation maps via a global max-pooling operation.}%
    \label{fig:approach}
\end{figure}

An overview of our approach is depicted in Figure~\ref{fig:approach}.
As can be seen in the figure, the input retinography is fed to a standard convolutional trunk.
Individual lesion activation maps are derived from this trunk using a convolutional layer.
Then, a vector of identified lesions is obtained from the activation maps using a global max-pooling (GMP) operation.
Finally, the diagnosis is made from the vector of identified lesions.
In this way, the diagnosis is ultimately derived from the lesion-specific activation maps, and can be easily explained by them.

In order to evaluate the performance of the proposed approach (AMD+Lesions or A+L) and quantify its advantages, we perform a comparison with the baseline classification-only approach (AMD-Only or A-O), which uses a standard classification network and does not have any lesion identification feedback.
This comparison allow us to assess the impact of the proposed setting, as well as of the lesion identification task, on the performance of the models in identifying AMD.

The A+L approach is presented in Section~\ref{subsec:ApL}, while A-O is presented in Section~\ref{subsec:AdO}.

\subsection{Proposed approach: AMD+Lesions (A+L)}%
\label{subsec:ApL}

To train the networks for both AMD diagnosis and lesion identification, we use a combined loss that jointly quantifies the error committed by the models in both tasks.
The different parts of this combined loss are described in detail below.

\subsubsection{Diagnostic loss}

For the diagnosis of AMD, two classes are considered: AMD and non-AMD.
Thus, during the training, the diagnostic error can be measured using a standard binary classification loss between the predicted diagnosis and the manual annotations.
In this case, we use Binary Cross-Entropy (BCE).
Formally, the diagnostic loss $\mathcal{L}_{\textit{\scriptsize diagnosis}}$ is defined as
\begin{equation}\label{eq:diagnosis_loss}
    \mathcal{L}_{\textit{\scriptsize diagnosis}} ~=~ \mathcal{L}_{BCE}\left(\textbf{f}(\textbf{r})_d, \textbf{d}\right)~~,
\end{equation}
where $\textbf{f}(\textbf{r})_d$ denotes the predicted network diagnosis for retinography $\textbf{r}$; $\textbf{d}$, the target AMD diagnosis; and $\mathcal{L}_{BCE}$, the BCE loss.
The formal definition of $\mathcal{L}_{BCE}$ for a single prediction is the following:
\begin{equation}\label{eq:BCE}
    \mathcal{L}_{BCE}\left(p, t\right) = -\left[t \cdot log\left(p\right)+\left(1-t\right) \cdot log\left(1-p\right)\right] \ ,
\end{equation}
where $p$ denotes the value predicted by the model and $t$ the corresponding target value.

\subsubsection{Lesion identification loss}

Since an input sample can present more than one type of lesion, we use a multi-label classification loss as lesion identification loss $\mathcal{L}_{\textit{\scriptsize lesions}}$.
Specifically, the loss is computed as the BCE between the vector of lesion predictions provided by the model and the vector of manually annotated lesions.
Formally, it is defined as
\begin{equation}\label{eq:lesions_loss}
    \mathcal{L}_{\textit{\scriptsize lesions}} ~=~ \frac{1}{N}\sum^{N}_{i=1}\mathcal{L}_{BCE}\left(\textbf{f}(\textbf{r})_{l_i}, \textbf{l}_i\right)~~,
\end{equation}
where $\textbf{f}(\textbf{r})_l$ denotes the vector of lesions predicted by the network for retinography $\textbf{r}$; $\textbf{l}$, the target lesion vector; $N$, the number of lesions; and $\mathcal{L}_{BCE}$, the BCE loss defined in Equation~\ref{eq:BCE}.

\subsubsection{Combined loss}

For the proposed approach (A+L), the diagnostic and lesion identification losses are combined together.
Thus, A+L models are simultaneously trained in the identification of AMD and the retinal lesions.
Specifically, the joint loss $\mathcal{L}_{\textit{\scriptsize A+L}}$ is defined as the direct sum of the diagnostic loss and the lesion identification loss.
Its formal definition is the following:
\begin{equation}
    \mathcal{L}_{\textit{\scriptsize A+L}} ~=~ \mathcal{L}_{\textit{\scriptsize diagnosis}}  ~+~  \mathcal{L}_{\textit{\scriptsize lesions}}~~,
\end{equation}
where $\mathcal{L}_{\textit{\scriptsize diagnosis}}$ is the diagnostic loss defined in Equation~\ref{eq:diagnosis_loss} and $\mathcal{L}_{\textit{\scriptsize lesions}}$ is the lesion identification loss defined in Equation~\ref{eq:lesions_loss}.

\subsection{Baseline approach: AMD-Only (A-O)}%
\label{subsec:AdO}

Since the baseline A-O approach focuses only on diagnosis, the A-O loss function $\mathcal{L}_{\textit{\scriptsize A-O}}$ coincides with the diagnostic loss:
\begin{equation}\label{eq:A-O}
    \mathcal{L}_{\textit{\scriptsize A-O}} ~=~ \mathcal{L}_{\textit{\scriptsize diagnosis}} ~=~ \mathcal{L}_{BCE}\left(\textbf{f}(\textbf{r})_d, \textbf{d}\right)~~,
\end{equation}
where $\mathcal{L}_{BCE}\left(\textbf{f}(\textbf{r})_d, \textbf{d}\right)$ is the loss function defined in Equation~\ref{eq:BCE}.

\subsection{Network architecture}

To implement the proposed approach A+L, we propose a particular architectural setting.
This setting is applied on top of a standard classification convolutional trunk.
For the experiments conducted in this work, we use the VGG~\cite{Simonyan_VGG_ICLR_2015} network architecture.
Since its publication, numerous works have applied VGG-based architectures to multiple problems involving natural images ~\cite{Girshick_TPAMI_2016,Kong_CVPR_2016,Kim_CVPR_2016,Cao_TPAMI_2021}.
Furthermore, VGG has been widely applied in medical imaging~\cite{Hu_Neuro_2018,Maninis_MICCAI_2016,Bychkov_SR_2018,Jia_TMI_2017} and, more specifically, ophthalmic imaging~\cite{Fang_TMI_2019,Pires_AIIM_2019,Apostolopoulos_PESM_2020}.
Previous works on AMD identification have reported positive results using this architecture~\cite{Grassmann_DL-AMD_O_2018}.
In the baseline approach, A-O, the original VGG architecture is used as it is, without the custom setting.
Differently, in the proposed approach, A+L, we use the VGG backend up to the final fully-connected part, which is replaced by our custom setting.

\begin{figure}[btp]
    \centering
    \includegraphics[width=0.98\textwidth]{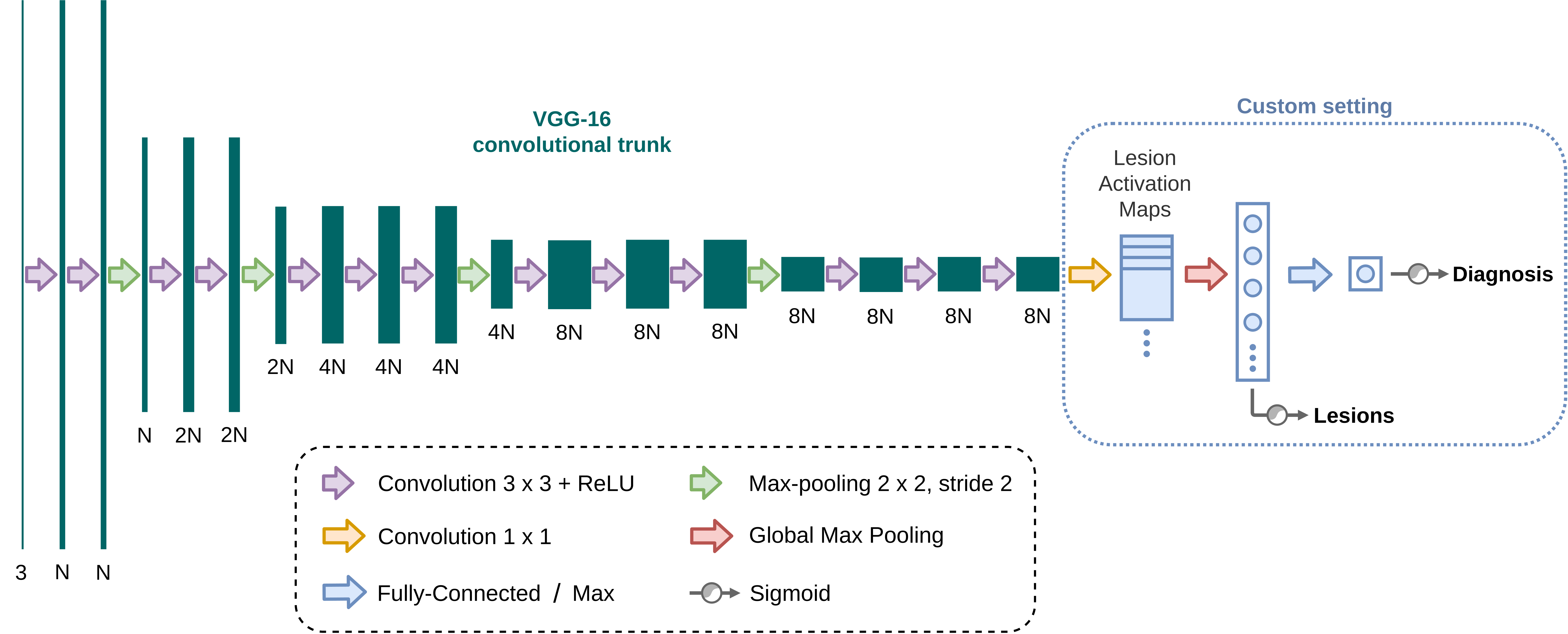}
    \caption{Proposed network architecture for the AMD+Lesions approach.}%
    \label{fig:VGG-16}
\end{figure}

Figure~\ref{fig:VGG-16} depicts an overview of the network architecture used in the proposed approach (A+L).
The convolutional part, up to the $1 \times 1$ convolution, coincides with that of the original VGG-16 except for one aspect: in our version, the original max-pooling at the end of this part is not included.
This has been done in order to obtain larger activation maps at the output of the convolutional trunk.
From this point onward, the rest of the original VGG was replaced by our architectural setting.
First, we use a $1 \times 1$ convolution of $N$ output channels (one per lesion) to generate the $N$ corresponding lesion activation maps.
In the proposed architecture, these maps will have 1/16 of the resolution of the original image.
Then, the lesion predictions are generated by applying a GMP operation to the maps.
We use GMP because we want to encode the presence of the lesions regardless of their size or position in the image (or activation map).
Lastly, to derive the diagnosis from the predictions of lesions, we present two different alternatives.

The first alternative, A+L FC, consists in adding a Fully Connected (FC) layer that takes the predicted vector of $N$ lesions as input and produces the diagnosis.
In this way, the network can freely weight the lesions when determining the final diagnosis.
The second alternative, A+L Max, consists in obtaining the diagnosis as the maximum value of the predicted vector of lesions.
This variant makes the explanation easier, since the final diagnosis is simply the maximum of the lesion activations.
However, it is less flexible than the former, since it assumes that the presence of \textit{any} lesion indicates the presence of AMD.
This characteristic does not hold for real screening scenarios, where there may be patients with many different pathologies characterized by many different lesions.
In such cases, it would be necessary to do a more detailed study of which lesions are related to the different diagnoses.
In both Max and FC variants, a sigmoid function is applied to the vector of lesions predictions and to the diagnosis in order to obtain the final outputs.

In any alternative, the diagnosis is derived from the lesion predictions, and the lesion predictions, from the lesion activation maps.
Thus, both outputs are ultimately derived from these maps.
This setting highly improves the model explainability, as it allows to better understand the final diagnosis of the model through the examination of the vector of lesion predictions and the visualization of the lesion activation maps.
These maps can be properly visualized as coarse lesion segmentation maps by applying the sigmoid function to them.
Furthermore, it should be noted that the proposed custom setting can be applied over most CNN architectures with minor modifications, since the only requirement of the module is to have an input consisting of $N$ feature maps.

\subsection{Data}%
\label{subsec:data}

For the experiments conducted in this work, we employed 4 different datasets: Age-related Macular Degeneration Lesions (AMDLesions), Automatic Detection challenge on Age-related Macular degeneration (ADAM)~\cite{Fu_IEEEDataport_2020}, Automated Retinal Image Analysis (ARIA)~\cite{Farnell_ARIA_JFI_2008} and STructured Analysis of the Retina (STARE)~\cite{Hoover_STARE_TMI_2000}.
ADAM, ARIA and STARE are public datasets.
In contrast, AMDLesions is a private dataset that was constructed for the purpose of performing this study.
Thus, this is the first work describing the dataset and reporting results for it.

\begin{figure}[tbp]
    \centering
    \subfloat[]{\includegraphics[height=3.2cm]
        {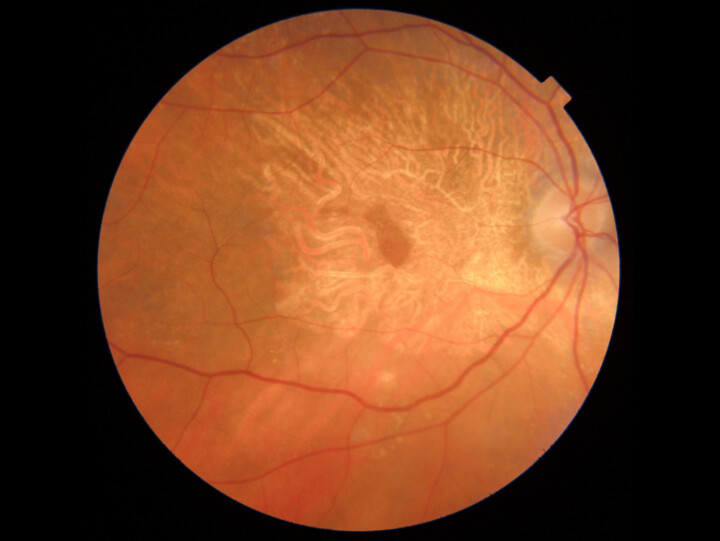}\label{subfig:AMDLesions_example}}
    \hfill
    \subfloat[]{\includegraphics[height=3.2cm]
        {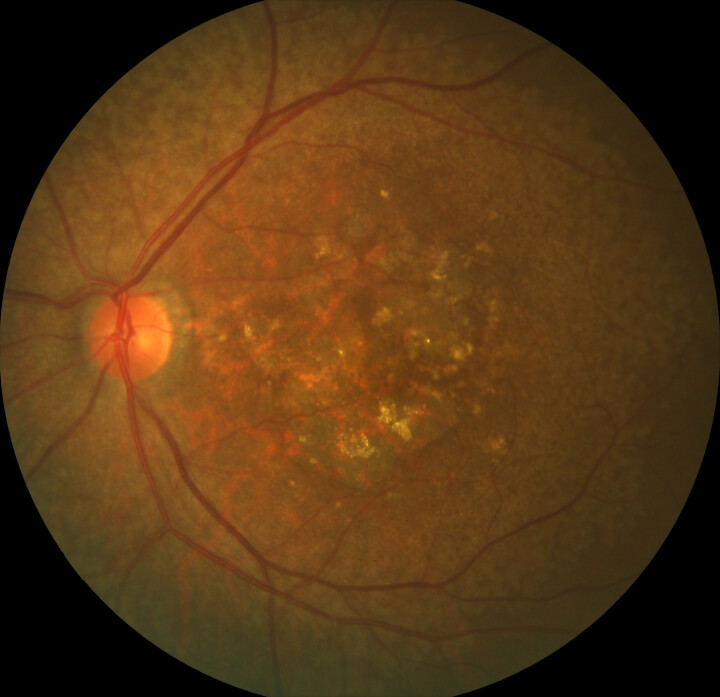}\label{subfig:ADAM_example}}
    \hfill
    \subfloat[]{\includegraphics[height=3.2cm]
        {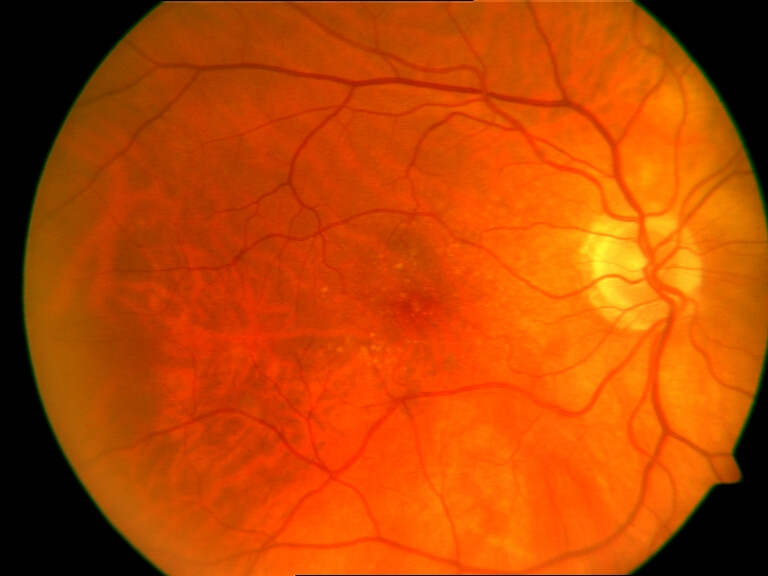}\label{subfig:ARIA_example}}
    \hfill
    \subfloat[]{\includegraphics[height=3.2cm]
        {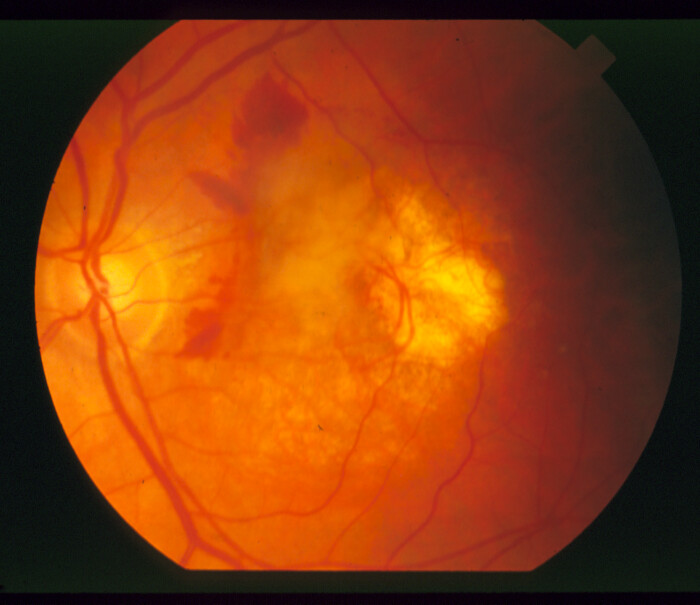}\label{subfig:STARE_example}}
    \caption{Example retinography images from (a)~AMDLesions, (b)~ADAM, (c)~ARIA and (d)~STARE datasets. All images are from patients diagnosed with AMD.}
    \label{fig:datasets_examples}
\end{figure}

Figure~\ref{fig:datasets_examples} shows examples of retinography images from the AMDLesions, ADAM, ARIA and STARE datasets.
All images are from patients with AMD.

\subsubsection{AMDLesions}

\begin{figure}[tbp]
    \centering
    \includegraphics[width=0.7\textwidth]{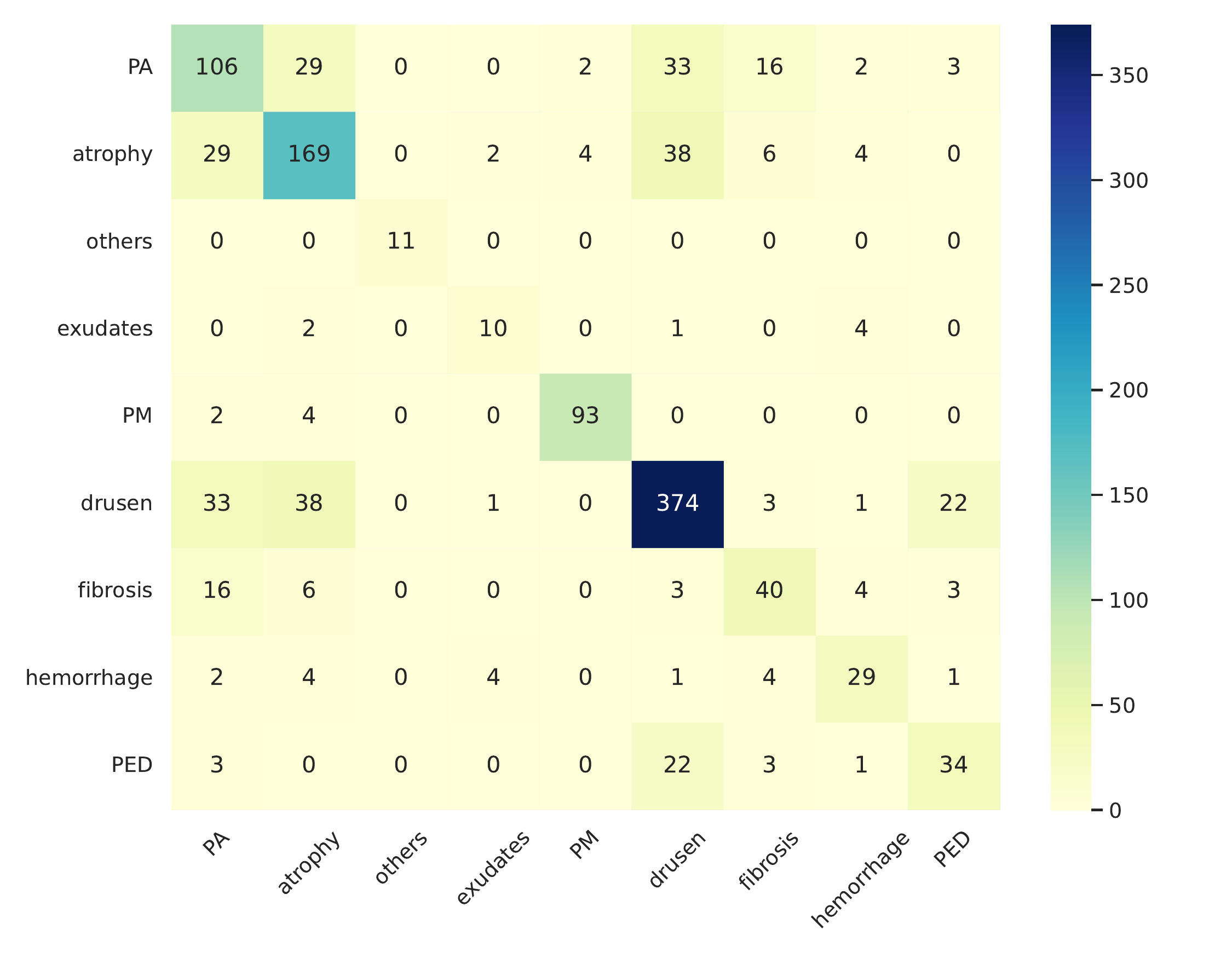}
    \caption{Distribution of lesion co-occurrence in the AMDLesions dataset. PA stands for Pigmentary Abnormalities; PM, for Pathological Myopia; and PED, for Pigment Epithelial Detachment.}%
    \label{fig:coincidence_matrix}
\end{figure}

The AMDLesions dataset is composed of 980 color fundus images from 491 different patients, with a 54\%-46\% proportion of females and males, respectively.
All the images were captured between November 2017 and April 2021 in the International Center for Advanced Ophthalmology (CIOA), Madrid, using a Triton (Topcon) fundus camera at 45\textdegree~of picture angle (equivalent 30\textdegree~[digital zoom]).
Poor quality images were discarded due to media opacity, as in the case of a very dense cataract or poor patient collaboration.
Most images (815 in total) are sized $1934 \times 2576$ pixels, while others (the remaining 165) are sized $1934 \times 1960$ pixels.
All of them are macula centered, with a completely circular region of interest (ROI).
Of the 980 retinography images, 271 are from healthy patients, and 709 are from patients with AMD.
All images include labels indicating the presence or absence of AMD disease.
In addition, for the positive cases, labels indicating which retinal lesions are visible in the image are also available.
Both AMD diagnosis and lesion annotations were made by a group of clinicians working in the field of retinal image analysis.
In total, 19 different types and subtypes of lesions were identified and annotated.
Of these, there were several with very few examples.
Thus, for the purpose of this work, the 19 lesion types and subtypes were grouped into 9 main categories: atrophy (169 images), drusen (374), exudates (10), fibrosis (40), hemorrhage (29), pathological myopia (PM) (93), pigmentary abnormalities (PA) (106), pigment epithelial detachment (PED) (34) and `others' (11).
The `atrophy' category comprises both regular atrophy and parapapillary atrophy; `drusen' includes both regular drusen and calcified drusen; and `others' comprises all the lesions that were found in less than 4 samples.
None of the lesions is mutually exclusive, so that there are multiple images that present more than one lesion.
In that regard, Figure~\ref{fig:coincidence_matrix} depicts the coincidence matrix of the lesions in the dataset.
The diagonal values indicate the total number of samples for each type of lesion, whereas the values outside the diagonal indicate the co-occurrences among different types of lesions.

\subsubsection{ADAM}%
\label{subsec:ADAM}

The ADAM dataset~\cite{Fu_IEEEDataport_2020}, also known as iChallenge-AMD, consists of 400 retinography images, from which 89 are from patients diagnosed with AMD.
The size of most images is $2124 \times 2056$ pixels, but there are also images whose size is $1444 \times 1444$ pixels.
All images have labels indicating whether or not the patient has AMD.
The reference standard for the positive diagnosis (i.e. the presence) of AMD is based on the retinography images themselves and other complementary information, such as visual field and OCT.
This complementary information, however, is not present in the dataset and was never released.
In addition to the AMD diagnosis labels, this dataset includes coarse segmentation maps of multiple lesions.
Segmentation maps are considered `coarse' because they do not contain a precise pixel-level segmentation of the lesions, but a segmentation of the area where the lesions are located.
Currently, this is the only public dataset that provides pixel-level annotations of different AMD-associated lesions.
The specific lesions for which such maps exist are drusen (61 images), exudates (38), hemorrhage (19) and scar (13).
There are also 17 images with unidentified lesions labeled as `others' (17).
In this dataset, the presence of a lesion does not necessarily imply the positive diagnosis of AMD, and the positive diagnosis of AMD does not necessarily imply the presence of an annotated lesion.
Along with this characteristic, it is worth mentioning that the dataset contains at least 125 images which belong to the same eye as others present in the dataset.
This circumstance is not mentioned in the dataset description, although it is essential when assessing the performance of the models, as images from the same patients should not be used both for training and for testing.
When partitioning the data, we have taken this circumstance into account.

\subsubsection{ARIA}

The ARIA dataset~\cite{Farnell_ARIA_JFI_2008} contains 143 retinography images from patients with diabetic retinopathy (59 images), with AMD (23), and without any disease (61).
All images are sized $768 \times 576$ pixels and have labels indicating to which of the above groups they belong.
Since we are only interested in the identification of AMD, we exclusively use the images from patients with AMD and from healthy patients.
Thus, we use a total of 83 images, of which 23 present signs of AMD.
From now on, when we mention ARIA, we will refer to this subset of the data.

\subsubsection{STARE}

The STARE dataset~\cite{Hoover_STARE_TMI_2000} is composed of 397 color fundus images from both healthy people and people with a medical condition.
The size of all images is $700 \times 605$ pixels.
Each image have text annotations indicating its diagnosis, as well as annotations of 39 possible manifestations (mainly lesions) visible in the image.
Other expert annotations, such as blood vessel segmentation maps, artery/vein labels, and the image coordinates of the optic nerve, are available for some images.
Similarly as for ARIA, we only use a subset of the dataset: the 36 images labeled as `normal' and the 46 images labeled as AMD.
As with ARIA, when we mention STARE, we are referring to this subset.

\subsection{Quantitative evaluation}%
\label{subsec:quantitative_evaluation}

To evaluate the potential and advantages of the proposed approach, we perform an evaluation consisting of three parts.
The first part is focused on assessing the performance of the proposed approach (A+L) and the baseline approach (A-O) in the identification of AMD.
This comparison allow us to assess the impact of the proposed setting, as well as of the lesion identification task, on the performance of the models in the identification of AMD.
The second part of the evaluation is focused on assessing the performance of the A+L models in the identification of lesions.
This part has the added utility of assessing how accurate is the explanation of the diagnosis via the identified lesions.
Finally, the third part of the evaluation is focused on measuring the capability of the A+L models to explain, via the lesion activation maps, the lesion identification and the final diagnosis.
For this end, A+L models are evaluated in the coarse segmentation of lesions, a task for which they have not been directly trained.

In the following paragraphs we describe in more detail how these different evaluations are performed.

\paragraph{1. Identification of AMD}
The quantitative evaluation of the different models in the identification of AMD is performed by directly comparing the predicted diagnosis with the manual annotations of the clinicians.
For each model, we compute the Receiver Operating Characteristic (ROC) curve, which plots True Positive Rate (TPR) against False Positive Rate (FPR).
This curve is built by computing the TPR and FPR at different values of the decision threshold.
In this way, it is not necessary to select a specific threshold for the evaluation, which would hinder the analysis of the results.
Also, to summarize the ROC curve, we compute the Area Under Curve (AUC) value in each case.
We will refer to the AUC of the ROC curve as AUC-ROC.

\paragraph{2. Identification of lesions}
This evaluation is included only for the models that were trained using the A+L approach, and it is focused on assessing their performance in the identification of lesions associated to AMD.
The evaluation procedure is similar to the one for the identification of AMD.
Specifically, we compute the ROC curve (and its corresponding AUC-ROC value) by directly comparing the predicted lesions with the image-level lesion annotations created by the clinicians.

\begin{figure}[tbp]
    \centering
    \includegraphics[width=0.49\textwidth]
        {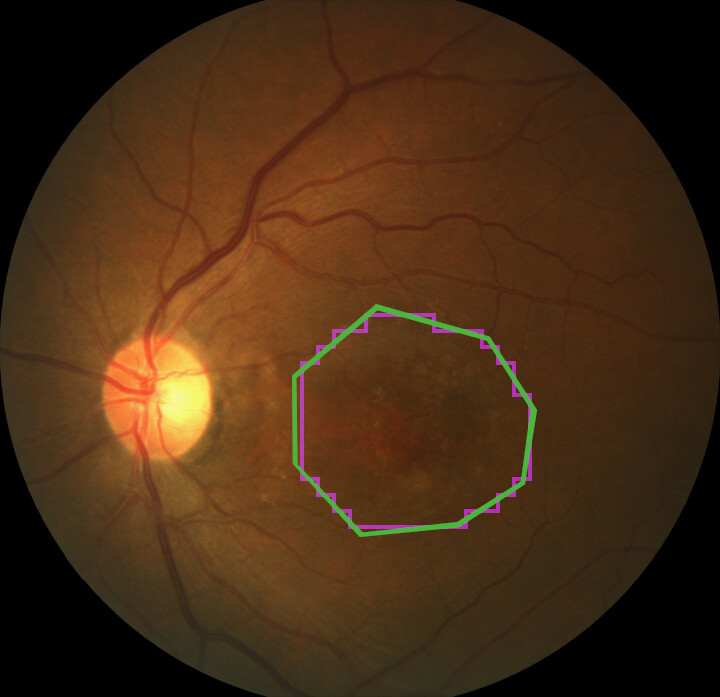}
    \caption{Example ADAM retinography with the contours of the original manual annotation (in green) and the downscaled annotation (in magenta) overlaid.}%
    \label{fig:A0043_ground_truths}
\end{figure}

\paragraph{3. Coarse segmentation of lesions}
The lesion segmentation evaluation, as evaluation \#2, is only intended for A+L models.
Its main objective is to quantify how accurate the explanations provided by the lesion activation maps are.
Specifically, the evaluation assesses the similarity of the coarse lesion segmentation maps obtained by the models (directly derived from the lesion activation maps) with the coarse segmentation maps of the lesions provided by the experts.
As mentioned in Section~\ref{subsec:ADAM}, a coarse segmentation does not provide a precise pixel-level segmentation of an element, but the area where the element is located.
Similarly to previous evaluation methods, we build the ROC curves (and compute their AUC-ROC value) by comparing the predictions of the model with the manual segmentation maps of the experts.
Given the difference between the resolution of the manual segmentation maps and the lesion activation maps, we downscaled the manual segmentation maps to perform the evaluation.
Figure~\ref{fig:A0043_ground_truths} shows an example ADAM retinography with the contours of the original manual annotation (in green) and the downscaled annotation (in magenta) overlaid.
In line with the segmentation challenge associated to the ADAM dataset~\cite{Fu_IEEEDataport_2020}, the coarse segmentation of lesions is evaluated only in those cases where manual segmentation maps exist.

\subsection{Experimental details}

All the models were trained and evaluated in the AMDLesions dataset.
Additionally, to reduce data bias, the models were also evaluated on three different public datasets: ADAM, ARIA and STARE.
To measure the robustness of the proposed approach in AMD identification, we evaluated the A+L models in a cross-dataset way, without fine-tuning, on the 3 public datasets; i.e. these whole datasets were treated as held out test data.
However, in order the comparison with other methods to be fair, we further evaluated the models after being fine-tuned in the corresponding target dataset. 
In order to consider the stochasticity of training deep neural networks and the data variability of the different datasets, we performed 4-fold cross-validation both for training and fine-tuning.
Folds were created randomly, yet ensuring that all the samples of a patient are in the same fold and that all the folds had a similar number of samples of each class.

Once trained, the models were evaluated in AMDLesions itself as well as in ADAM, ARIA and STARE.
Due to the different label availability, we used different datasets for each evaluation.
To evaluate the models in the identification of AMD, we used all the datasets.
To evaluate them in the identification of lesions, we used AMDLesions and ADAM.
And lastly, to evaluate them in the coarse segmentation of lesions, we used ADAM.
For fine-tuned models, only evaluation \#1 (identification of AMD) is performed on each dataset.
The results of these models resulting from evaluation \#1 in the ADAM dataset are compared with other state-of-the-art works in AMD identification.
As no state-of-the-art method in AMD diagnosis provides lesion-specific activation maps, we do not include a comparison with other methods concerning explainability.

Since we use 4-fold cross-validation, we obtained the mean AUC-ROC as the mean of the AUC-ROCs of the folds.
In each case, we also computed the standard deviation.
To depict the curves, we built the mean ROC curve of each alternative by merging the operating points of the curves of the different folds.

Furthermore, in some cases, for determining if the difference between the results of the presented approaches was statistically significant, we performed a two-tailed Student's \textit{t}-test.

\subsubsection{Training details}

Both for training and for testing, we rescale the images from AMDLesions and ADAM to a fixed width of 720 pixels, similar to the original width of ARIA and STARE.
Thus, all images have a similar resolution.

In order to mitigate data scarcity, we artificially increase the variability of the training samples through online data augmentation.
Thus, in each training epoch, random transformations are applied to the original input images.
These transformations include vertical and horizontal flipping, subtle random intensity and color variations and slight affine transformations, namely shearing, scaling and rotation.

To optimize the loss functions used to train the models, we use the Adam optimization algorithm~\cite{Kingma_Adam_ICLR_2015}.
The values of the different parameters of the algorithm were set as follows.
The initial learning rate ($\alpha$) was set to $\alpha = 1 \times 10^{-5}$, and the decay rates for first ($\beta_1$) and second order moments ($\beta_2$) were set to $\beta_1 = 0.9$ and $\beta_2 = 0.999$, respectively.
The values for $\beta_1$ and $\beta_2$ are the same as those proposed by Kingma and Ba in~\cite{Kingma_Adam_ICLR_2015}.
The learning rate $\alpha$ remains constant throughout the entire training, which has a fixed duration of 100 epochs.
This value was set by taking into consideration the evolution of the learning curves during training.
To fine-tune the networks in the target datasets, we use the same hyperparameters except for the number of epochs, which is set to 15.

For training, the parameters of the original convolutional layers of the networks are initialized to the parameter values of their corresponding ImageNet-pretrained model.
In this case, added convolutional and linear layers are initialized using the He et al.~\cite{He_Initialization_ICCV_2015} initialization method with Uniform distribution.
Differently, for fine-tuning, the parameters are initialized to the parameter values of the corresponding AMDLesions-trained model.

\subsubsection{Cross-dataset evaluation details}%
\label{subsubsec:cross-dataset_evaluation}

\begin{figure}[tbp]
    \centering
    \includegraphics[width=0.34\textwidth]{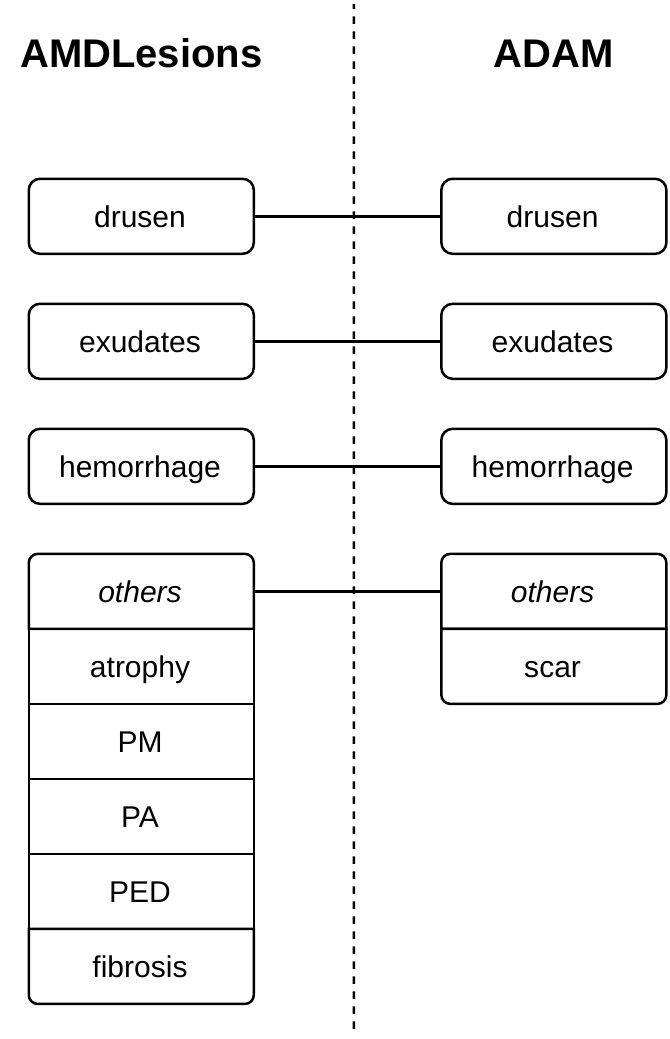}
    \caption{Mapping between AMDLesions and ADAM lesions.}%
    \label{fig:mapping}
\end{figure}

All the datasets have labels indicating the presence of AMD.
Thus, cross-dataset evaluation regarding the identification of AMD is straightforward.
Distinctly, the set of available lesion labels is different in each dataset.
In particular, AMDLesions has labels for 9 different lesions, ADAM, for 5, and ARIA and STARE, for none.
Furthermore, only 3 lesions of AMDLesions and ADAM coincide.
This causes the outputs of the models trained in AMDLesions to be different from the classes available in ADAM.
Thus, to evaluate the models trained on AMDLesions in ADAM, it is necessary to define a mapping between the lesions of this dataset and AMDLesions.
The mapping we have defined is depicted in Figure~\ref{fig:mapping}.
As can be seen in the figure, the matching of drusen, exudates and hemorrhage is direct.
Furthermore, all the lesions of AMDLesions that are not in ADAM are added to the `others' group, and vice versa.
Thus, when evaluating in ADAM, the prediction for `others' is computed as the maximum of the predictions of the model for fibrosis, atrophy, PM, PA, PED and `others' itself.
Similarly, the ground truth value for `others' is calculated as the maximum of the ground truth values of ADAM for scar and `others'.

\section{Results and discussion}%
\label{sec:results_and_discussion}

\subsection{Identification of AMD}

\begin{figure}[tbp]
    \centering
    \includegraphics[width=0.48\textwidth]{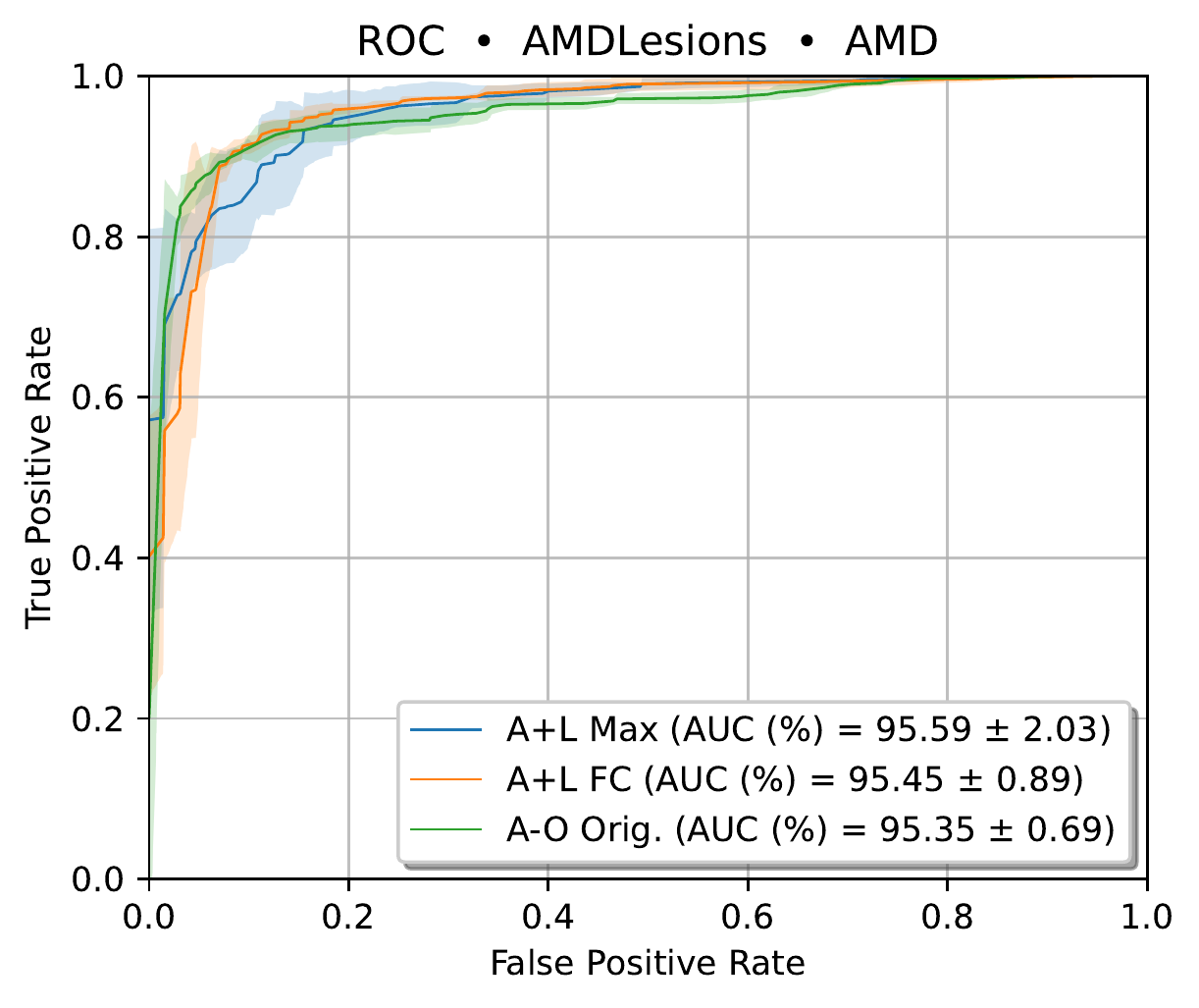}
    \includegraphics[width=0.48\textwidth]{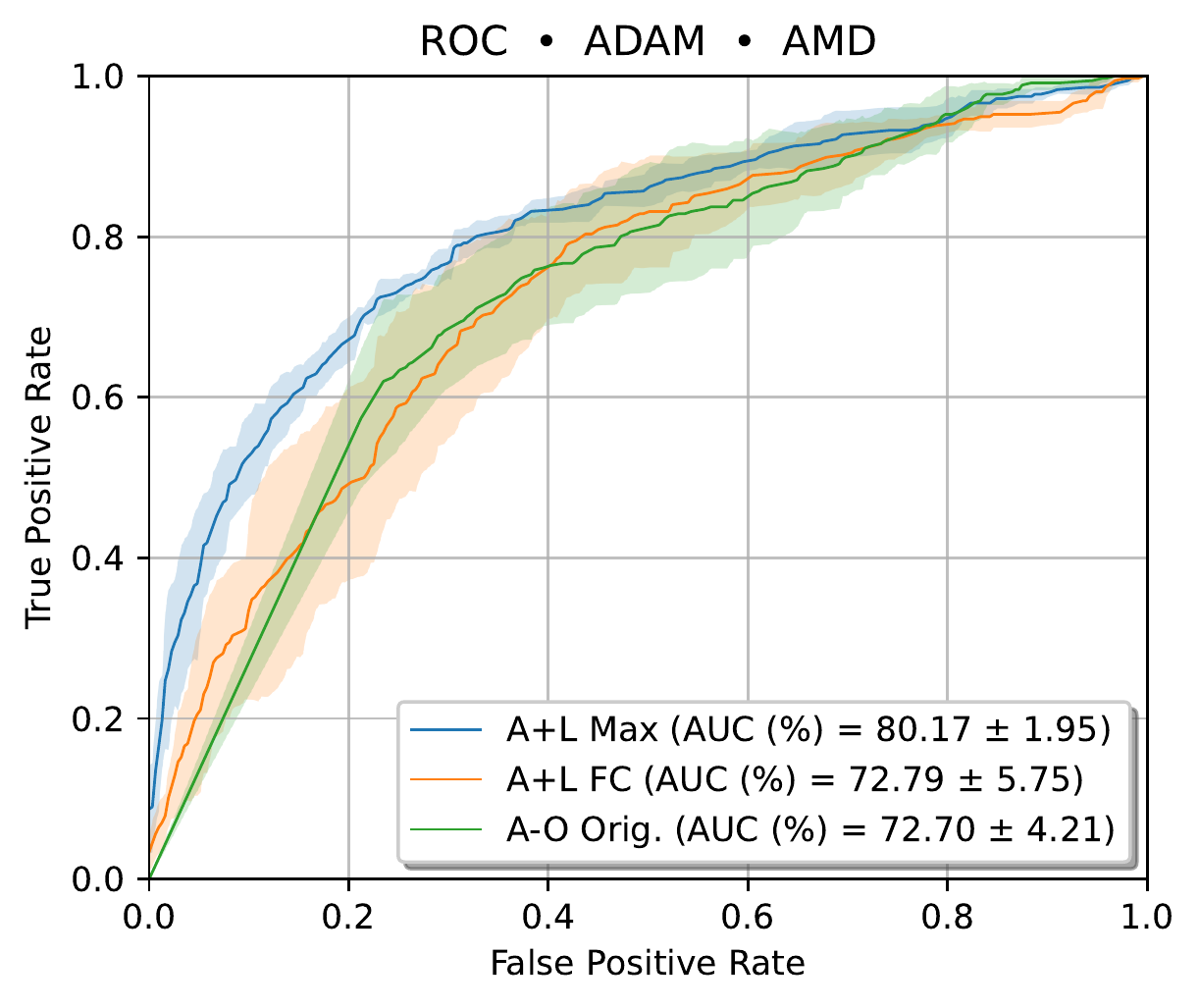}
    \includegraphics[width=0.48\textwidth]{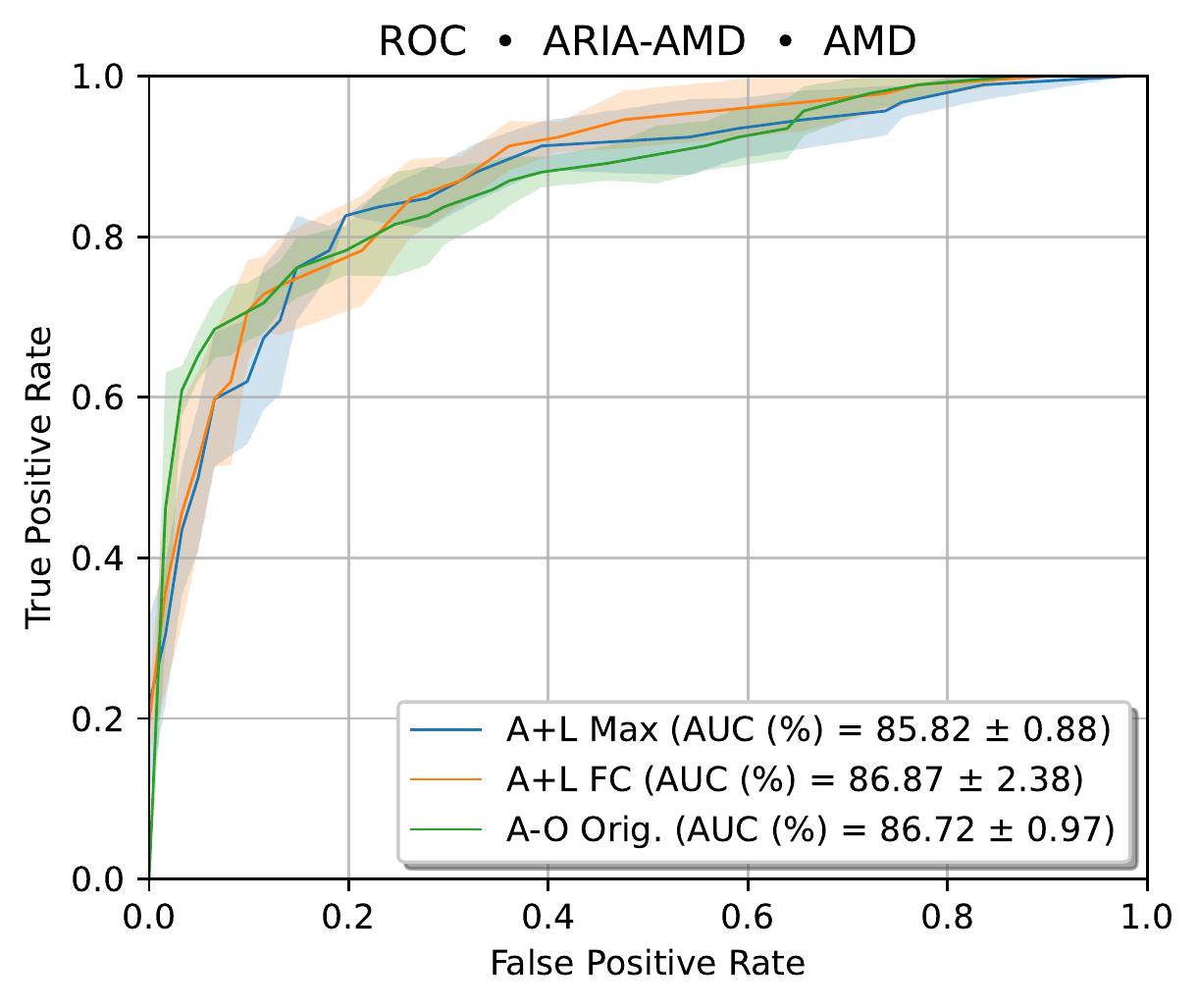}
    \includegraphics[width=0.48\textwidth]{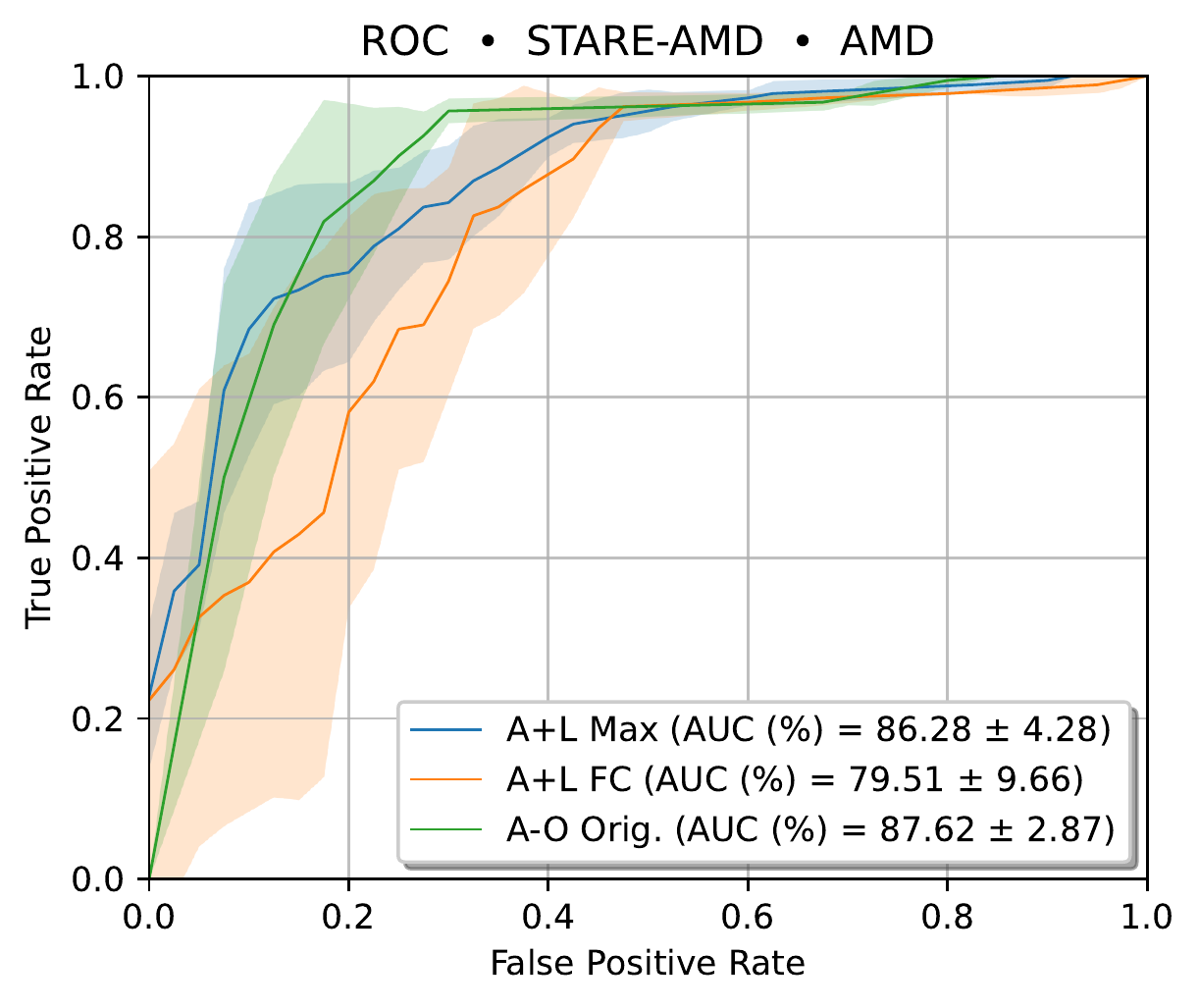}
    \caption{Mean ROC curves in AMD identification for the different A+L and baseline (A-O) approaches in AMDLesions, ADAM, ARIA and STARE datasets.}%
    \label{fig:ROCs_AMD}
\end{figure}

\begin{figure}[tbp]
    \centering
    \includegraphics[width=0.48\textwidth]
        {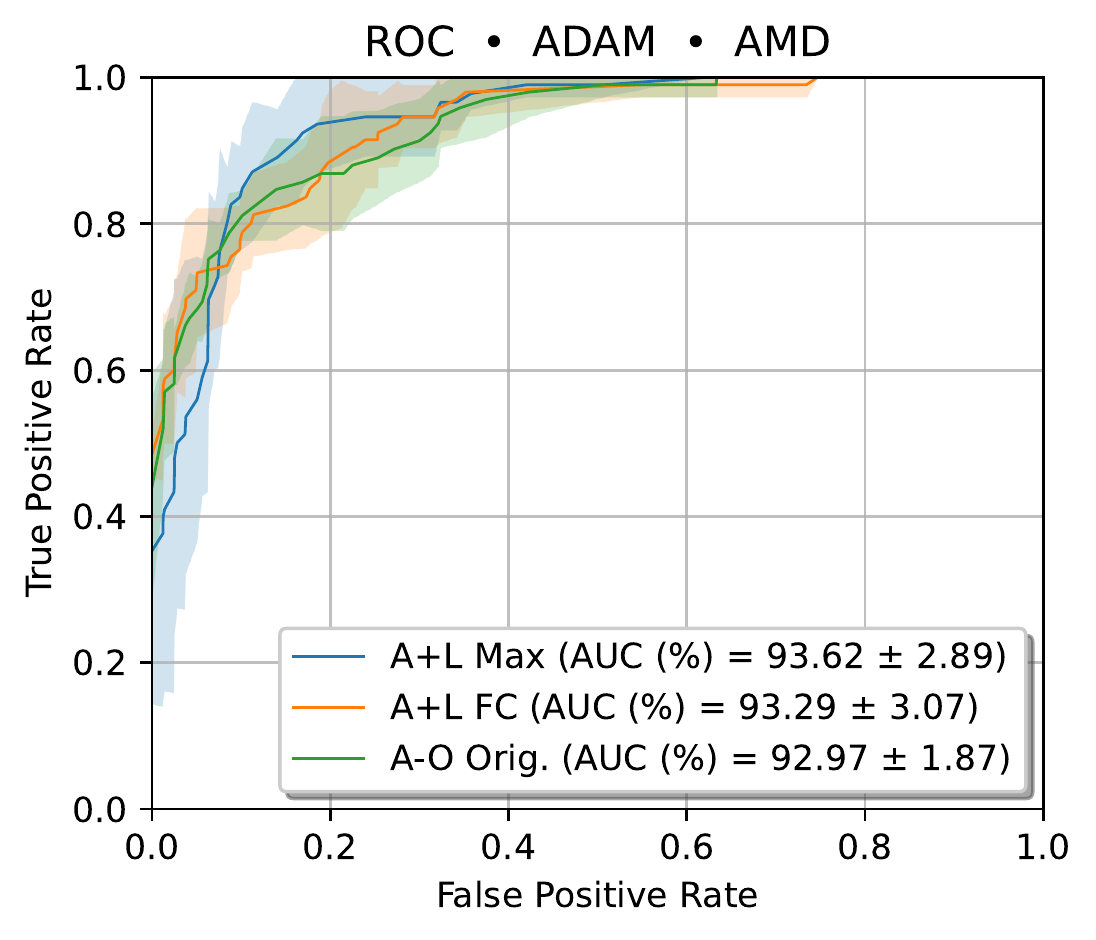}
    \includegraphics[width=0.48\textwidth]
        {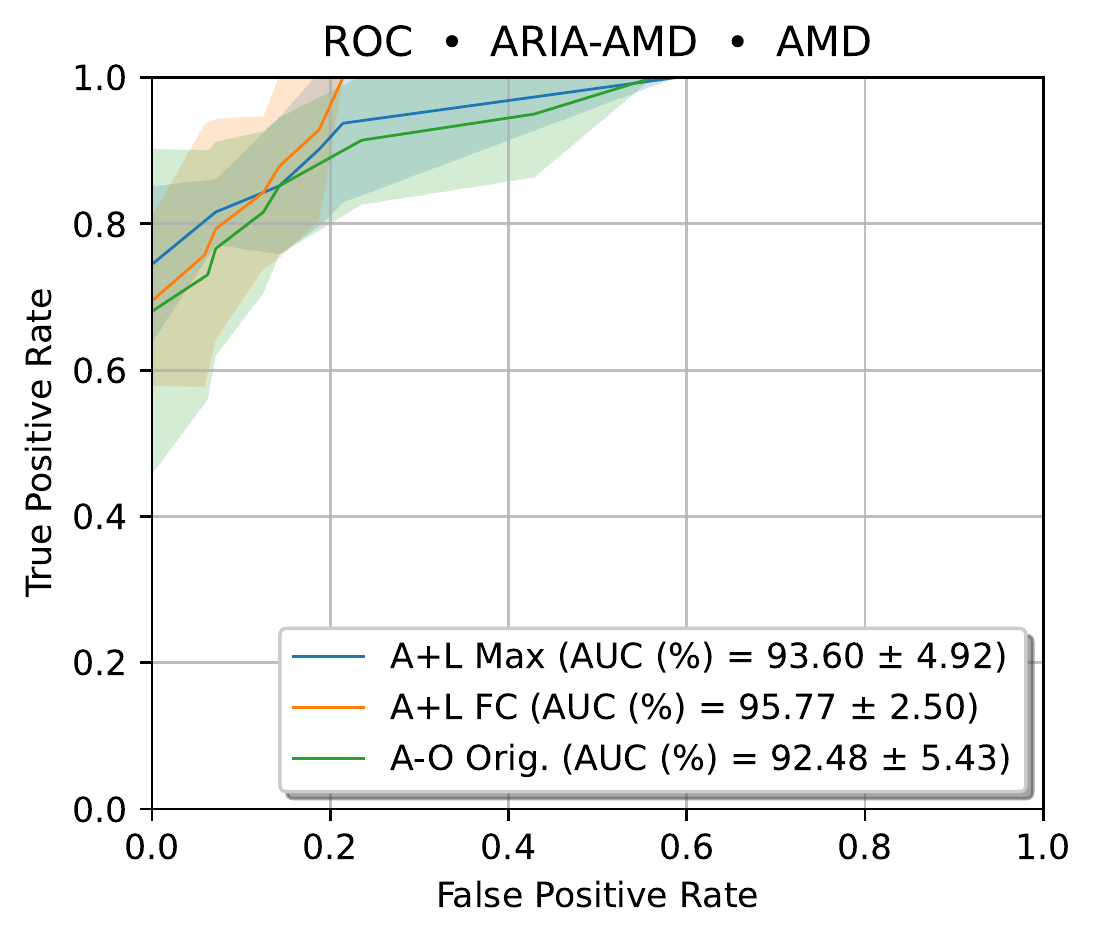}
    \includegraphics[width=0.48\textwidth]
        {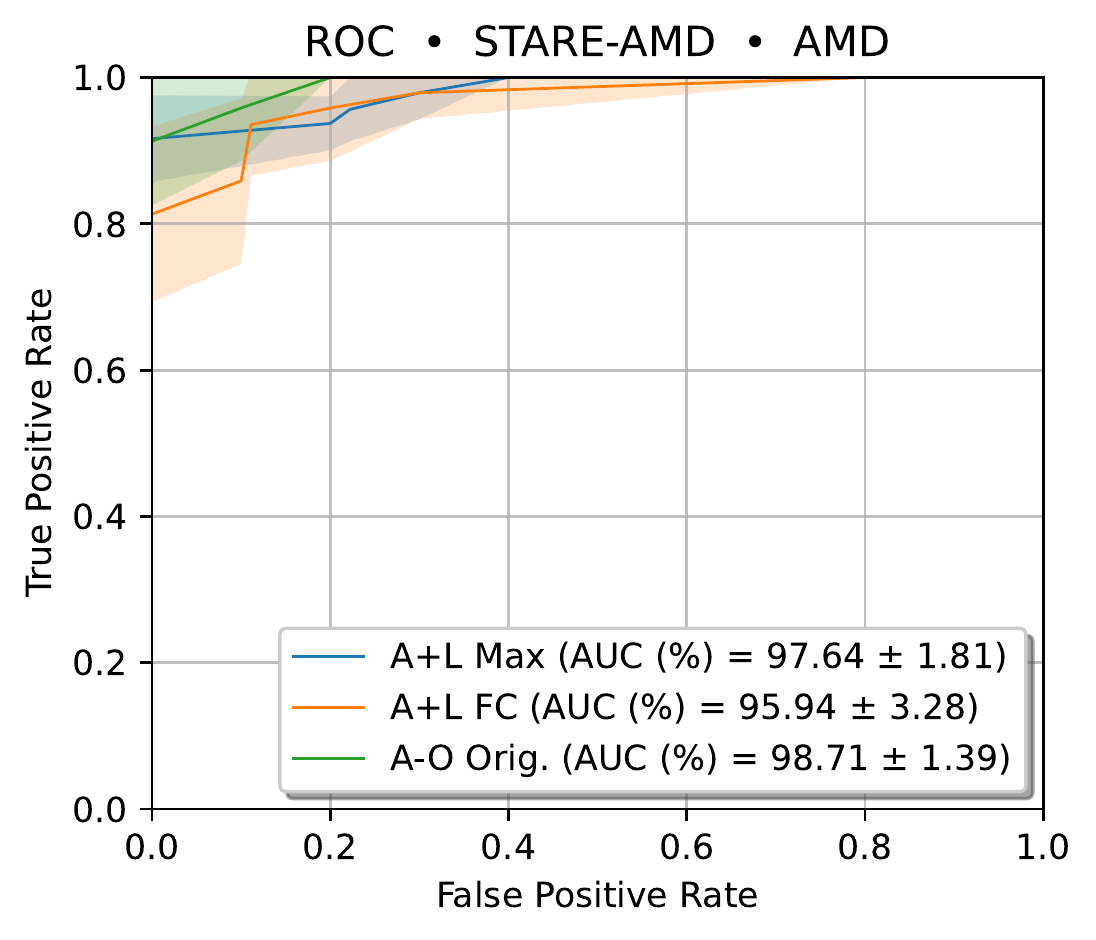}
    \caption{Mean ROC curves in AMD identification for the A+L and baseline (A-O) approaches in ADAM, ARIA and STARE datasets. All models were fine-tuned on the target datasets.}%
    \label{fig:ROCs_AMD_fine_tuned}
\end{figure}

Figure~\ref{fig:ROCs_AMD} depicts the mean ROC curves for AMD identification for both the A+L approach and the baseline A-O approach, without fine-tuning, in AMDLesions, ADAM, ARIA and STARE datasets.
None of the models were fine-tuned on the target datasets.
Similarly, Figure~\ref{fig:ROCs_AMD_fine_tuned} depicts the mean ROC curves for AMD identification for both the baseline and the A+L approaches, with fine-tuning, in AMDLesions, ADAM, ARIA and STARE datasets.
Complementarily, Table~\ref{tab:amd_identification} shows the mean AUC-ROC values of all the models in those datasets.
\begin{table}[tbp]
\centering
\caption{Mean AUC-ROC values in AMD identification of the A+L and A-O approaches in AMDLesions, ARIA, ADAM and STARE. Bold denotes the best mean value for each dataset.}%
\label{tab:amd_identification}
\begin{tabular}{llll}
\toprule
Dataset    & \multicolumn{3}{c}{AUC-ROC (\%)}              \\
\cmidrule{2-4}
           & A+L Max     & A+L FC      & A-O Original \\
\midrule

AMDLesions & $\mathbf{95.59\pm2.03}$ & $95.45\pm0.89$ & $95.35\pm0.69$  \\

\midrule

ARIA*   & $85.82\pm0.88$ & $\mathbf{86.87\pm2.38}$ & $86.72\pm0.97$  \\
ADAM*       & $\mathbf{80.17\pm1.95}$ & $72.79\pm5.75$ & $72.70\pm4.21$  \\
STARE*  & $86.28\pm4.28$ & $79.51\pm9.66$ & $\mathbf{87.62\pm2.87}$  \\

\midrule
    ARIA
    & $93.60 \pm 4.92$
    & $\mathbf{95.77 \pm 2.50}$
    & $92.48 \pm 5.43$  \\

    ADAM
    & $\mathbf{93.62 \pm 2.89}$
    & $93.29 \pm 3.07$
    & $92.97 \pm 1.87$  \\

    STARE
    & $97.64 \pm 1.81$
    & $95.94 \pm 3.28$
    & $\mathbf{98.71 \pm 1.39}$  \\

\bottomrule
\multicolumn{4}{l}{*Cross-dataset evaluation: trained on AMDLesions.}
\end{tabular}
\end{table}
The table shows the results of the models both with and without fine-tuning on the target datasets.

As can be seen both in Figures~\ref{fig:ROCs_AMD} and~\ref{fig:ROCs_AMD_fine_tuned} and Table~\ref{tab:amd_identification}, the results of the A+L models are very similar to those of the baseline approach with the original VGG-16 architecture.
Given the mean AUC-ROC values and the standard deviations provided in Table~\ref{tab:amd_identification}, no alternative can be said to be significantly superior to others.
The only exception is the A+L Max alternative without fine-tuning, which, in ADAM, is significantly better than the other two non-fine-tuned models ($p<0.02$).

The results from Figures~\ref{fig:ROCs_AMD} and~\ref{fig:ROCs_AMD_fine_tuned} and Table~\ref{tab:amd_identification} also show that the AMD identification performance of all fine-tuned models greatly surpasses the performance of their non-fine-tuned counterparts.
This improvement occurs for all models in all the datasets for which cross-dataset evaluation was performed: ADAM, ARIA and STARE.
The large gain in performance of fine-tuned models can be explained by the significant differences in the appearance of the images from the 3 datasets.
These differences can be seen at a glance in Figure~\ref{fig:datasets_examples} (Section~\ref{subsec:data}).
This issue---the performance drop in a cross-dataset scenario---is not unique to our work, but a known limitation of deep learning-based methodologies facing training and test data with dissimilar statistics~\cite{Zhang_ACMCS_2019,Galdran_NSR_2022,Zhang_TMI_2020,Qiao_CVPR_2020}.
In medical imaging, due to the acute data scarcity, this problem is particularly common~\cite{Zhang_ACMCS_2019}.
Still, most AUC-ROC values of the non-fine-tuned models in ADAM, ARIA and STARE are above 80\%.
Taking into account the inherent limitations of the datasets and the models, the results are satisfactory.

Looking at the results in Table~\ref{tab:amd_identification}, A+L Max seems to be the most stable A+L alternative, particularly in the cross-dataset scenario.

In light of the results, it can be stated that the proposed approach, particularly using the A+L Max variant, equals or surpasses the baseline approach (A-O) in AMD identification, despite being focused on additional valuable tasks related to diagnosis.

\subsubsection{Comparison with the state of the art}

Table~\ref{tab:sota_classification_results} shows the AMD identification results of the proposed A+L models and several state-of-the-art methods in the ADAM dataset.
In particular, the state-of-the-art methods are the top 5 methods of the ADAM challenge~\cite{Fang_ADAM_TMI_2022}.
\begin{table}[bp]
\centering
\caption{AMD identification results of the proposed A+L models and the top 5 methods of the ADAM challenge~\cite{Fang_ADAM_TMI_2022} in the ADAM dataset. Note that all the results are in ADAM, but not exactly in the same data, as the challenge test set is not public. For our method, the results correspond to the 4-fold evaluation on the training set, while those of the other methods to the evaluation in the final test set of the challenge. Bold denotes the highest mean AUC-ROC value.}%
\label{tab:sota_classification_results}
\begin{tabular}{@{\extracolsep{8pt}}ll}
\toprule

    Method
    & AUC-ROC (\%) \\

\midrule

    VUNO EYE TEAM
    & \textbf{97.14} \\

    ForbiddenFruit
    & 95.92 \\

    Zasti\_AI
    & 95.81 \\

    Muenai\_Tim
    & 93.99 \\

    \textit{A+L Max}
    & $93.62 \pm 2.89$ \\

    \textit{A+L FC}
    & $93.29 \pm 3.07$ \\

    ADAM-TEAM
    & 92.87 \\

\bottomrule
\end{tabular}
\end{table}
Since the final test set of the challenge is not public, we fine-tuned and evaluated our model solely in the ADAM training set.
Specifically, we performed 4-fold cross-validation in this set with randomly created folds.
Thus, provided AUC-ROC values correspond to the mean AUC-ROC values of the 4 folds.
For the rest of the methods, we show the AUC-ROC values they obtained in the definitive test set of the challenge finals~\cite{Fang_ADAM_TMI_2022}.

As can be seen in Table~\ref{tab:sota_classification_results}, our method obtains competitive results in the identification of AMD in the ADAM dataset.
Specifically, in this task, the A+L Max and A+L FC approaches rank 4th and 5th, respectively, among the methods of the 11 teams invited to the finals (out of 610 participating teams).
It is worth noting that all the challenge methods are focused solely on the identification of AMD.
Differently, the main aim of the A+L approach is to provide an explainable method that, along with the identification of AMD, provides the identification of AMD-associated lesions and their corresponding lesion activation maps, without the need of pixel-level annotations.
The explainability and the extra information provided by the A+L models further emphasizes the value of their results and the adequacy of the proposed approach.

\subsection{Identification of lesions}

\begin{figure}[tbp]
    \centering
    \includegraphics[width=0.32\textwidth]
        {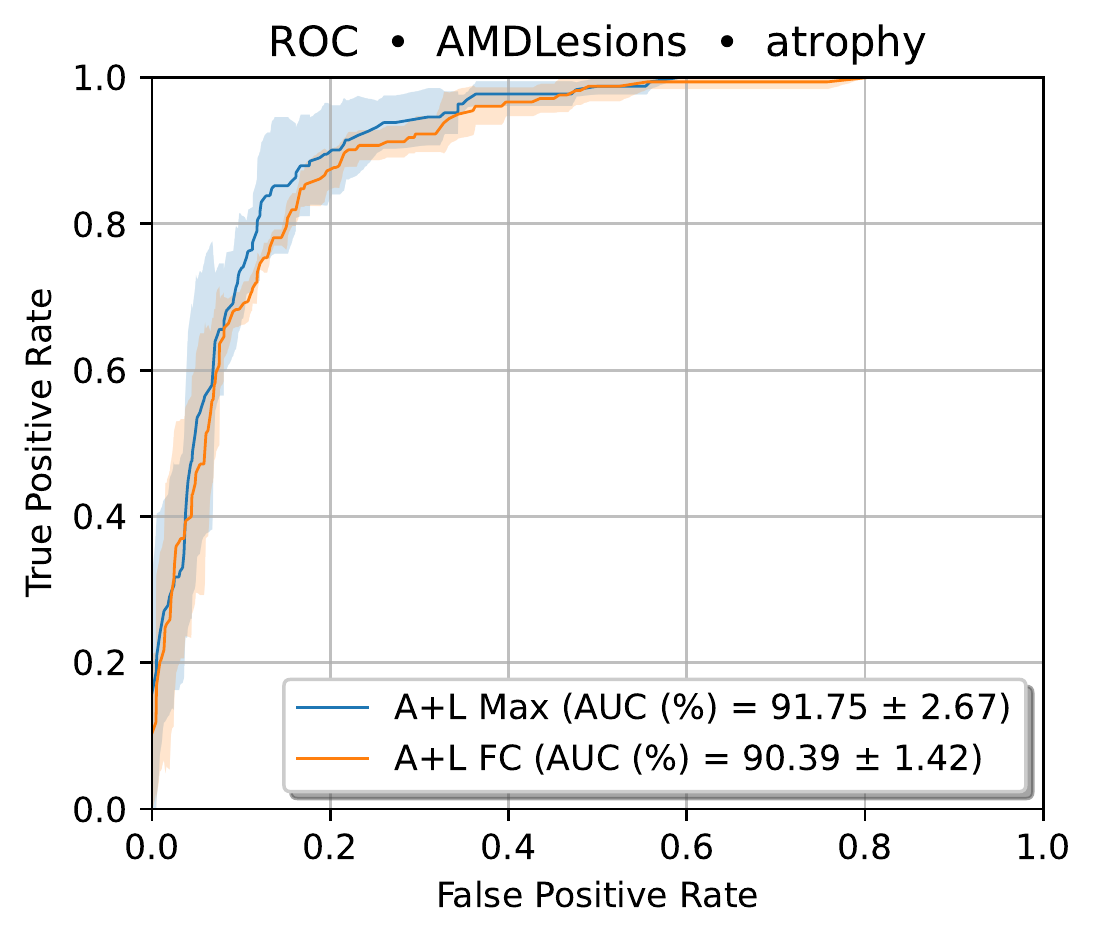}
    \includegraphics[width=0.32\textwidth]
        {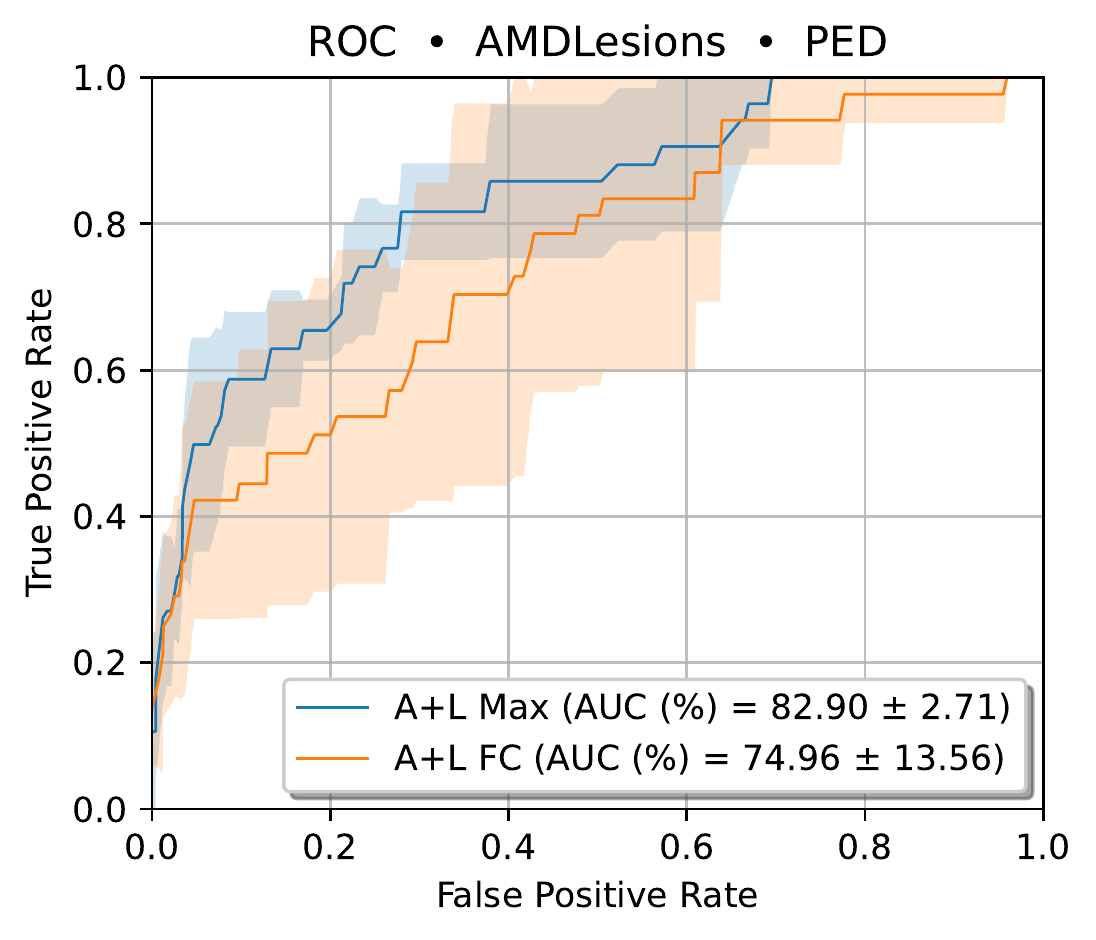}
    \includegraphics[width=0.32\textwidth]
        {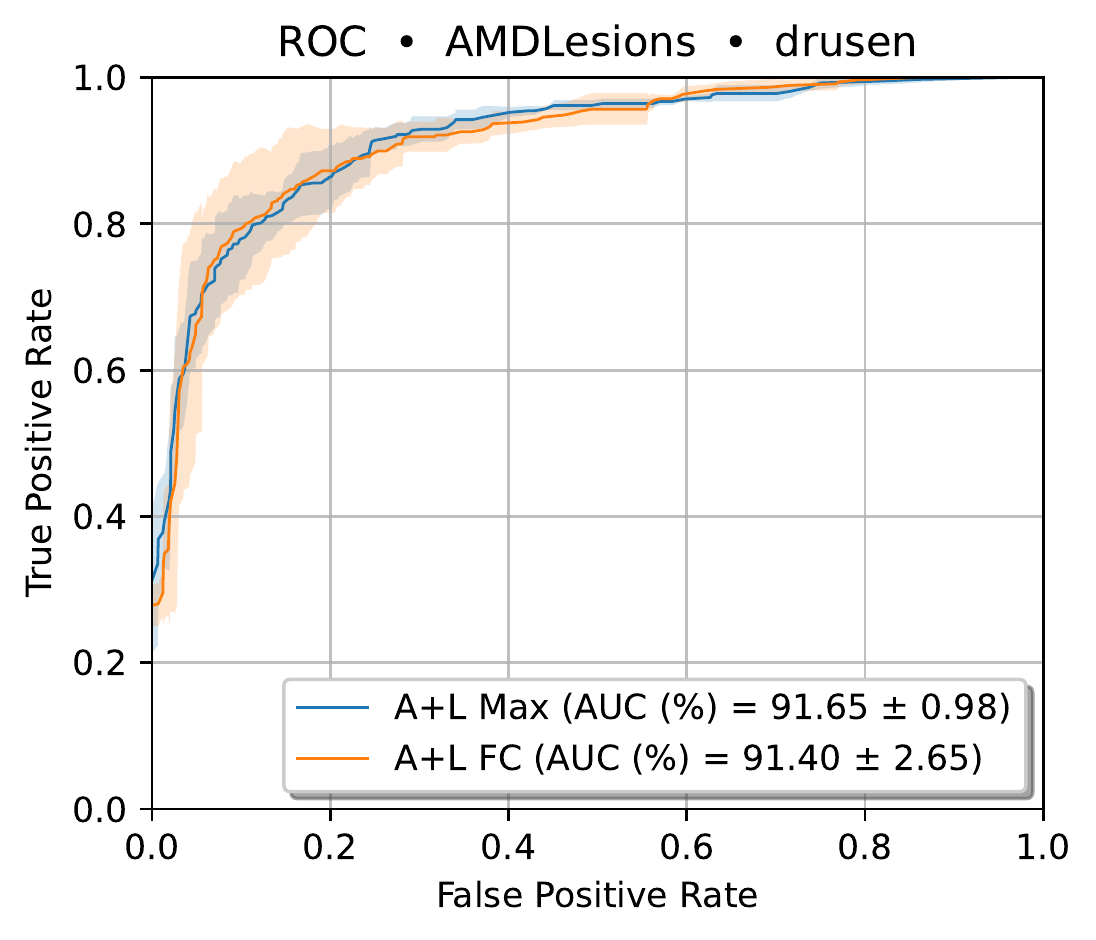}
    \includegraphics[width=0.32\textwidth]
        {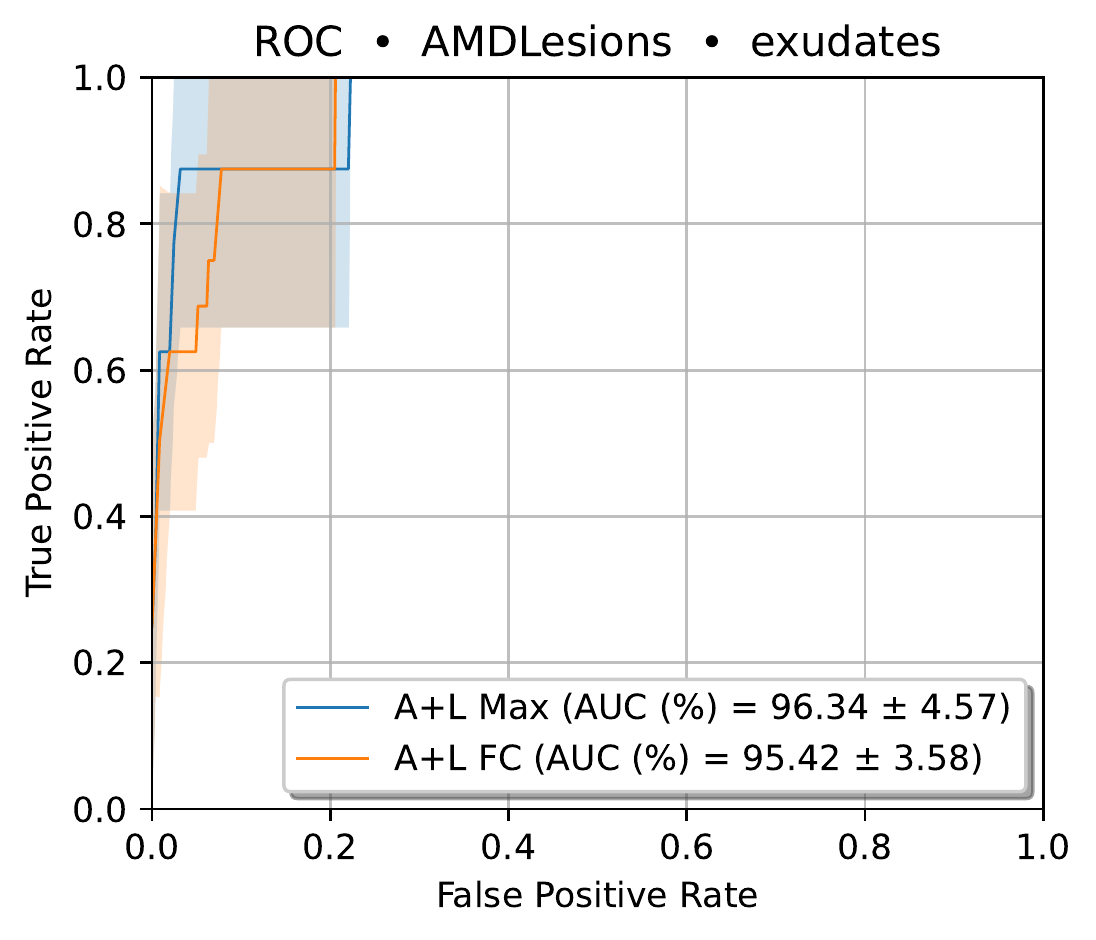}
    \includegraphics[width=0.32\textwidth]
        {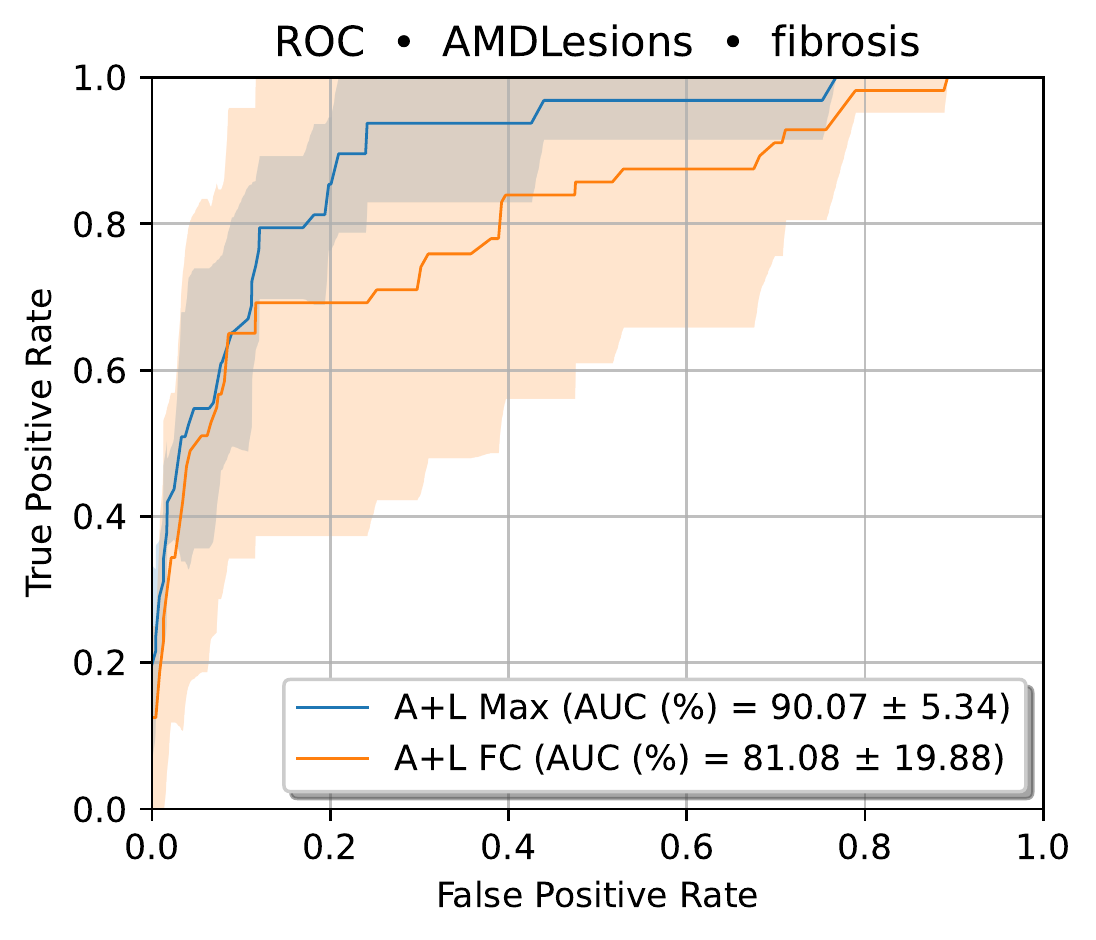}
    \includegraphics[width=0.32\textwidth]
        {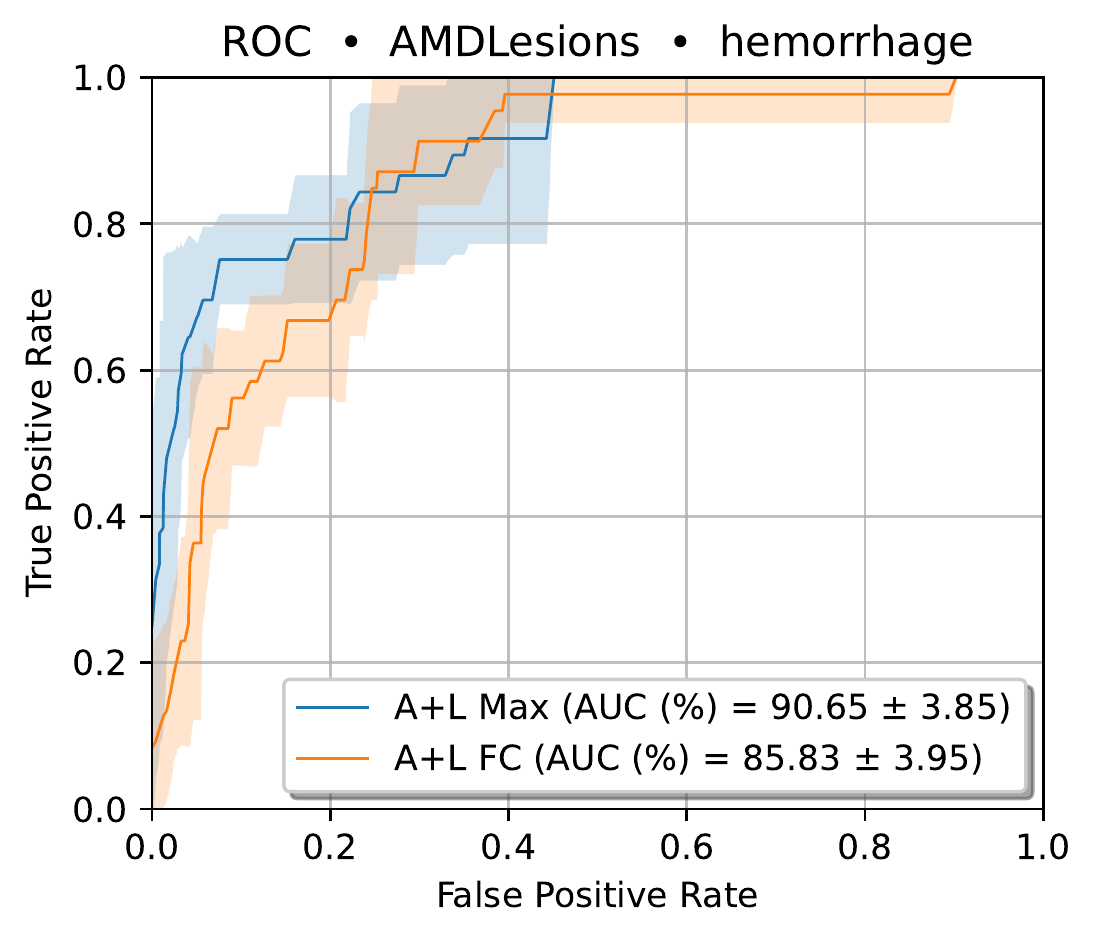}
    \includegraphics[width=0.32\textwidth]
        {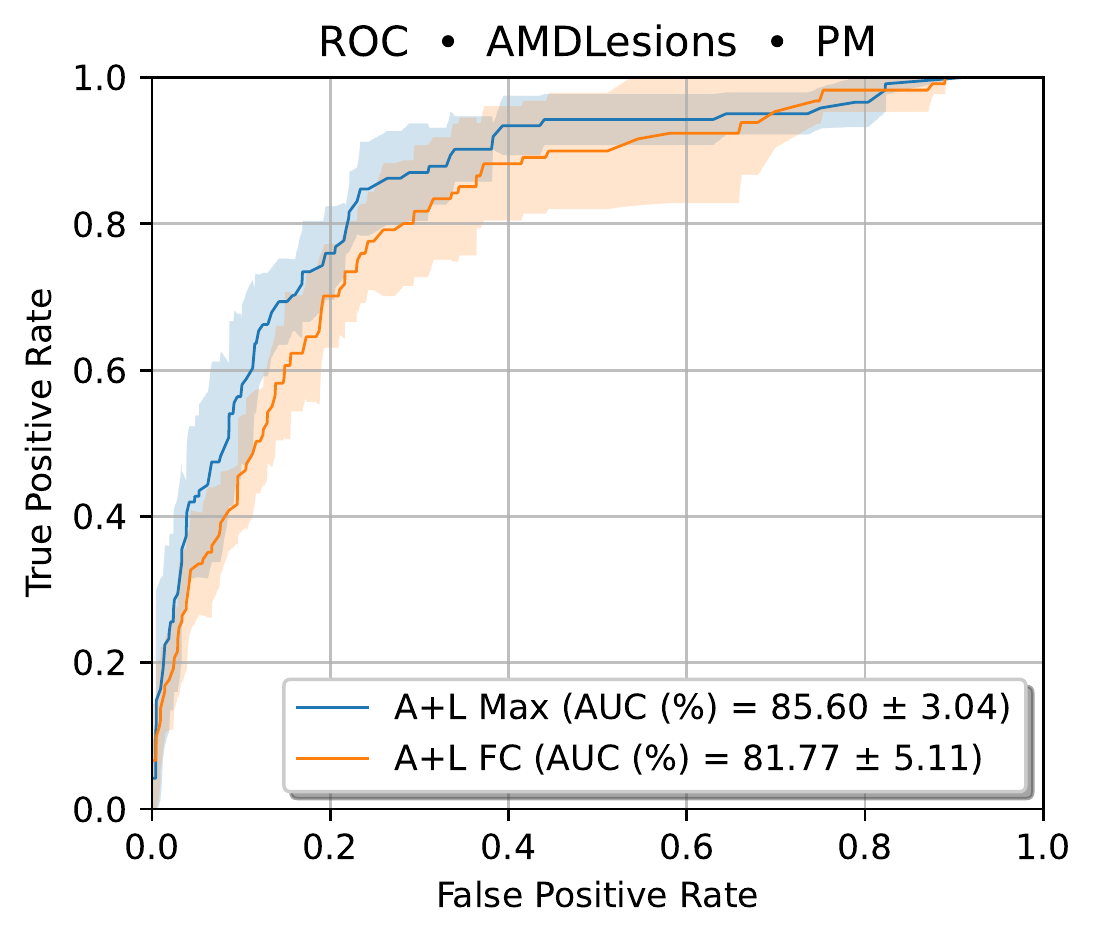}
    \includegraphics[width=0.32\textwidth]
        {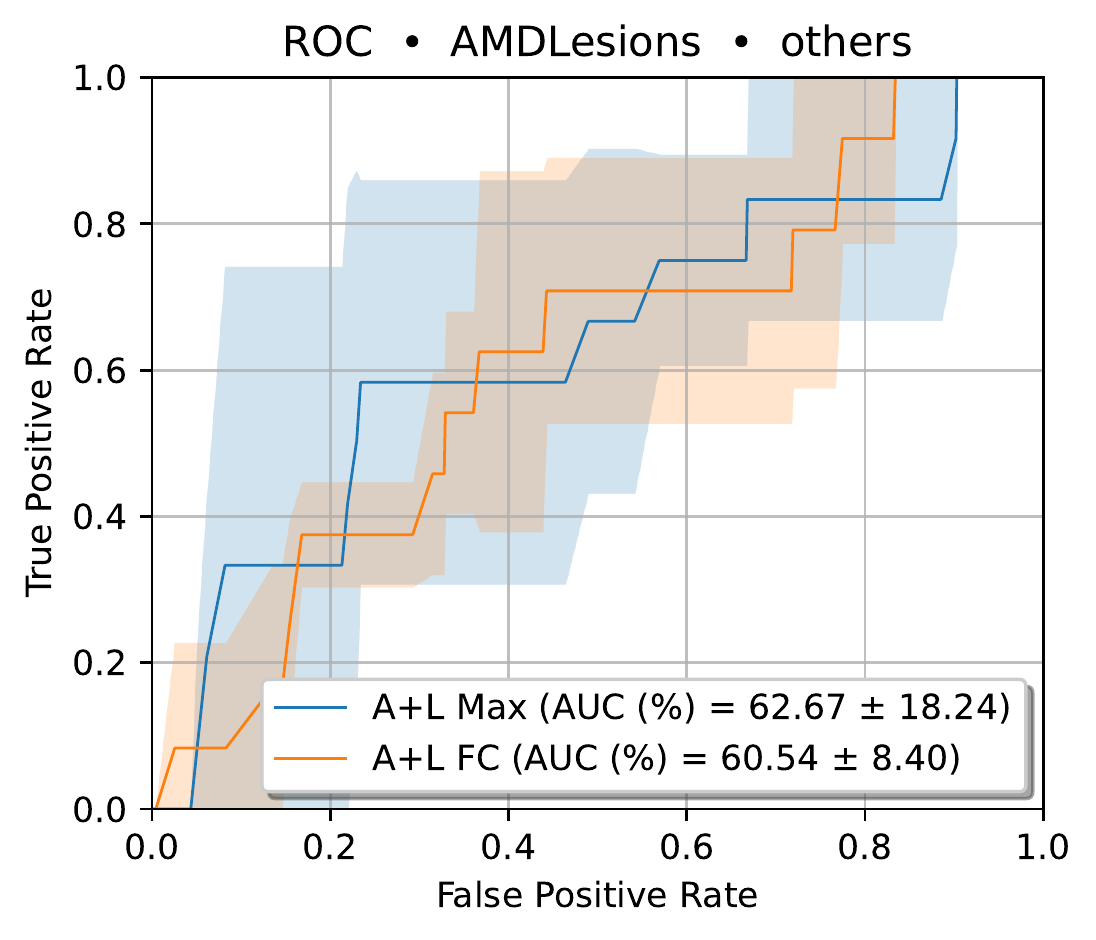}
    \includegraphics[width=0.32\textwidth]
        {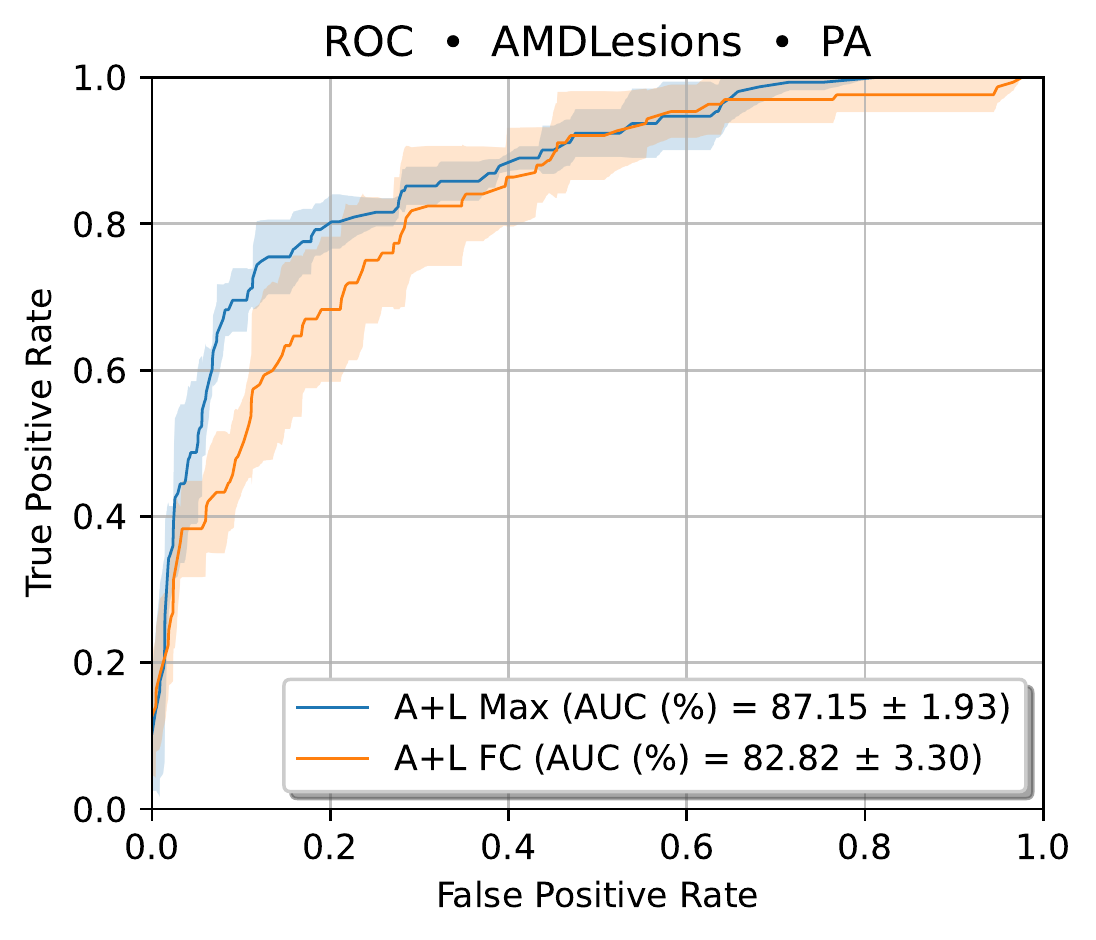}
    \caption{Mean ROC curves in lesion identification for the different A+L approaches in AMDLesions.}%
    \label{fig:ROCs_Lesions_AMDLesions}
\end{figure}

\begin{figure}[tbp]
    \centering
    \includegraphics[width=0.32\textwidth]
        {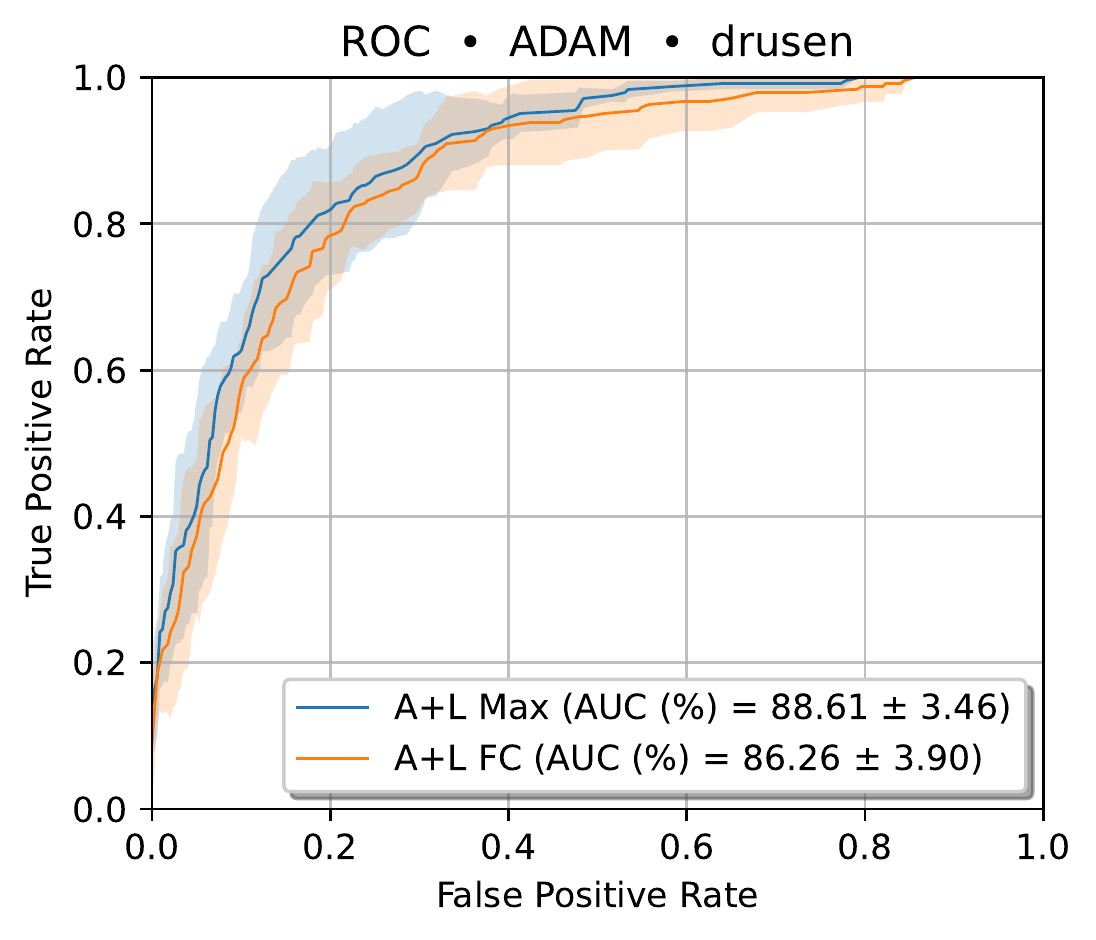}
    \includegraphics[width=0.32\textwidth]
        {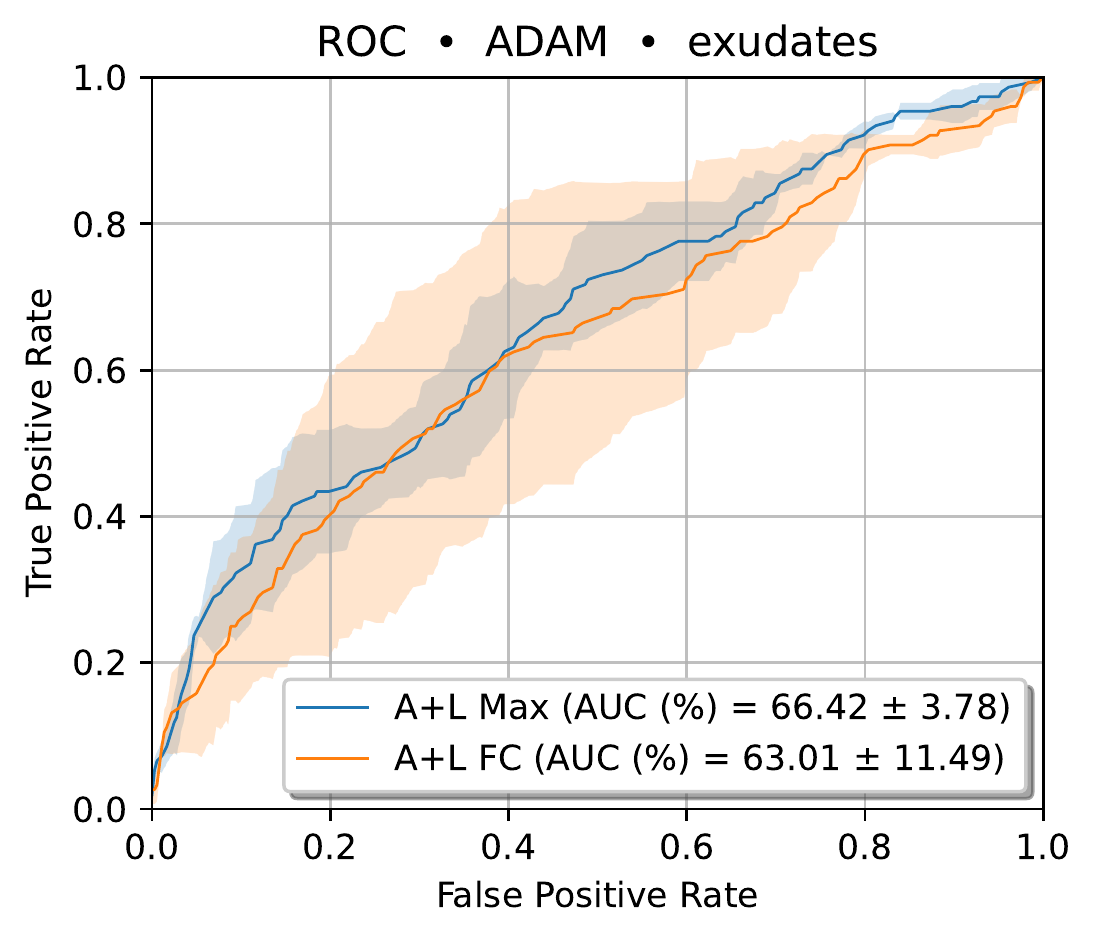} \\
    \includegraphics[width=0.32\textwidth]
        {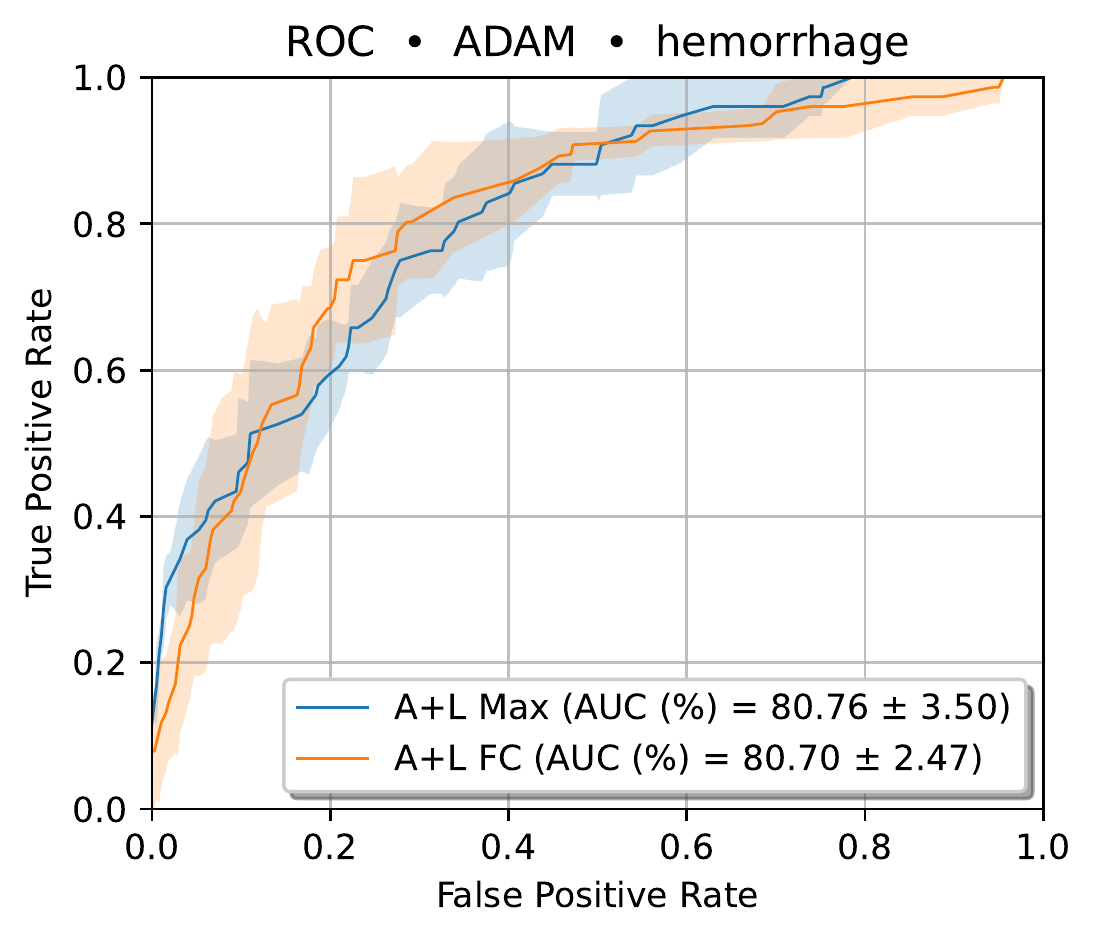}
    \includegraphics[width=0.32\textwidth]
        {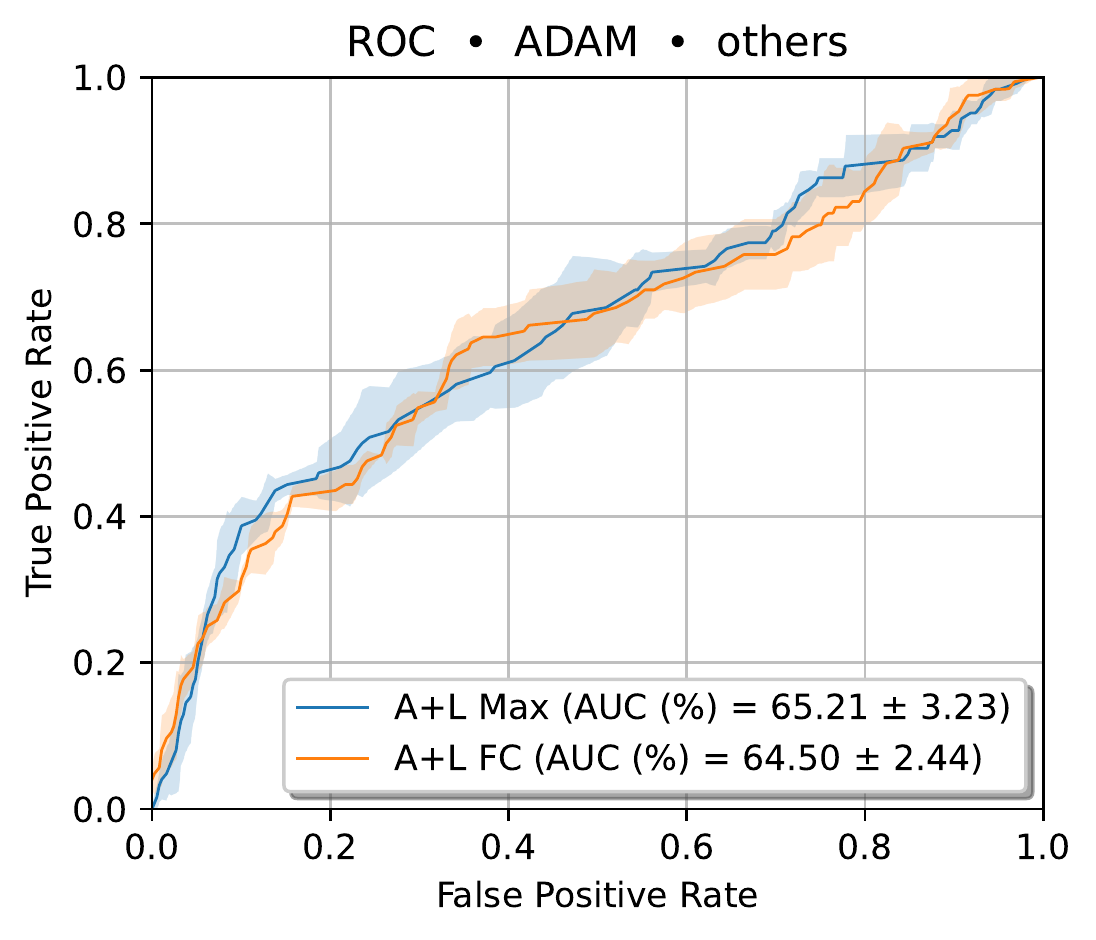}
    \caption{Mean ROC curves in lesion identification of the A+L alternatives in the ADAM dataset. Models were trained on AMDLesions.}%
    \label{fig:ROCs_Lesions_identification_ADAM}
\end{figure}

In Figure~\ref{fig:ROCs_Lesions_AMDLesions}, we depict the mean ROC curves for the identification of lesions in the AMDLesions dataset for the A+L models.
In addition, Table~\ref{tab:lesion_identification_AMDLesions} reports the mean AUC-ROC values and the standard deviations of the models in the same task.
\begin{table}[tbp]
\centering
\caption{Mean AUC-ROC values and standard deviations for lesion identification in AMDLesions. Bold denotes the best mean value for each lesion.}%
\label{tab:lesion_identification_AMDLesions}
\begin{tabular}{lll}
\toprule
Lesion        & \multicolumn{2}{c}{AUC-ROC (\%)} \\
\cmidrule{2-3}
             & A+L Max  & A+L FC \\
\midrule
atrophy      & $\mathbf{91.75\pm2.67}$  & $90.39\pm1.42$  \\
drusen       & $\mathbf{91.65\pm0.98}$  & $91.40\pm2.65$  \\
exudates     & $\mathbf{96.34\pm4.57}$  & $95.42\pm3.58$  \\
fibrosis     & $\mathbf{90.07\pm5.34}$  & $81.08\pm19.88$  \\
hemorrhage   & $\mathbf{90.65\pm3.85}$  & $85.83\pm3.95$  \\
PM           & $\mathbf{85.60\pm3.04}$  & $81.77\pm5.11$  \\
PA           & $\mathbf{87.15\pm1.93}$  & $82.82\pm3.30$  \\
PED          & $\mathbf{82.90\pm2.71}$  & $74.96\pm13.56$  \\
others       & $\mathbf{62.67\pm18.24}$ & $60.54\pm8.40$  \\
\bottomrule
\end{tabular}
\end{table}
Figure~\ref{fig:ROCs_Lesions_identification_ADAM} depicts the mean ROC curves of the A+L models trained in AMDLesions for lesion identification in the ADAM dataset, while Table~\ref{tab:lesion_identification_ADAM} reports the corresponding mean AUC-ROC values.
The details for this cross-dataset evaluation are described in Section~\ref{subsubsec:cross-dataset_evaluation}.
\begin{table}[tbp]
\centering
\caption{Mean AUC-ROC values and standard deviations for lesion identification in ADAM. Bold denotes the best mean value for each lesion.}%
\label{tab:lesion_identification_ADAM}
\begin{tabular}{lll}
\toprule
Lesion        & \multicolumn{2}{c}{AUC-ROC (\%)} \\
\cmidrule{2-3}
             & A+L Max  & A+L FC \\
\midrule
drusen       & $\mathbf{88.61\pm3.46}$ & $86.26\pm3.90$  \\
exudates     & $\mathbf{66.42\pm3.78}$ & $63.01\pm11.49$  \\
hemorrhage   & $\mathbf{80.76\pm3.50}$ & $80.70\pm2.47$  \\
others       & $\mathbf{65.21\pm3.23}$ & $64.50\pm2.44$  \\
\bottomrule
\end{tabular}
\end{table}

As can be observed in Figure~\ref{fig:ROCs_Lesions_AMDLesions} and Table~\ref{tab:lesion_identification_AMDLesions}, the proposed approach allows the identification of most lesions in AMDLesions.
In that regard, both A+L Max and A+L FC provide particularly accurate results for drusen, atrophy and exudates, whereas A+L Max also provides similarly good results for fibrosis and hemorrhage.
This is highly convenient, since the clinicians particularly focus on these lesions during the diagnostic process.
This is because the localization and quantification of drusen determines the grade of development of the disease, while the presence of atrophy is directly related to 90\% of cases of late AMD~\cite{Ferris_Ophth_2013}.
Moreover, exudates are a common sign of neovascular AMD~\cite{Tan_AP_2017}.
In contrast with these satisfactory results, we found that the mean AUC-ROC for the `others' group does not surpass the 65\%, and that the performance for this class highly depends on the evaluated fold (as indicated by the high standard deviations [$\sigma$]: $\sigma>8$ for A+L FC and $\sigma>18$ for A+L Max).
This lower performance can be explained by the limited examples that are available for this class as well as the high intra-class variability (11 images containing 5 different types of lesions to be distributed in 4 folds). 
With so few examples, it is difficult for the models to be able to learn the representative features of the lesions.
Even more so in cases where these features are very diverse---as is the case of `others'.
It is probable that increasing the number of examples of these under-represented classes would result in a significant gain in the identification performance.

In ADAM, the mean AUC-ROC values for drusen and `others' (see Table~\ref{tab:lesion_identification_ADAM}) are similar to those of AMDLesions.
Regarding hemorrhage, the AUC-ROC values are slightly lower.
However, considering that it is a cross-dataset evaluation, the results are also satisfactory.
Conversely, the results for exudates show a more significant drop in performance with respect to AMDLesions.
In this case, the small number of samples available in AMDLesions (only 10) seems to compromise the generalization ability of the model.
This may be due to the limited diversity that is provided by only a few samples of exudates.
As was the case with other lesions, it is very likely that a larger number of examples for `exudates' in the training datasets would make the results improve significantly.

Looking at the results of A+L models separately, the Max variant performs better in most cases.
However, the differences are not significant.
In any case, allowing the models to freely weight the lesion predictions to obtain the diagnosis (FC variant) has no observable benefit in these tasks.

In sum, both A+L variants provide an adequate performance in the identification of most lesions, even in a cross-dataset scenario.
Moreover, lesions for which a low performance has been observed always present a very low number of training samples (e.g. exudates) and, in the case of `others', a substantial intra-class variability.
In these cases, it is expected that the addition of more training samples would significantly enhance the performance of the models.

In contrast to the traditional A-O approach, the lesion information provided by A+L helps to better understand the decisions made by the model.
In this case, the diagnosis can be explained by examining the lesion predictions.
Furthermore, this information also complements the diagnosis by allowing the clinicians to easily assess the severity of the disease.
As indicated in Section~\ref{sec:introduction}, the distinction between lesions is crucial in determining the developmental stage of AMD.
For example, it is very different to be affected by AMD and have only drusen than to present both drusen and atrophy.
In the former case, the disease is either at the early stage or at the intermediate stage, while in the latter it is very likely to be at the late stage.
This difference completely changes the clinical approach to the disease.

As a possible drawback of our proposal, it can be pointed out that image-level lesion labels are necessary for training the networks.
However, the collection of these labels would be, in most cases, quite straightforward, as the information is usually present in medical records.
This is because the lesion identification is an indispensable part of the diagnostic and monitoring processes.
Thus, in contrast to other tasks, such as lesion segmentation, the identification of lesions does not require a great effort on the part of the clinicians.
In some cases, the datasets could even be constructed \textit{a posteriori}, avoiding the common ad hoc implication of experts in the labeling process.

\subsection{Coarse lesion segmentation}

\begin{figure}[tbp]
    \centering
    \includegraphics[width=0.32\textwidth]
        {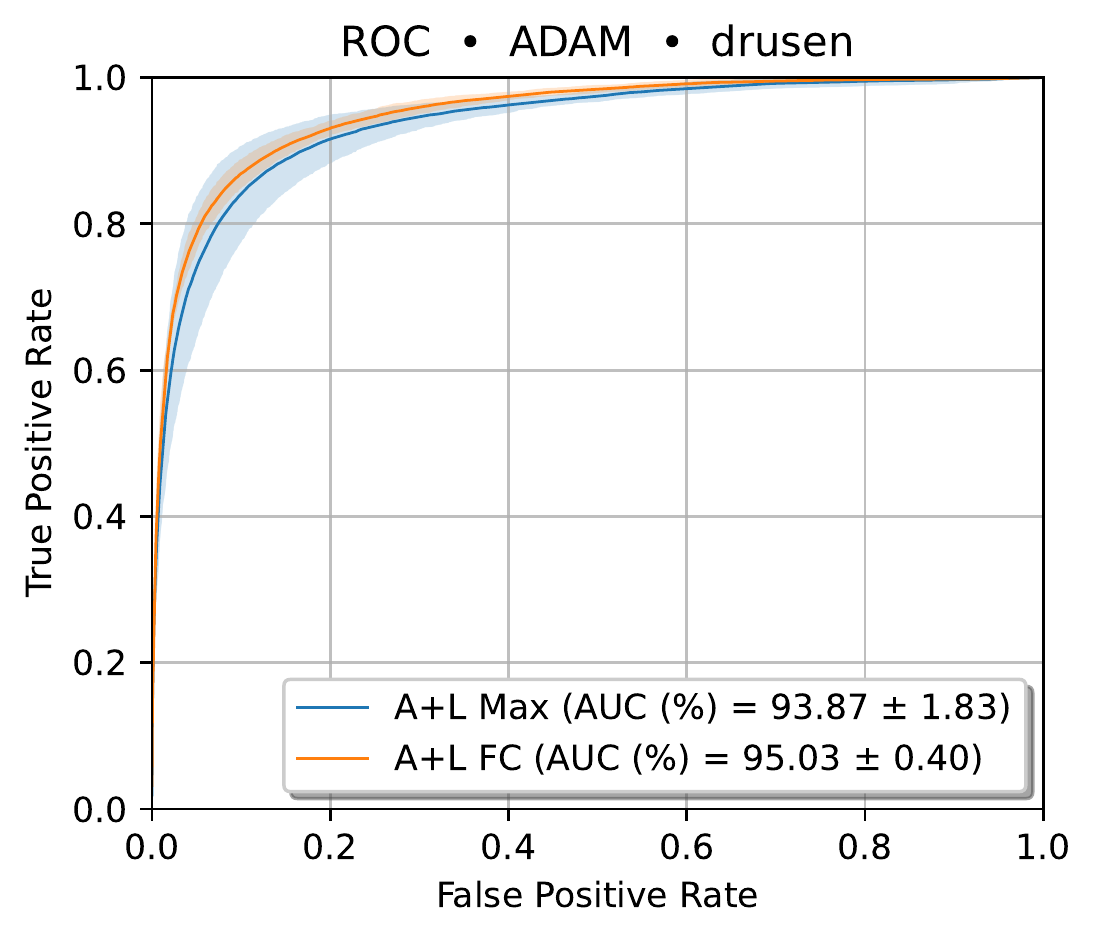}
    \includegraphics[width=0.32\textwidth]
        {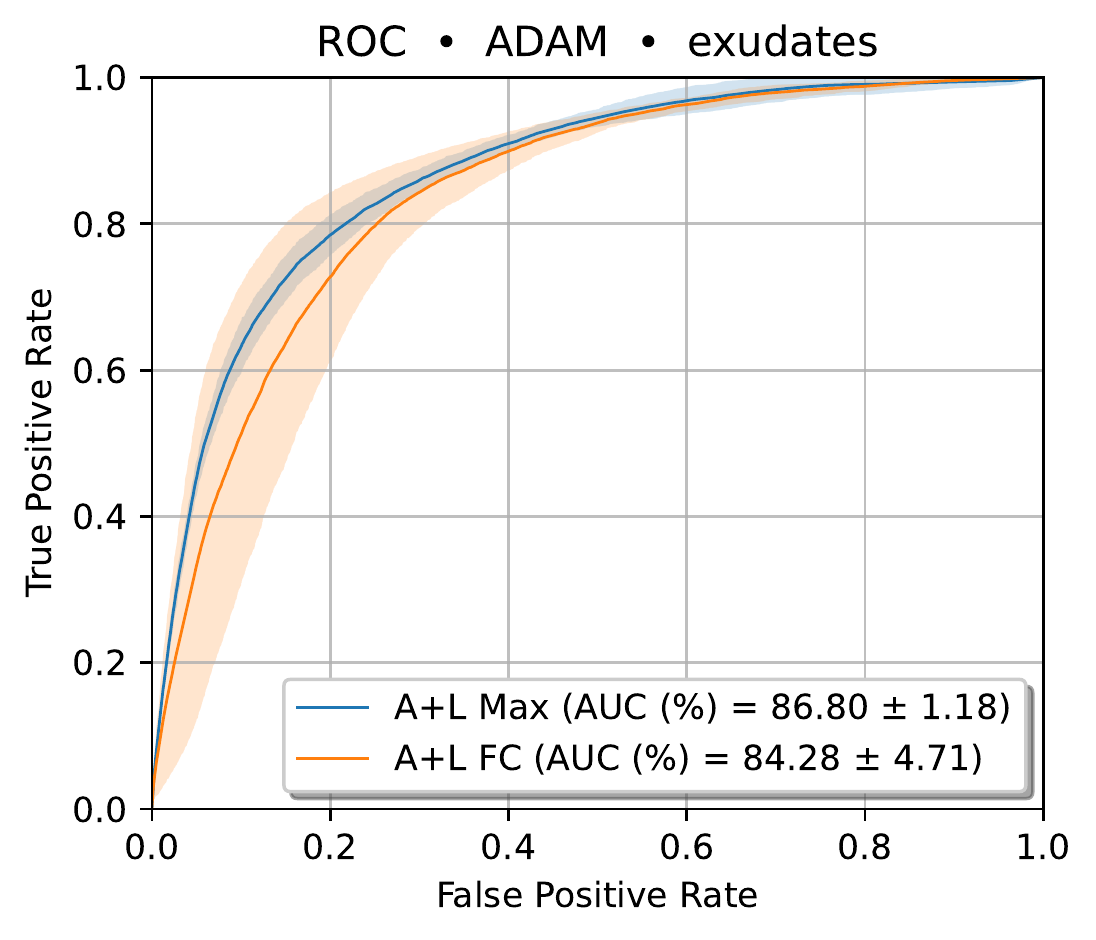} \\
    \includegraphics[width=0.32\textwidth]
        {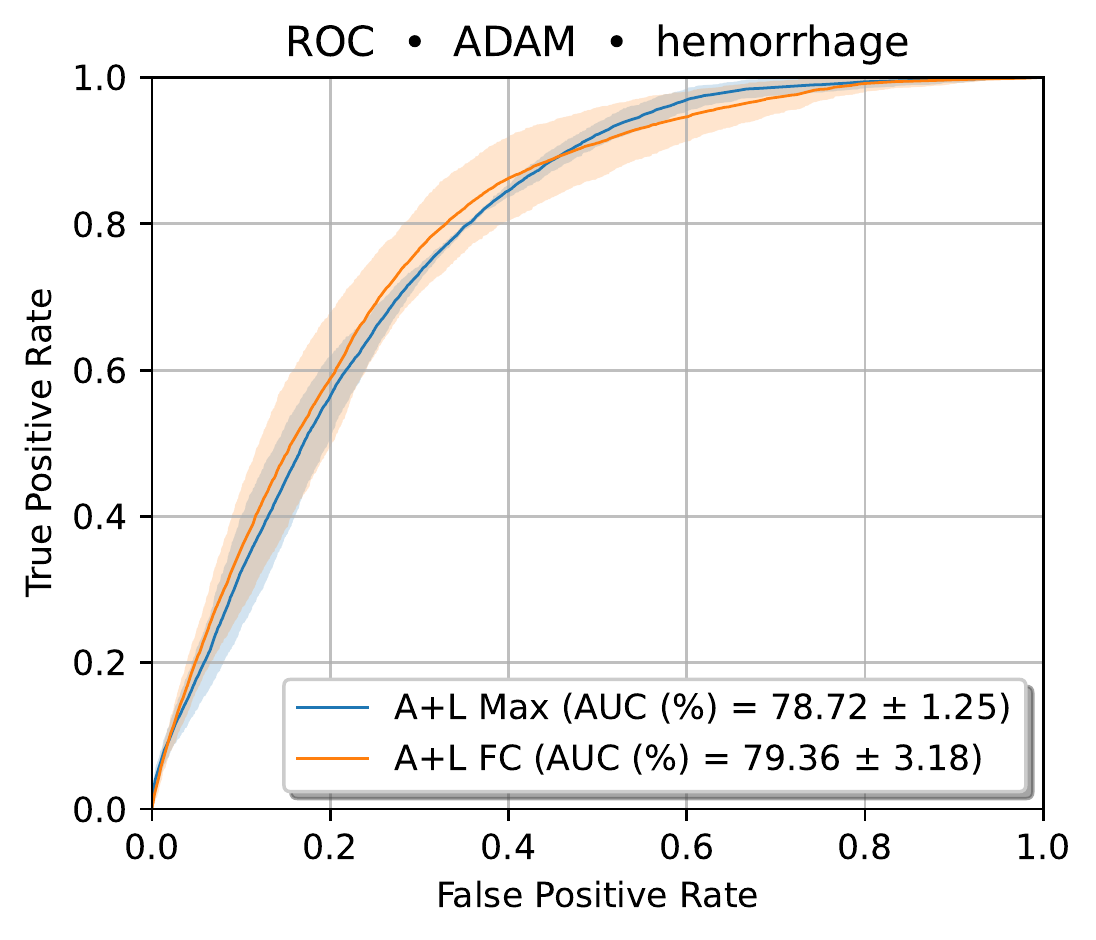}
    \includegraphics[width=0.32\textwidth]
        {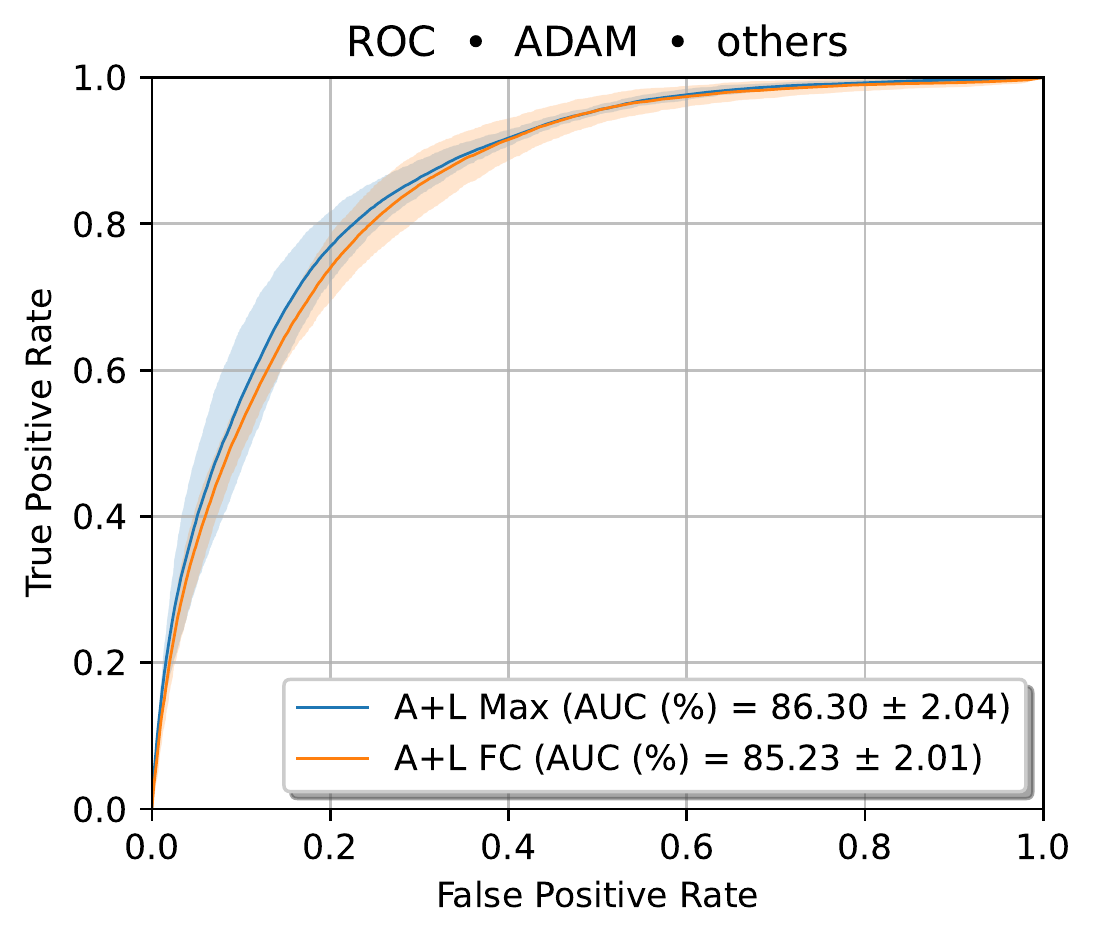}
    \caption{Mean ROC curves in coarse lesion segmentation of the A+L alternatives in the ADAM dataset. Models were trained on AMDLesions.}%
    \label{fig:ROCs_Lesions_segmentation_ADAM}
\end{figure}

Figure~\ref{fig:ROCs_Lesions_segmentation_ADAM} depicts the mean ROC curves of A+L models for the coarse segmentation of lesions in the AMDLesions dataset.
Complementarily, Table~\ref{tab:AUCs_lesion_segmentation_ADAM} reports the mean AUC-ROC values and the standard deviations of the models in the same task.
\begin{table}[tbp]
\centering
\caption{Mean AUC-ROC values and standard deviations for lesion segmentation in ADAM. Bold denotes the best mean value for each lesion.}%
\label{tab:AUCs_lesion_segmentation_ADAM}
\begin{tabular}{lll}
\toprule
Lesion        & \multicolumn{2}{c}{AUC-ROC (\%)} \\
\cmidrule{2-3}
             & A+L Max  & A+L FC \\
\midrule
drusen       & $93.87\pm1.83$ & $\mathbf{95.03\pm0.40}$  \\
exudates     & $\mathbf{86.80\pm1.18}$ & $84.28\pm4.71$  \\
hemorrhage   & $78.72\pm1.25$ & $\mathbf{79.37\pm3.18}$  \\
others       & $\mathbf{86.30\pm2.05}$ & $85.23\pm2.01$  \\
\bottomrule
\end{tabular}
\end{table}

\begin{figure}[tbp]
    \captionsetup[subfigure]{labelformat=empty}
    \centering
    \textbf{ADAM} \\\medskip
    \subfloat[Drusen]{\includegraphics[width=0.49\textwidth]
        {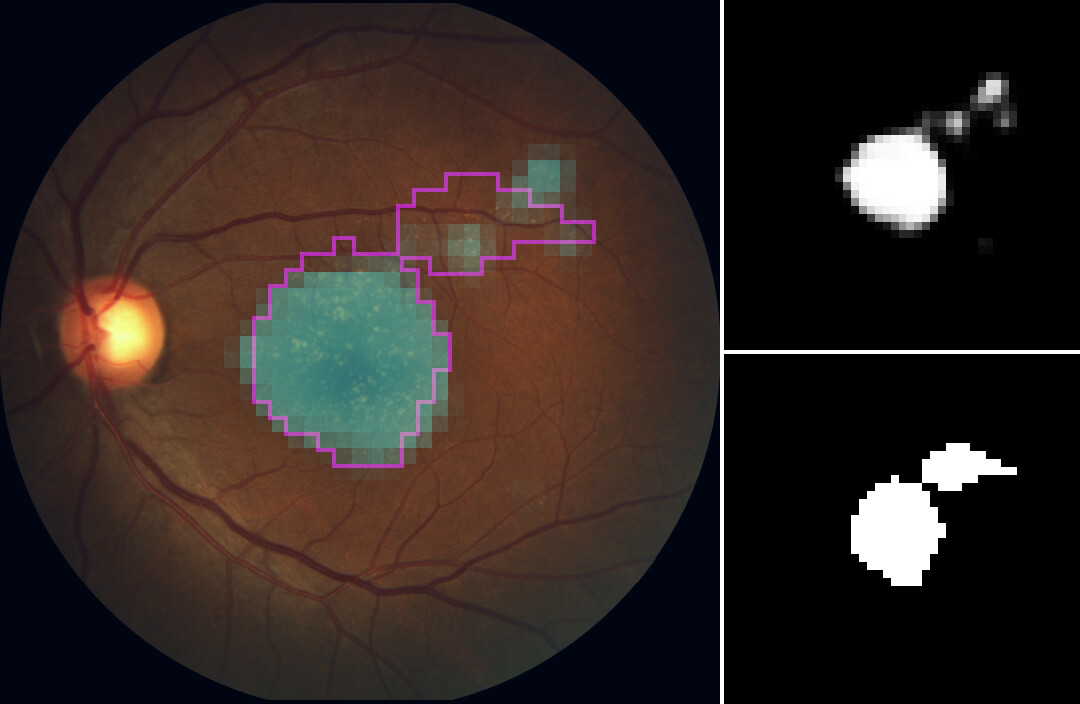}}
    \hfill
    \subfloat[Others]{\includegraphics[width=0.49\textwidth]
        {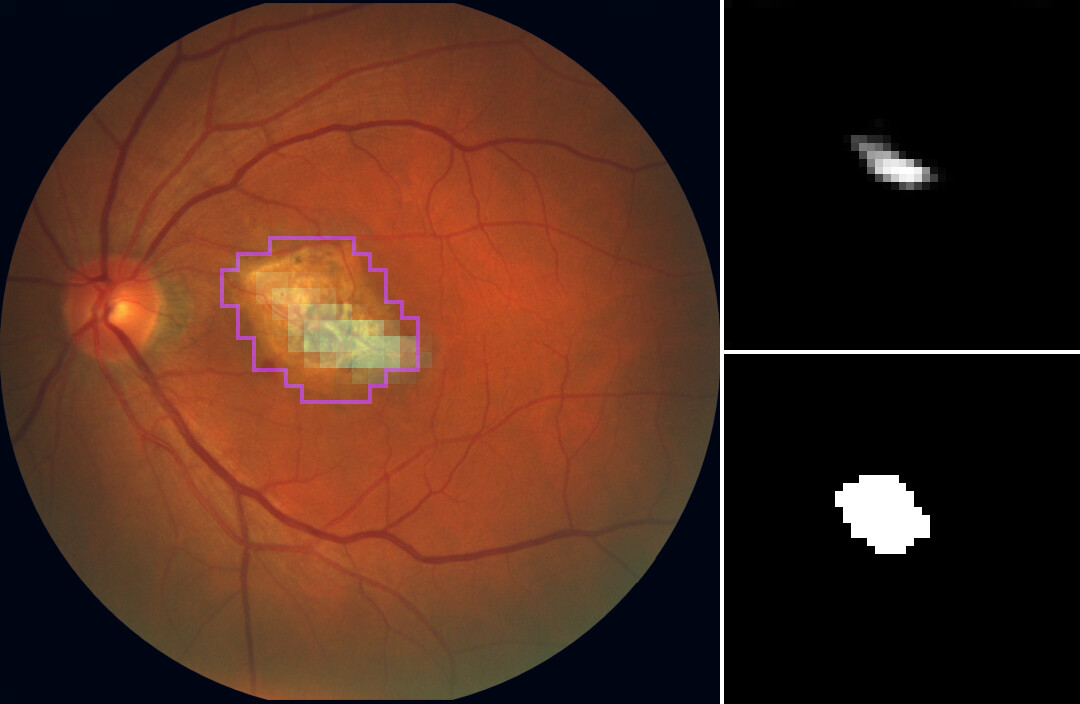}} \\\medskip
    \subfloat[Exudates]{\includegraphics[width=0.49\textwidth]
        {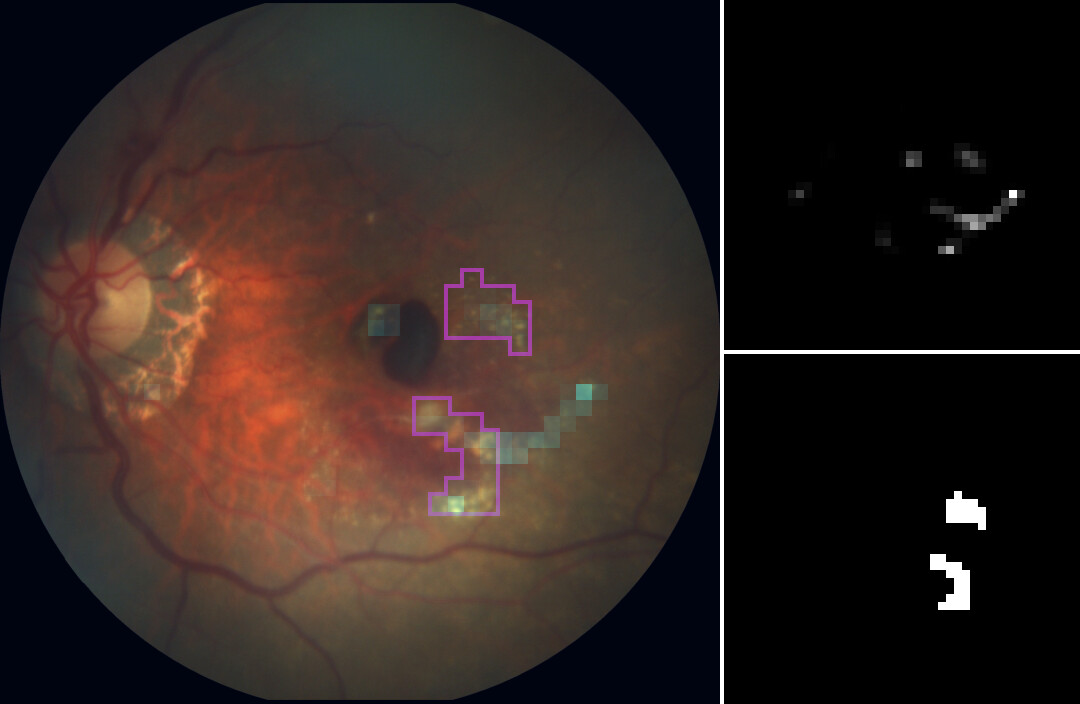}}
    \hfill
    \subfloat[Hemorrhage]{\includegraphics[width=0.49\textwidth]
        {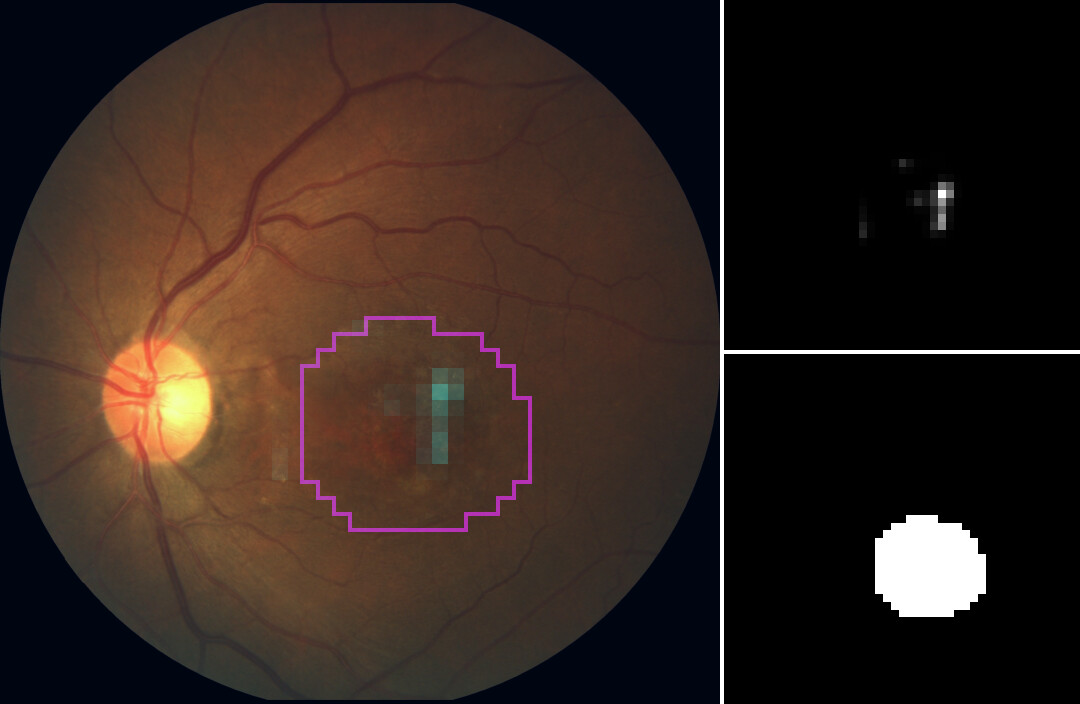}}
    \caption{Examples of lesion activation maps provided by the A+L FC models for multiple ADAM images. In each case, left image depicts the activation map of the lesion from the caption over the original retinography, as well as the contour of the corresponding segmentation ground truth employed in the evaluation (in magenta). Both the activation map (top) and the ground truth (bottom) are depicted separately on the right.}%
    \label{fig:activations_ADAM_FC}
\end{figure}

\begin{figure}[tbp]
    \captionsetup[subfigure]{labelformat=empty}
    \centering
    \textbf{ADAM} \\\medskip
    \subfloat[Drusen]{\includegraphics[width=0.49\textwidth]
        {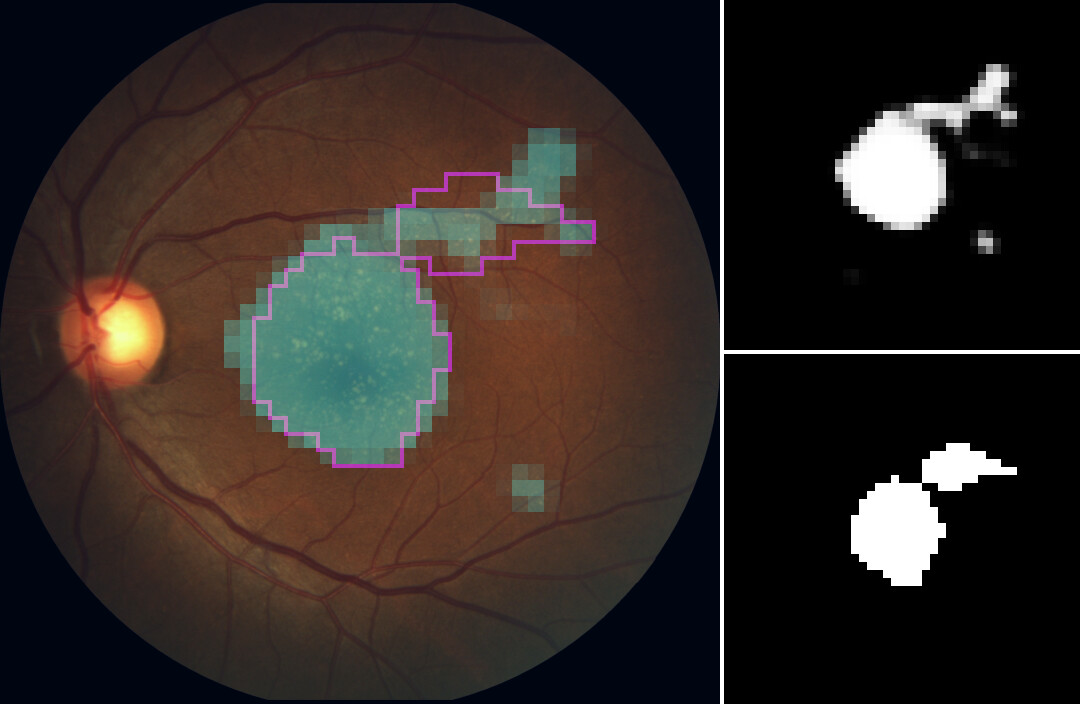}}
    \hfill
    \subfloat[Others]{\includegraphics[width=0.49\textwidth]
        {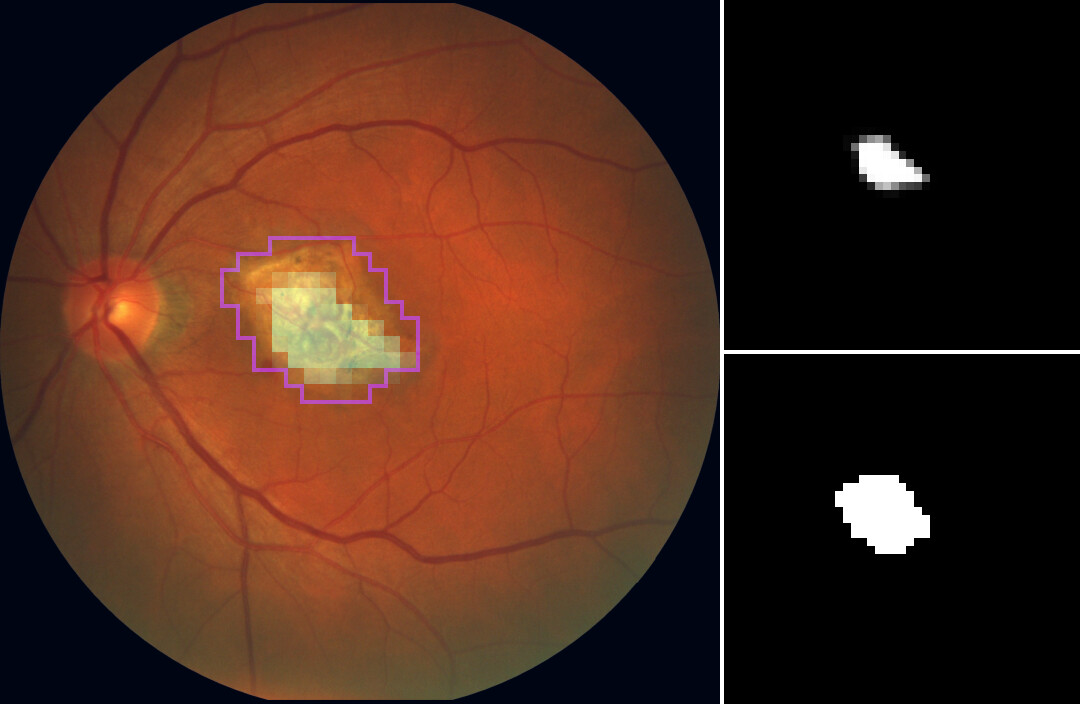}} \\\medskip
    \subfloat[Exudates]{\includegraphics[width=0.49\textwidth]
        {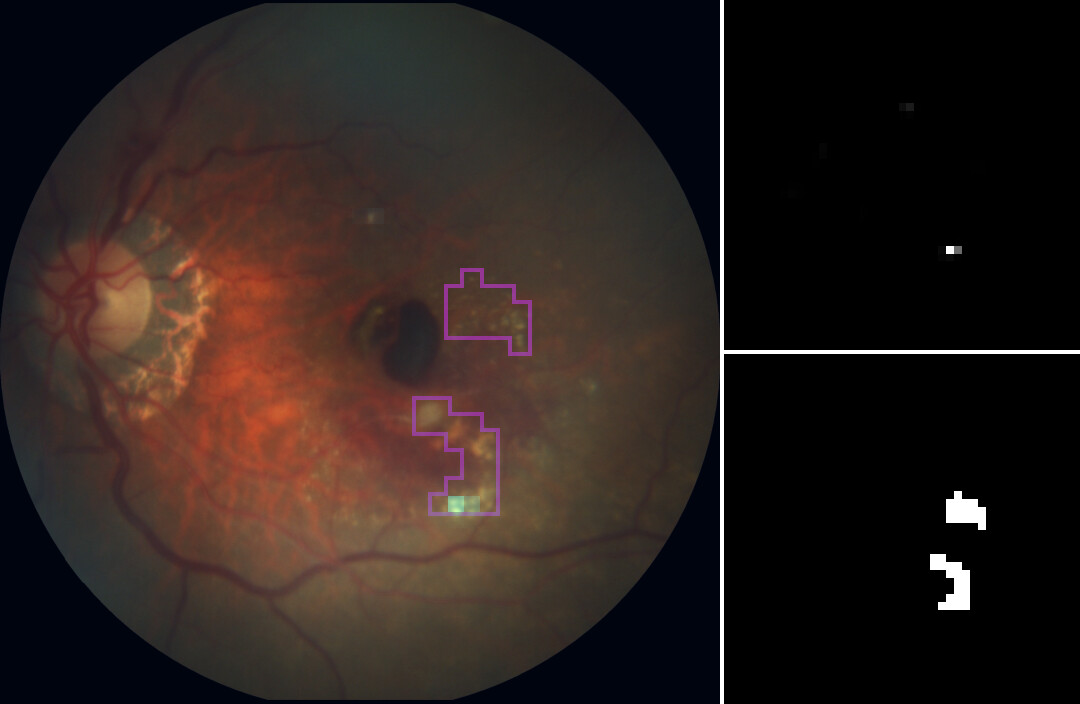}}
    \hfill
    \subfloat[Hemorrhage]{\includegraphics[width=0.49\textwidth]
        {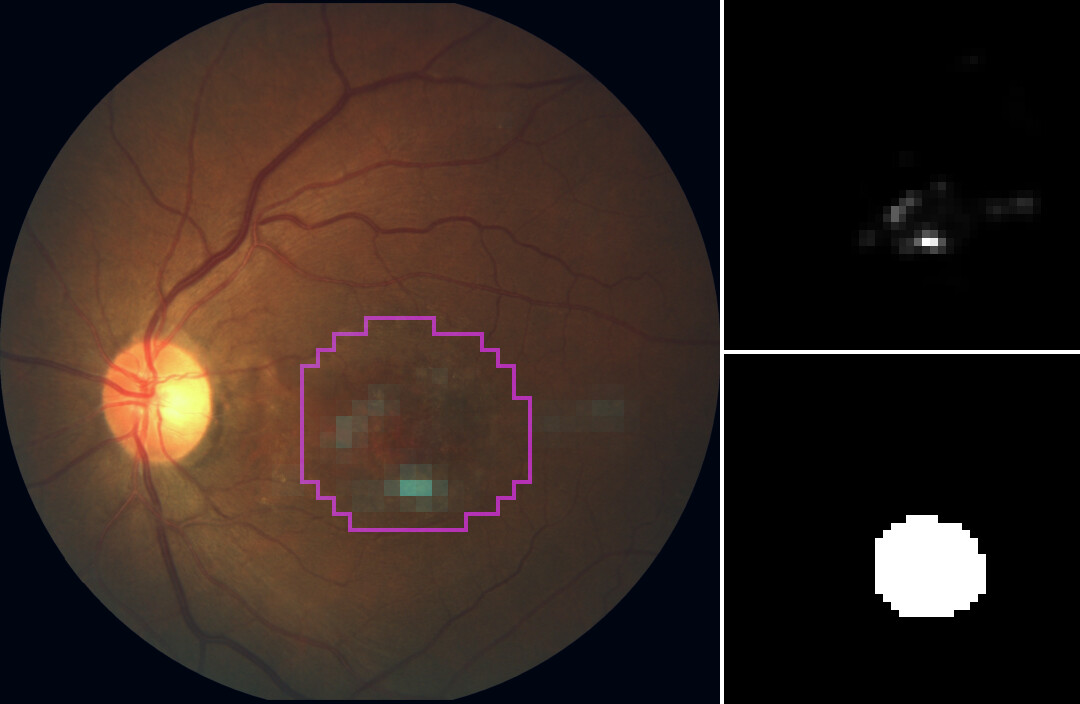}}
    \caption{Examples of lesion activation maps provided by the A+L Max models for multiple ADAM images. In each case, left image depicts the activation map of the lesion from the caption over the original retinography, as well as the contour of the corresponding segmentation ground truth employed in the evaluation (in magenta). Both the activation map (top) and the ground truth (bottom) are depicted separately on the right.}%
    \label{fig:activations_ADAM_Max}
\end{figure}

\begin{figure}[tbp]
    \captionsetup[subfigure]{labelformat=empty}
    \centering
    \textbf{AMDLesions} \\\medskip
    \subfloat[Hemorrhage]{\includegraphics[width=0.49\textwidth]
        {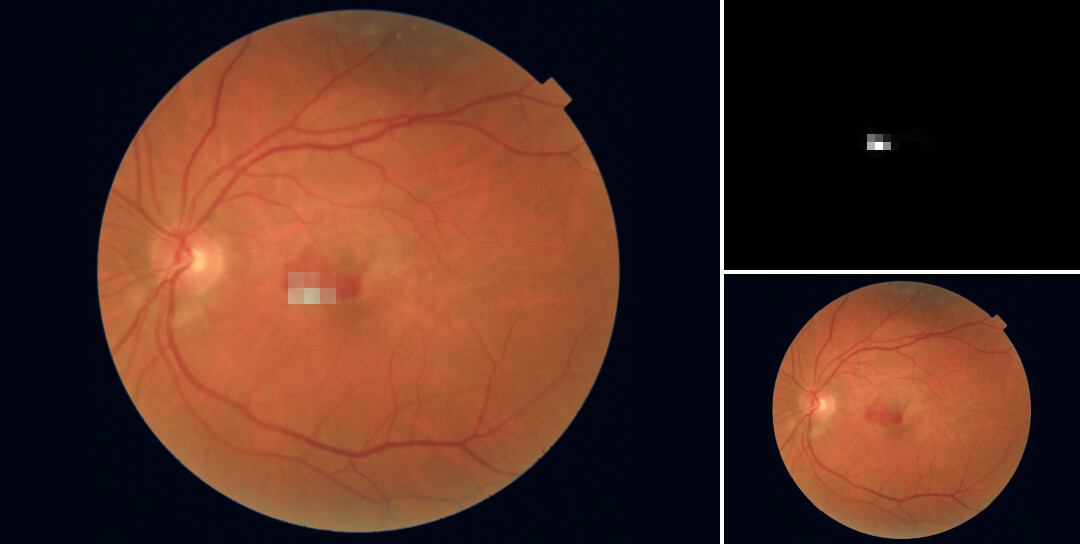}}
    \hfill
    \subfloat[Atrophy]{\includegraphics[width=0.49\textwidth]
        {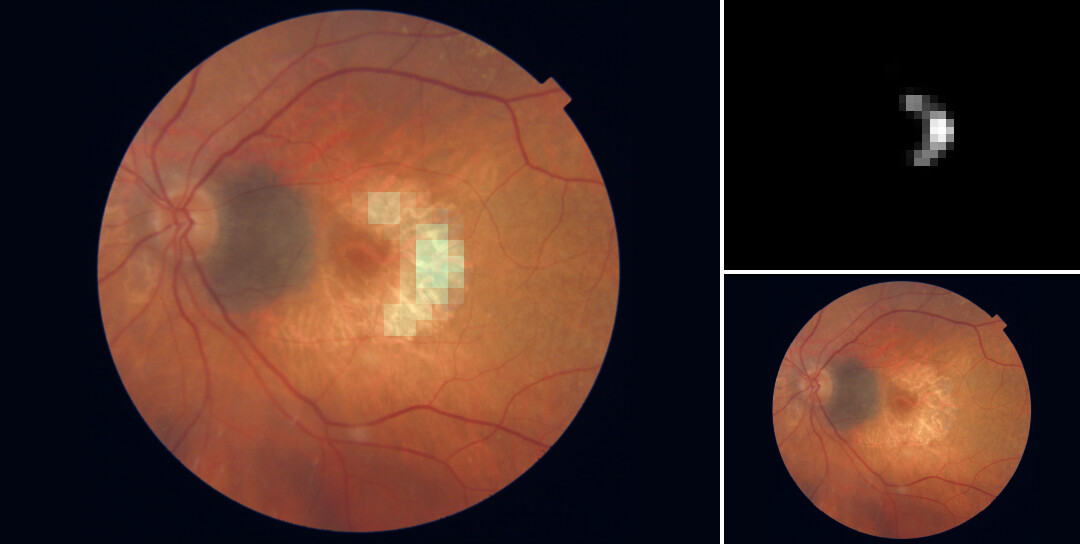}} \\\medskip
    \textbf{ARIA} \\\medskip
    \subfloat[Drusen]{\includegraphics[width=0.49\textwidth]
        {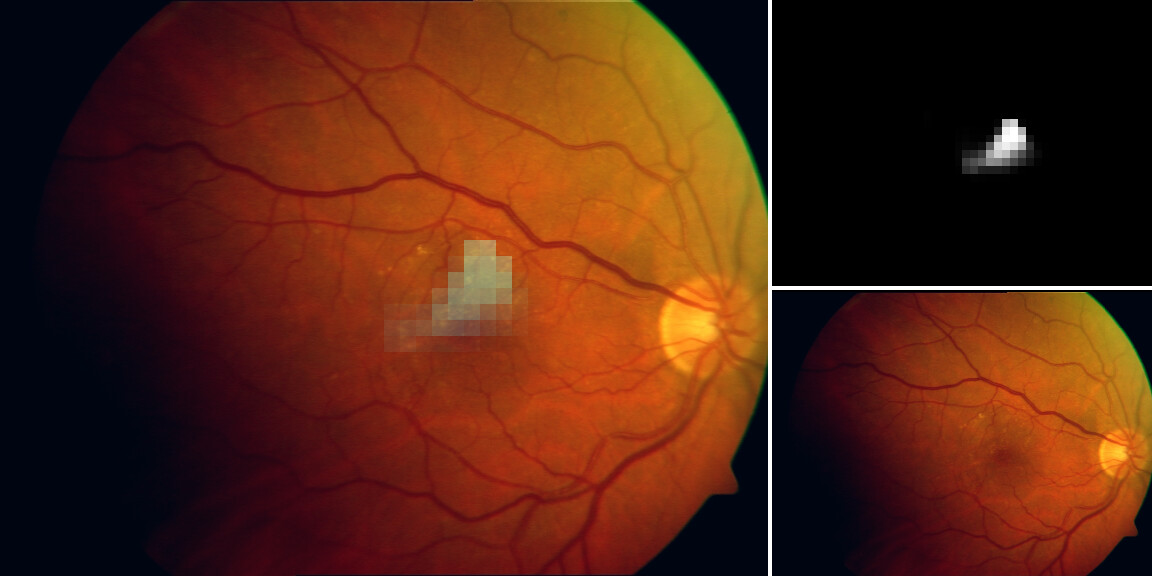}}
    \hfill
    \subfloat[Atrophy]{\includegraphics[width=0.49\textwidth]
        {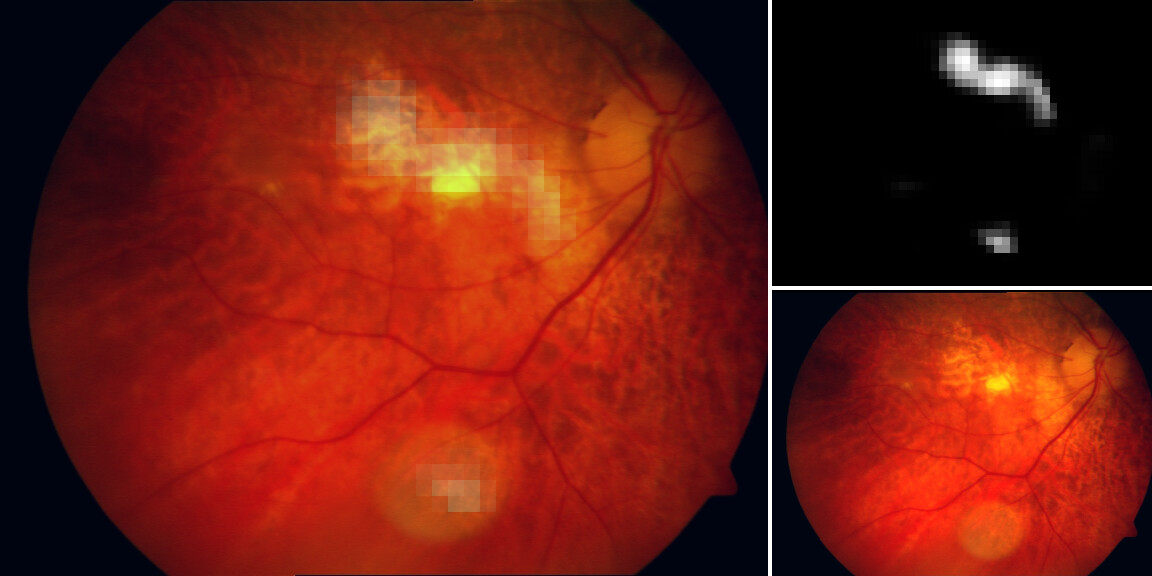}} \\\medskip
    \textbf{STARE} \\\medskip
    \subfloat[Drusen]{\includegraphics[width=0.49\textwidth]
        {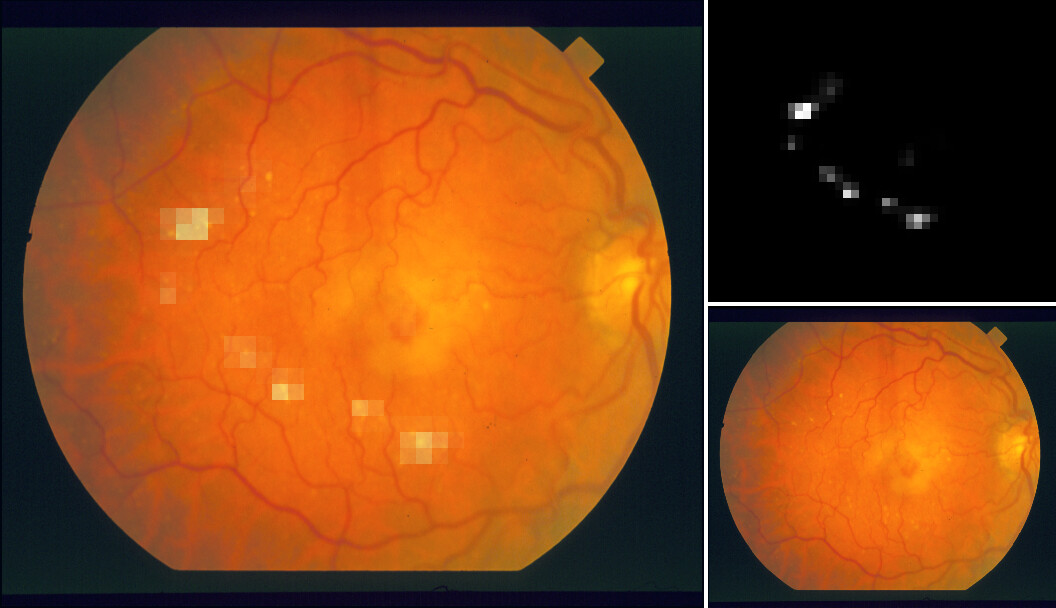}}
    \hfill
    \subfloat[PA]{\includegraphics[width=0.49\textwidth]
        {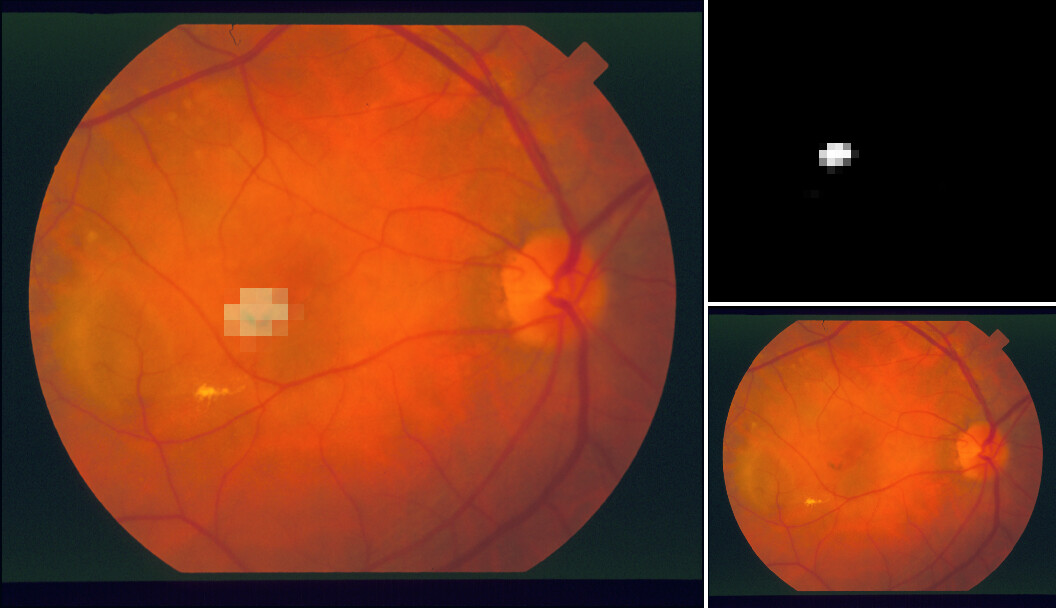}}
    \caption{Examples of lesion activation maps provided by the A+L FC models for various AMDLesions, ARIA and STARE images. In each case, left image depicts the activation map of the lesion from the caption over the original retinography. Both the activation map (top) and the original image (bottom) are depicted separately on the right.}%
    \label{fig:activations_only_FC}
\end{figure}

\begin{figure}[tbp]
    \captionsetup[subfigure]{labelformat=empty}
    \centering
    \textbf{AMDLesions} \\\medskip
    \subfloat[Hemorrhage]{\includegraphics[width=0.49\textwidth]
        {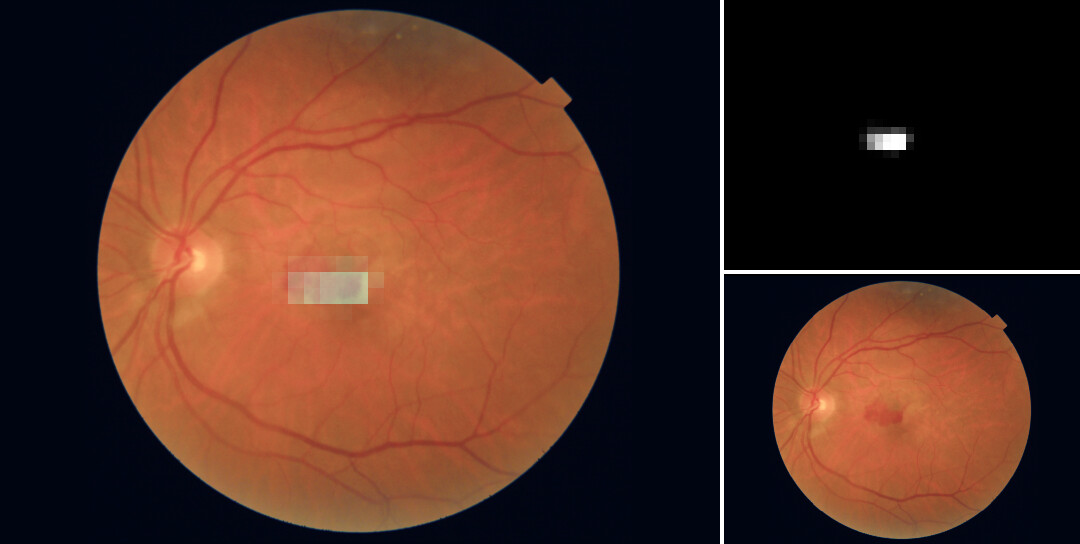}}
    \hfill
    \subfloat[Atrophy]{\includegraphics[width=0.49\textwidth]
        {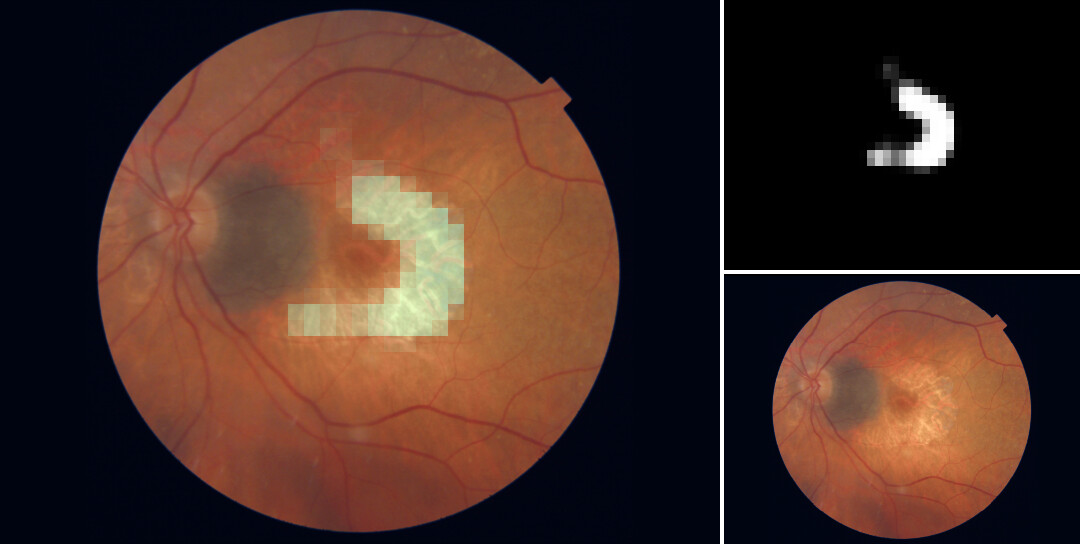}} \\\medskip
    \textbf{ARIA} \\\medskip
    \subfloat[Drusen]{\includegraphics[width=0.49\textwidth]
        {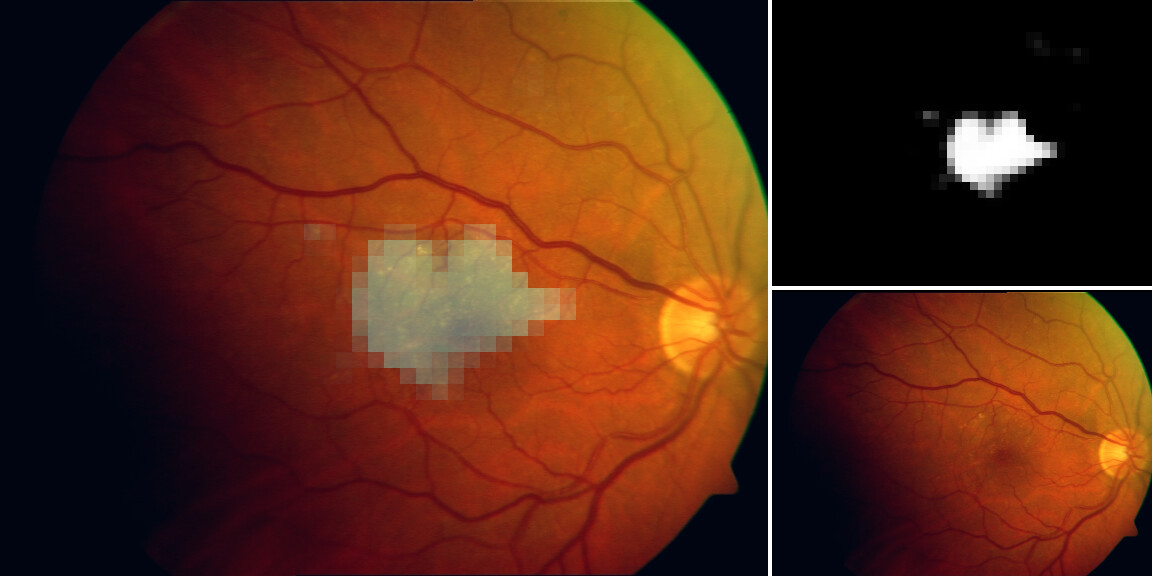}}
    \hfill
    \subfloat[Atrophy]{\includegraphics[width=0.49\textwidth]
        {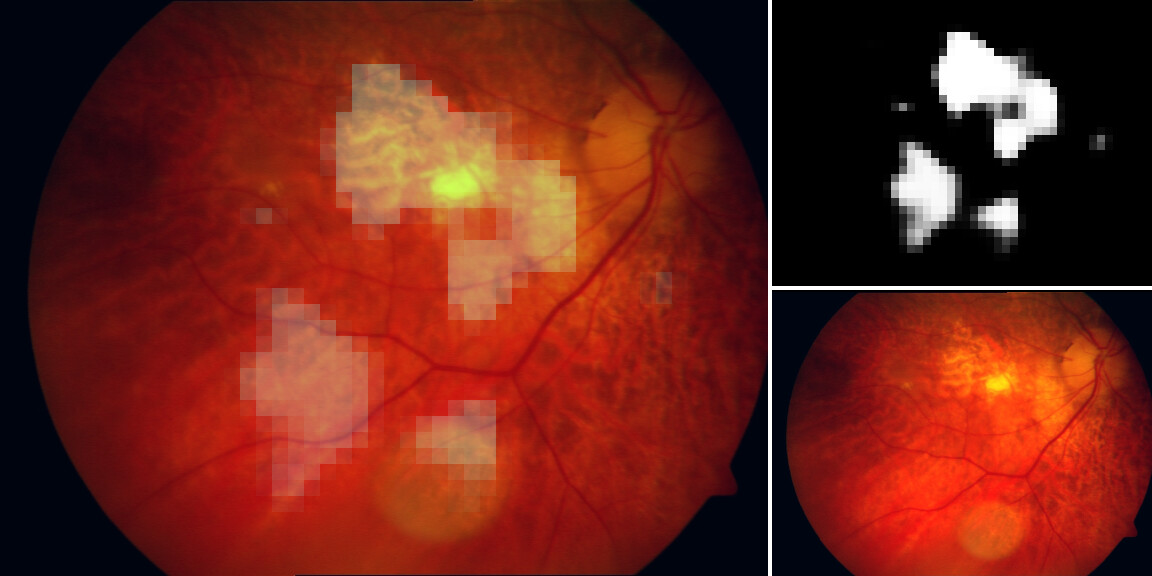}} \\\medskip
    \textbf{STARE} \\\medskip
    \subfloat[Drusen]{\includegraphics[width=0.49\textwidth]
        {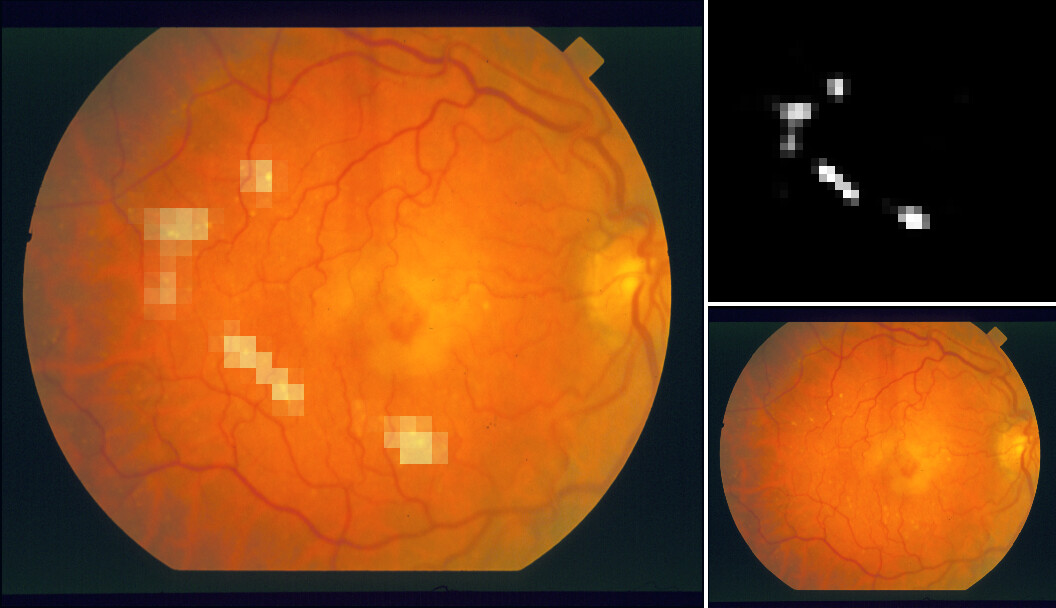}}
    \hfill
    \subfloat[PA]{\includegraphics[width=0.49\textwidth]
        {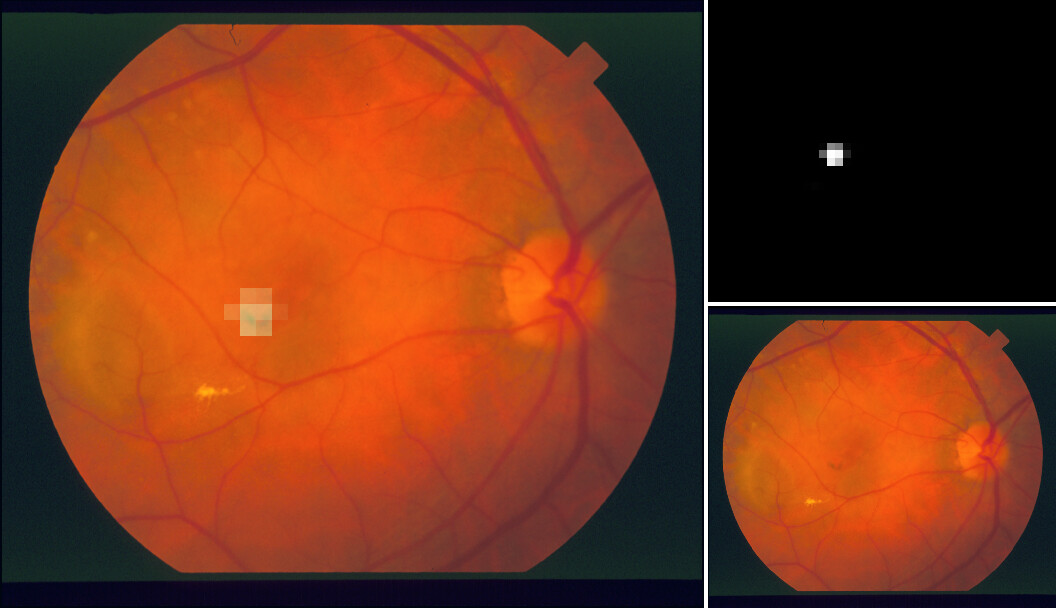}}
    \caption{Examples of lesion activation maps provided by the A+L Max models for various AMDLesions, ARIA and STARE images. In each case, left image depicts the activation map provided by the model for the lesion from the caption over the original retinography. Both the activation map (top) and the original image (bottom) are depicted separately on the right.}%
    \label{fig:activations_only_Max}
\end{figure}

Along with the quantitative results, we present several examples of the lesion activation maps provided by the A+L models for the different datasets.
Figures~\ref{fig:activations_ADAM_FC} and~\ref{fig:activations_ADAM_Max} present some examples from ADAM for A+L FC and A+L Max, respectively.
Additionally, Figures~\ref{fig:activations_only_FC} and~\ref{fig:activations_only_Max} present some examples from AMDLesions, ARIA and STARE for both variants---FC and Max, respectively.

The quantitative results from Figure~\ref{fig:ROCs_Lesions_segmentation_ADAM} and Table~\ref{tab:AUCs_lesion_segmentation_ADAM} show that, in general, the models present a satisfactory performance.
Also, they clearly show that the best results in coarse segmentation are always achieved for drusen.
As in the case of lesion identification, it is very likely that the difference in performance between drusen and the rest of the lesions is due to the scarcity of training data for the latter.
In particular, in the training dataset (AMDLesions) there are 374 images for drusen, while there are only 10, 29 and 11 images for exudates, hemorrhage and `others', respectively. Moreover, it is worth noting that we use 4-fold cross validation, so that the number of effective training samples is even more reduced.
Regarding the high intra-class variability of `others', in this case it does not seem to penalize the performance of this class in comparison to exudates or hemorrhage.
In fact, the lowest AUC-ROC in coarse segmentation is achieved for hemorrhage instead.

\begin{figure}[tbp]
    \captionsetup[subfigure]{labelformat=empty}
    \centering
    \subfloat[Drusen]{\includegraphics[height=0.38\textwidth]
        {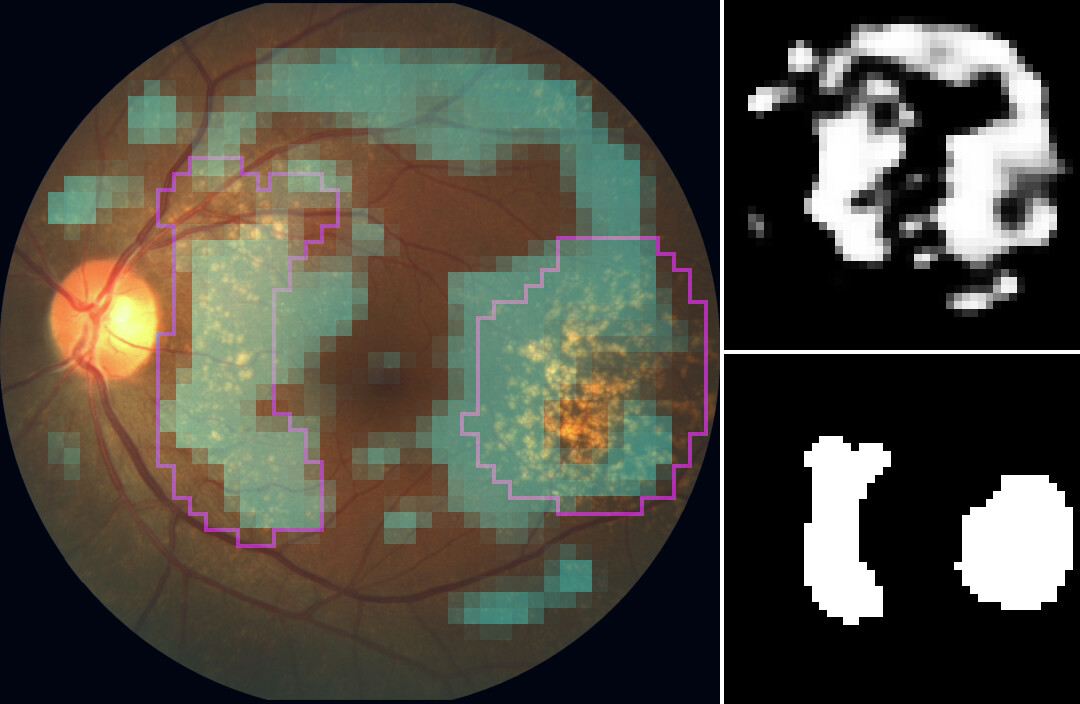}}
    \hfill
    \subfloat[Drusen]{\includegraphics[height=0.38\textwidth]
        {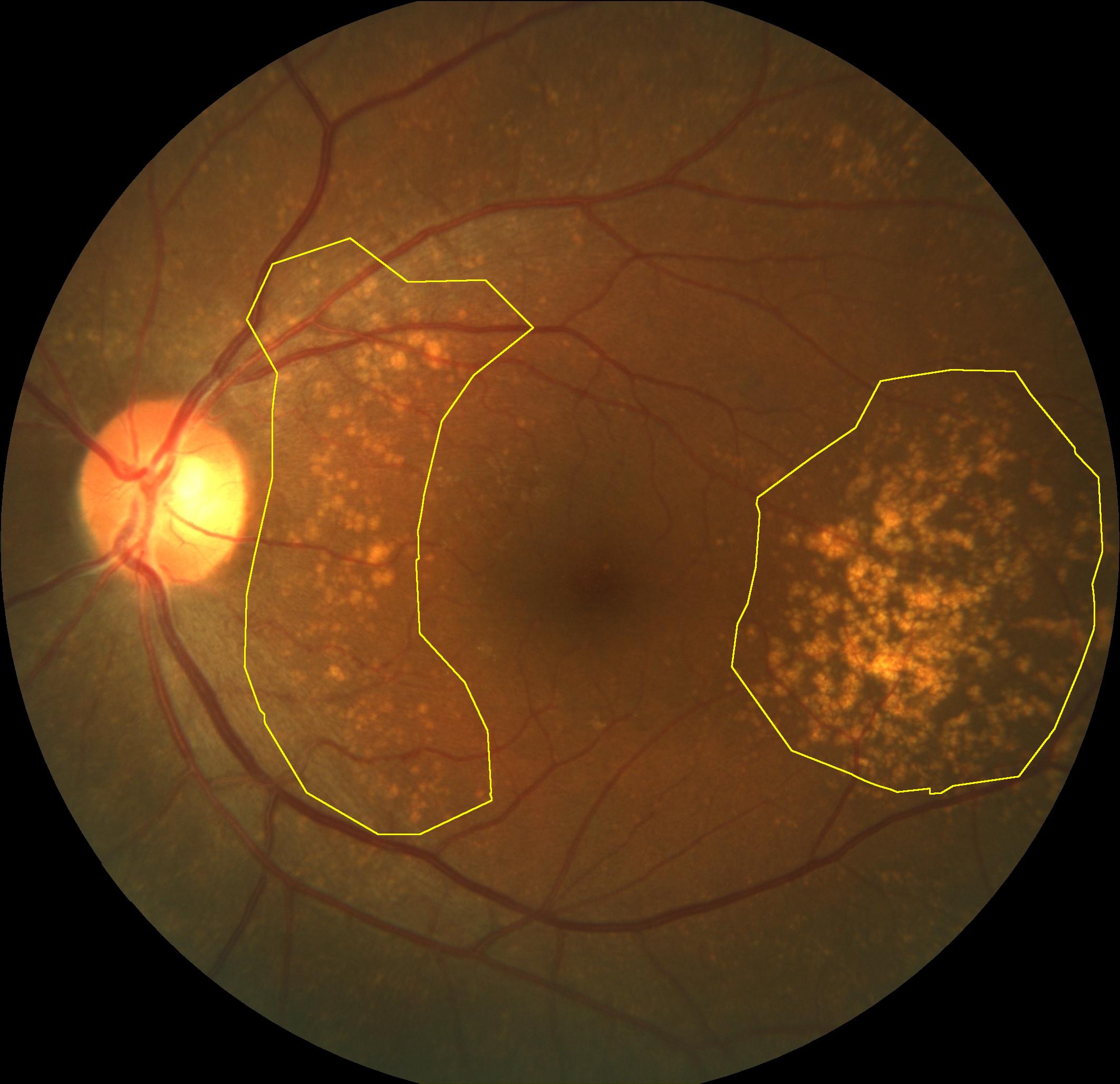}} \\\bigskip
    \subfloat[Drusen]{\includegraphics[height=0.38\textwidth]
        {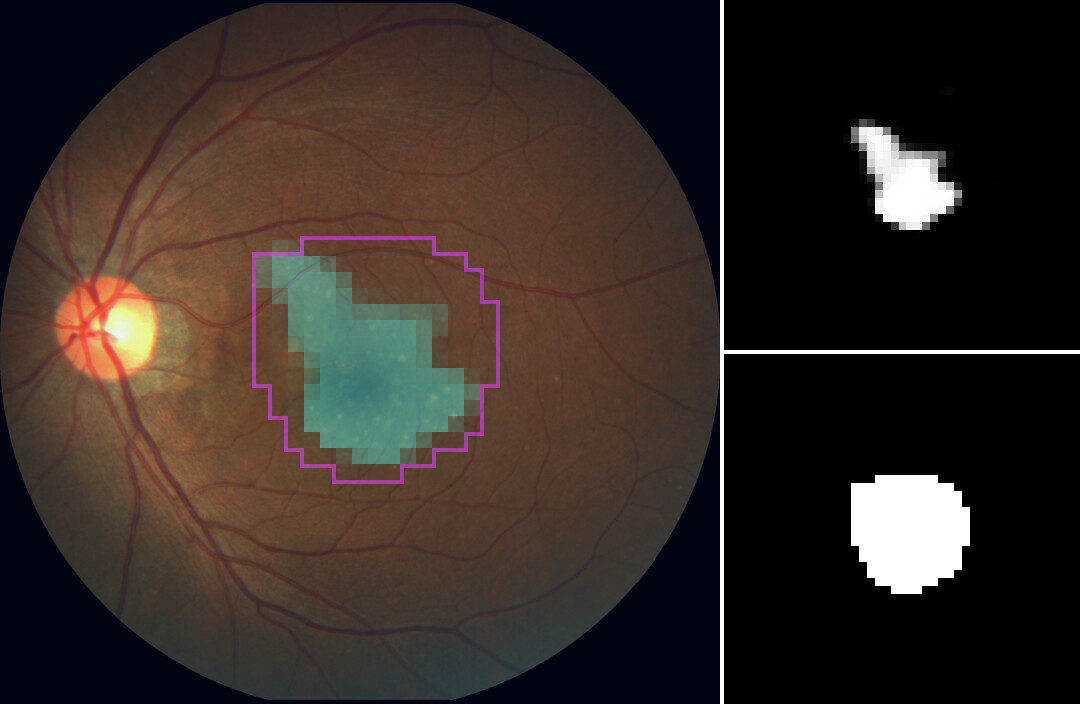}}
    \hfill
    \subfloat[Drusen]{\includegraphics[height=0.38\textwidth]
        {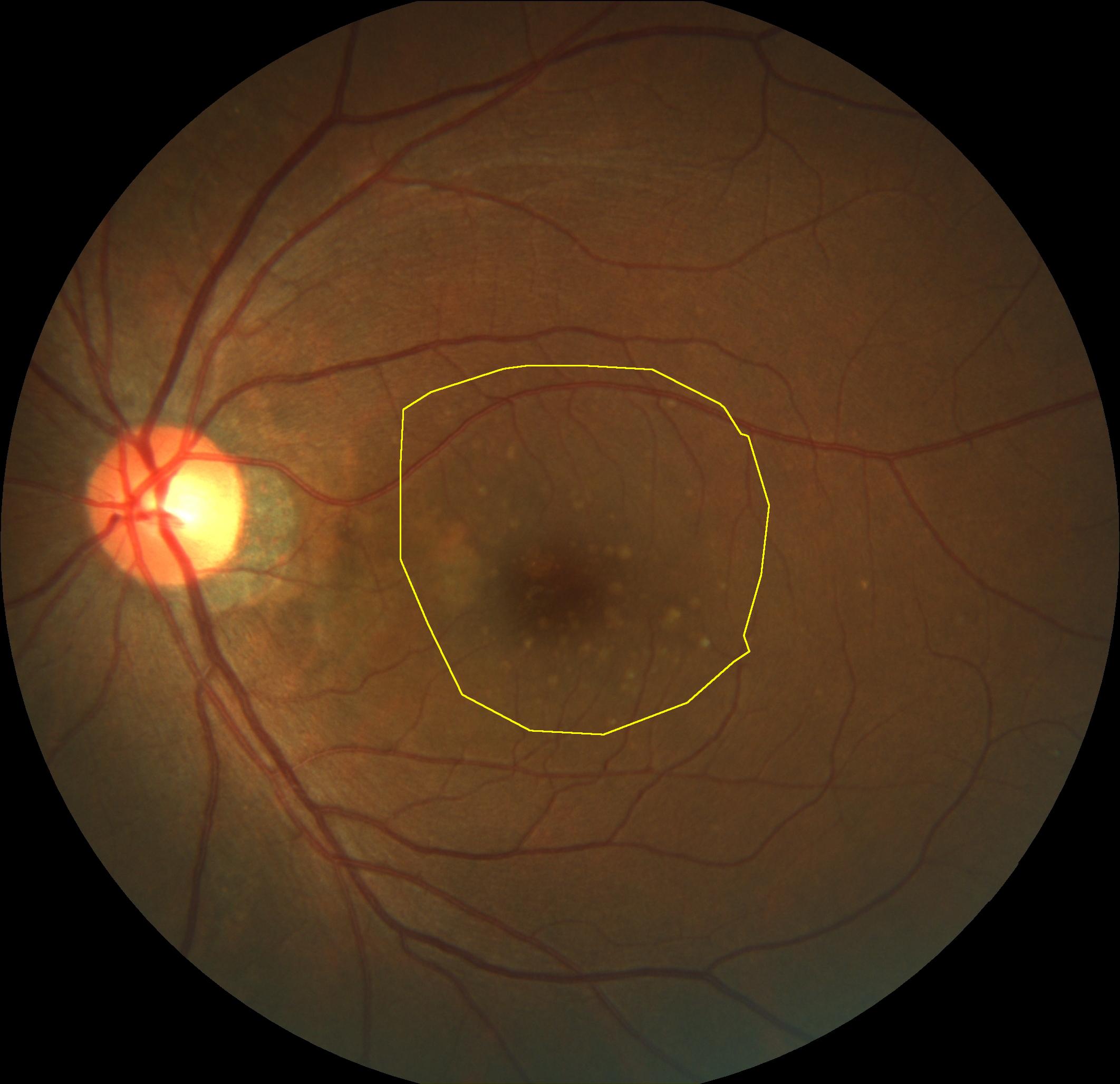}}
    \caption{Examples of lesion activation maps provided by the A+L Max models for 2 ADAM images. In each case, left image depicts the activation map of the lesion from the caption over the original retinography, as well as the contour of the corresponding segmentation ground truth employed in the evaluation (in magenta). Both the activation map (top) and the ground truth (bottom) are depicted separately on the right. Additionally, we include the same retinography with the contour of the original lesion segmentation ground truth overlaid.}%
    \label{fig:imprecise_GT}
\end{figure}

Lastly, there is one additional factor that can potentially penalize all the lesions in the segmentation evaluation: the low exactitude of the manual annotations of ADAM.
Figure~\ref{fig:imprecise_GT} depicts further examples from ADAM of the coarse lesion activation maps obtained by the A+L Max variant, along with the original manual annotations.
For the top image, the A+L model has detected many lesions outside the manually annotated areas.
However, despite not being marked in the ground truth, these detections are arguably correct.
Further examples of this type can be seen in the left images of Figure~\ref{fig:activations_ADAM_FC}.
The bottom example of Figure~\ref{fig:imprecise_GT} represents the opposite scenario.
In this case, the entire area near the macula has been labeled as hemorrhage.
However, there are a large number of pixels in that area where the lesion is not really discernible.
Thus, the coarse segmentation map predicted by the A+L model does not fill the area specified in the ground truth.
This is also the case with the bottom-right images of Figures~\ref{fig:activations_ADAM_FC} and~\ref{fig:activations_ADAM_Max}.
These two circumstances are found in several images of the dataset, penalizing the quantitative results herein presented.

Notwithstanding, the qualitative evaluation of the lesion activation maps proves that the models are able to locate lesion areas of several images in an approximate way.
This is achieved by using only image-level labels.
Some examples of satisfactory weakly-supervised segmentation maps are shown in Figures~\ref{fig:activations_ADAM_FC} and \ref{fig:activations_ADAM_Max}.
Additionally, these figures also reflect a limitation of the proposed method.
Given that the lesion predictions are directly generated by applying a GMP operation on the activation maps, a high activation in one single point of a map is enough to successfully mark the presence of the corresponding lesion.
Consequently, a single high activation can significantly reduce the lesion identification error for a positive sample.
This means that the network is not guided to detect complete lesion areas during training.
This effect is clearly visible in the bottom images of Figure~\ref{fig:activations_ADAM_Max}.
Although the identification error for these images is low, the lesion area that is activated is far from complete.
This effect is more frequent for lesions with few training examples.

In AMDLesions, ARIA and STARE, since there is no available lesion segmentation ground truth, a quantitative evaluation is not possible.
However, the qualitative analysis of the lesion activation maps provided by the A+L models (see Figures~\ref{fig:activations_only_FC} and~\ref{fig:activations_only_Max}) show that the lesion areas detected by these models are habitually correct.

In summary, the experiments and the evaluations conducted demonstrate that the proposed A+L approach enables the identification of AMD and its associated lesions with satisfactory performance.
Furthermore, the evaluation of the coarse segmentation maps of the individual lesions---directly derived from the lesion activation maps---clearly indicate that the explanations provided by the models are meaningful, as they usually locate the pathological areas within the image correctly.
All these outputs are achieved using only image-level lesion labels relatively easy to obtain.
This is particularly convenient, since in medical imaging, due to the difficulty of annotations, the scarcity of annotated data is especially pronounced.
Also, with the proposed setting, the AMD diagnosis is directly derived from the identified lesions, and these, from the lesion activation maps, which highly enhances the explainability of the learned model.

\section{Conclusions}%
\label{sec:conclusions}

In this work, we have proposed an explainable deep learning approach for the simultaneous identification of AMD and its associated retinal lesions in color fundus images.
The proposed approach uses an slightly adapted CNN that directly links the predicted diagnosis to the identified lesions and allows the generation of weakly-supervised lesion activation maps.
With the proposed setting, the lesion predictions derive directly from the lesion activation maps, and therefore also the final diagnosis.
Thus, both the lesion predictions and the diagnosis can be explained by the lesion activation maps.
This setting is highly intuitive, as it mimics the manual process followed by clinicians, consisting of localizing and then classifying the retinal lesions.
Furthermore, it is not dependent on the network architecture, and can be applied, with minor modifications, over any CNN for image classification.

The complementary lesion information, in addition to provide an explanation for the decisions of the model, can also be used by the clinicians to assess the severity of AMD, as it provides the location and classification of the retinal lesions.
This approach represents an important advance with respect to the current state-of-the-art approaches for AMD diagnosis, which are focused solely on screening and do not incorporate any explainability mechanisms.
This highly limits the applicability of previous methods.
Additionally, the proposed method is the first that simultaneously obtains lesion predictions, diagnostic predictions and lesion-specific activation maps using only image-level labels.
In this regard, in contrast to previous works exploring explainability mechanisms for the diagnosis of retinal diseases, our proposal presents the advantage of providing lesion-specific activation maps instead of global activation maps.

To validate our proposal, we collected a private dataset of color fundus images with expert-annotated labels for the diagnosis of AMD and the presence of its associated retinal lesions (AMDLesions).
We performed an exhaustive experimentation in this and other three additional public datasets: ADAM, ARIA and STARE.
The trained networks were evaluated for three different tasks: diagnosis of AMD, lesion identification and coarse lesion segmentation.
This last evaluation aimed to validate the quality of the visual explanations provided by the lesion activation maps.
For the diagnosis of AMD, we compared our approach (A+L) with the baseline approach (A-O), which is solely focused on the identification of AMD and uses a standard classification CNN.
In addition, we compared the AMD identification performance of the proposed approach with that of several state-of-the-art methods in the ADAM reference dataset.
The methods that were compared are focused solely on AMD identification.
The evaluation in the four different datasets demonstrates that the proposed approach provides satisfactory results in the identification of AMD and its associated lesions.
Furthermore, the comparison with the state-of-the-art methods in AMD identification shows that the results of A+L models are highly competitive, while the models are much more explainable and provide extra useful outputs.
The information resulting from lesion identification, along with the lesion activation maps, conveniently complements the diagnosis, and it is useful to better understand the decision made by the model.
What is also relevant, the collection of the training data that is needed for the approach does not imply much extra effort from clinicians, since the identification of lesions can be habitually found in the medical records.
This is because the lesion identification is part of the diagnostic process in the clinical practice.
In light of the results herein presented, we think that the proposed methodology makes relevant advances in terms of explainability, and that it could be successfully applied in several diagnostic scenarios.
An example could be the diagnosis of diabetic retinopathy. This disease, like AMD, is also frequently diagnosed by color fundus imaging, and it is characterized by the presence of multiple lesions of different types.

Notwithstanding, our approach presents two main points for further improvement.
First, the generation of the activation maps.
With the proposed approach, the network has no incentive to activate large lesion areas, resulting in incomplete activation maps.
It is very likely that the addition of such an incentive would greatly mitigate this issue.
Second, the computation of the diagnosis from the lesions.
The proposed setting is valid for single-pathology studies.
However, it would be interesting to extend it to multi-pathology studies, more similar to real screening scenarios.
Both issues represent interesting fields for further research.

\section*{Declaration of competing interest}

The authors declare that they have no known competing financial interests or personal relationships that could have appeared to influence the work reported in this paper.

\section*{Acknowledgments}

This work was funded by Instituto de Salud Carlos III, Government of Spain, and the European Regional Development Fund (ERDF) of the European Union (EU) through the \mbox{DTS18/00136} research project;
Ministerio de Ciencia e Innovación, Government of Spain, through \mbox{RTI2018-095894-B-I00} and \mbox{PID2019-108435RB-I00} research projects;
Axencia Galega de Innovación (GAIN), Xunta de Galicia, ref. \mbox{IN845D 2020/38};
%
%
Consellería  de  Cultura,  Educación e Universidade, Xunta de Galicia, through Grupos de Referencia Competitiva, ref. \mbox{ED431C 2020/24}, 
the predoctoral grant ref. \mbox{ED481A 2021/140}, and the postdoctoral grant ref. \mbox{ED481B-2022-025};
CITIC, Centro de Investigación de Galicia ref. \mbox{ED431G 2019/01}, is funded by Consellería de Educación, Universidade e Formación Profesional, Xunta de Galicia, through the ERDF (80\%) and Secretaría Xeral de Universidades (20\%).


\bibliography{references}

\end{document}